\documentclass[preprint,11pt]{elsarticle}
\usepackage{deluxetable}
\usepackage{amssymb}
\usepackage{enumitem}

\newcommand{\be}{\begin{equation}}
\newcommand{\ee}{\end{equation}}
\newcommand{\ba}{\begin{eqnarray}}
\newcommand{\ea}{\end{eqnarray}}

\journal{Physics Reports}

\topmargin = -\topskip
\advance\topmargin by -\headsep
\begin{document}

\begin{frontmatter}

\title{Holographic Dark Energy}

\author[1]{Shuang Wang\corref{*}}
\ead{wangshuang@mail.sysu.edu.cn}
\cortext[*]{corresponding author}
\author[2]{Yi Wang\corref{**}}
\ead{phyw@ust.hk}
\cortext[**]{co-corresponding author}
\author[1]{Miao Li}
\ead{limiao9@mail.sysu.edu.cn}
\address[1]{School of Physics and Astronomy, Sun Yat-Sen University,\\ Guangzhou 510275, P. R. China}
\address[2]{Department of Physics, The Hong Kong University of Science and Technology,\\ Clear Water Bay, Kowloon, Hong Kong, P. R. China}

\begin{abstract}
We review the paradigm of holographic dark energy (HDE),
which arises from a theoretical attempt of applying the holographic principle (HP) to the dark energy (DE) problem.
Making use of the HP and the dimensional analysis, we derive the general formula of the energy density of HDE.
Then, we describe the properties of HDE model, in which the future event horizon is chosen as the characteristic length scale.
We also introduce the theoretical explorations and the observational constraints for this model.
Next, in the framework of HDE, we discuss various topics,
such as spatial curvature, neutrino, instability of perturbation, time-varying gravitational constant, inflation, black hole and big rip singularity.
In addition, from both the theoretical and the observational aspects, we introduce the interacting holographic dark energy scenario,
where the interaction between dark matter and HDE is taken into account.
Furthermore, we discuss the HDE scenario in various modified gravity (MG) theories,
such as Brans-Dicke theory, braneworld theory, scalar-tensor theory, Horava-Lifshitz theory, and so on.
Besides, we introduce the attempts of reconstructing various scalar-field DE and MG models from HDE.
Moreover, we introduce other DE models inspired by the HP, in which different characteristic length scales are chosen.
Finally, we make comparisons among various HP-inspired DE models, by using cosmological observations and diagnostic tools.
\end{abstract}


\end{frontmatter}


\tableofcontents

\vfill\eject

\

\section{Introduction}
\label{sec:1}

Inspired by the investigation of black hole thermodynamics \cite{Bekenstein1973,Hawking:1974sw},
Gerard 't Hooft proposed for the first time the famous holographic principle (HP) \cite{Hooft1993}.
As a modern version of ``Plato's cave'',
the HP states that all of the information contained in a volume of space can be represented as a hologram,
which corresponds to a theory locating on the boundary of that space.
Soon after, Leonard Susskind gave a precise string-theory interpretation of this principle \cite{Susskind1995}.
Moreover, in 1997, Juan Maldacena \cite{Maldacena:1997re} proposed the famous AdS/CFT correspondence,
which is the most successful realization of the HP.
Now it is widely believed that the HP should be a fundamental principle of quantum gravity.

So far, the idea of HP has been applied to various fields of physics.
For examples, in the field of nuclear physics,
the AdS/QCD correspondence has been proposed to study the problems of quark-gluon plasma \cite{Liu:2006he};
in the field of condensed matter physics,
the AdS/CMT correspondence has been proposed to explore the problems of superconductivity and superfluid \cite{Hartnoll:2009sz};
in the field of theoretical physics, the AdS/CFT correspondence has lead to the idea of holographic entanglement entropy \cite{Takayanagi:2012kg};
in the field of cosmology, the AdS/CFT correspondence has also been used to discuss the nature of de Sitter space and inflation \cite{Strominger:2001pn}.
These scientific progresses show that the HP has great potential to solve many long-standing issues in various physical fields.

On the other side, since its discovery in 1998 \cite{Riess:1998cb,Perlmutter:1998np},
dark energy (DE) has become one of the central problems in theoretical physics and modern cosmology.
The current observations favor a cosmological constant $\Lambda$
being the origin of DE driving the present epoch of the accelerated expansion of our universe
and a dark matter (DM) component giving rise to galaxies and their large scale structures (LSS) distributions.
This cosmological model is called $\Lambda$CDM, to indicate the nature of its main components
\footnote{The DM is often assumed to have negligible pressure and temperature and is termed Cold.
That is why it is always called Cold Dark Matter (CDM).}.
Although favored by the observations, the $\Lambda$CDM model suffers from two cosmological constant problems
\cite{Peebles:2002gy,Padmanabhan:2002ji,Copeland:2006wr,Frieman:2008sn,Caldwell:2009ix,Silvestri:2009hh,Li:2011sd,Bamba:2012cp,Li:2012dt}.
To solve these theoretical puzzles, numerous DE models have been proposed in the past 18 years;
unfortunately, so far the nature of DE still remains a complete mystery.
It is believed that the DE problem may be in essence an issue of quantum gravity.
Therefore, as the most fundamental principle of quantum gravity,
HP may play an important role in solving the DE problem.

In 2004, after applying the HP to the DE problem,
one of the present authors (Miao Li) proposed a new DE model, called holographic dark energy (HDE) model \cite{Li:2004rb}.
In this model, the energy density of DE $\rho_{de}$ only relies on two physical quantities on the boundary of the universe:
one is the reduced Planck mass $M_p \equiv \sqrt{1/8\pi G}$, where $G$ is the Newton constant;
another is the cosmological length scale $L$, which is chosen as the future event horizon of the universe \cite{Li:2004rb}.
The HDE model is the first theoretical model of DE inspired by the HP,
and is in good agreement with the current cosmological observations at the same time.
This makes HDE a very competitive candidate of DE.

In recent ten years, the paradigm of HDE has drawn a lot of attention and has been widely studied:
\begin{enumerate}
\item Many theoretical mechanisms, such as entanglement entropy, holographic gas, Casmir energy, entopic force and action principle,
are proposed to explain the origin of HDE;
\item A series of topics, including spatial curvature, neutrino, perturbation, time-varying gravitational constant,
black hole, inflation, and big rip, are discussed in the HDE cosmology;
\item The interaction between DM and HDE are studied from both the theoretical and the observational sides;
\item The properties of HDE are revisited in various modified gravity theories,
such as the Brans-Dicke, the scalar-tensor, the DGP, the braneworld and the Horava-Lifshitz theories;
\item The HDE are used to reconstruct various scalar-field DE and modified gravity models;
\item A series of theoretical models have also been proposed, where different forms of $L$ are taken into account;
\item A lot of efforts have been made to constrain these HDE models by using various cosmological observations.
\end{enumerate}

\begin{table*}
\centering
\caption{List of commonly used acronyms}
\label{tab:1}
\centering
\begin{tabular}{cc}
\hline
Acronym & Meaning \\
\hline

$\Lambda$CDM   & Cosmological Constant $\Lambda$ + Cold Dark Matter \\
ADE            & Agegraphic Dark Energy \\
AIC            & Akaike Information Criterion \\
AP             & Alcock-Packzynki \\
BAO            & Baryon Accoustic Oscillation \\
BIC            & Bayesian Information Criterion \\
BE             & Bayesian Evidence \\
CL             & Confidence Level \\
CMB            & Cosmic Microwave Background \\
CPL            & Chevalliear-Polarski-Linder \\
DE             & Dark Energy \\
DGP            & Dvali-Gabadadze-Porrati \\
DM             & Dark Matter \\
EoS            & Equation of State \\
FA             & Flux Averaging \\
FLRW           & Friedmann-Lema\^{\i}tre-Robertson-Walker \\
GC             & Galaxy Clusters \\
GR             & General Relativity \\
GRB            & Gamma Ray Burst \\
GW             & Gravitational Wave \\
HDE            & Holographic Dark Energy \\
HP             & Holographic Principle \\
IDE            & Interacting Dark Energy \\
IHDE           & Interacting Holographic Dark Energy \\
IR             & Infrared \\
JLA            & Joint Light-curve Analysis \\
LCF            & light-curve fitters \\
LSS            & Large Scale Structure \\
MCMC           & Markov Chain Monte-Carlo \\
MG             & Modified Gravity \\
RDE            & Ricci Dark Energy \\
RSD            & Redshift Space Distortion \\
SL             & Sandage-Loeb \\
SN             & Supernova \\
SNIa           & Type Ia Supernova \\
UV             & Ultraviolet  \\
wCDM           & Constant w + Cold Dark Matter \\
WL             & Weak Lensing \\

\hline

\end{tabular}
\end{table*}


\begin{table*}
\centering
\caption{List of observational facilities}
\label{tab:2}
\centering
\begin{tabular}{cc}
\hline
Acronym & Meaning \\
\hline

BOSS           & Baryon Oscillation Spectroscopic Survey \\
COBE           & Cosmic Background Explorer \\
DES            & Dark Energy Survey \\
ESSENCE        & Equation of State: Supernovae trace Cosmic Expansion \\
Euclid         & Euclid Satellite \\
HST            & Hubble space telescope \\
LSST           & Large Synoptic Survey Telescope \\
Pan-STARSS     & Panoramic Survey Telescope and Rapid Response System \\
Planck         & Planck Satellite \\
SDSS           & Sloan Digital Sky Survey \\
SKA            & Square Kilometre Array \\
SNLS           & Supernova Legacy Survey \\
WFIRST         & Wide Field Infrared Survey Telescope \\
WMAP           & Wilkinson Microwave Anisotropy Probe \\

\hline

\end{tabular}
\end{table*}

In this paper, we will review all the topics mentioned above.
For the convenience of readers, we list all the commonly used acronyms and observational facilities in table \ref{tab:1} and table \ref{tab:2}.
Throughout the review, we assume today's scale factor $a_0 = 1$, so the redshift $z = a^{-1} - 1$;
the subscript ``0'' always indicates the present value of the corresponding quantity,
and we use the metric convention ($-$,+,+,+) and the natural units $c=\hbar=1$.

\

\section{Basic Knowledge of Modern Cosmology}
\label{sec:2}

In this section we briefly introduce the basic knowledge of modern cosmology,
such as the Friedmann-Lema\^{\i}tre-Robertson-Walker (FLRW) cosmology, the DE, and the cosmic probes of DE.

\

\subsection{FLRW Cosmology}
\label{sec:2.1}

There are two cornerstones for modern cosmology.
The first is Einstein's general relativity (GR), which tell us that the Einstein equation
\be \label{eq:Ein}
  G_{\mu\nu} \equiv R_{\mu\nu}-\frac{1}{2}g_{\mu\nu}R = 8\pi G T_{\mu\nu}
\ee
is valid on cosmological scale.
Here $G_{\mu\nu}$ is the Einstein tensor, $R_{\mu\nu}$ is the Ricci tensor, $R$ is the Ricci scalar,
$g_{\mu\nu}$ is the metric, and $T_{\mu\nu}$ is the energy-momentum tensor.
For the perfect fluid, $T_{\mu\nu} = (\rho+p)u_{\mu}u_{\nu}+g_{\mu\nu}p$,
where $\rho$ and $p$ are the total energy density and the total pressure of all the components in the universe, respectively.

The second is the cosmological principle (also called Copernican principle),
which tells us that the universe is homogeneous and isotropic on large scales.
Then, the universe can be described by the FLRW metric
\be \label{eq:FLRW}
  ds^2=dt^2-a^2(t)\left[\frac{dr^2}{1-kr^2}+r^2d\Omega_2^2\right].
\ee
Here $t$ is the cosmic time, $a(t)$ is the scale factor,
$r$ is the spatial radius coordinate, $\Omega_2$ is the 2-dimensional unit sphere volume,
and the quantity $k$ characterizes the curvature of 3-dimensional space,
where $k=-1,0,1$ correspond to open, flat and closed universe, respectively.

Inserting Eq. (\ref{eq:FLRW}) to Eq. (\ref{eq:Ein}),
one can derive two Friedmann equations
\be \label{eq:Fri1}
  3M_{p}^{2}H^{2} = \rho - \frac{3M_{p}^{2}k}{a^{2}},
\ee
\be \label{eq:Fri2}
  \frac{\ddot{a}}{a} = -\frac{\rho + 3p}{6M_{p}^{2}}.
\ee
The above dot denotes the derivative with respect to cosmic time $t$.
In addition, $H \equiv \dot{a}/a$ is the Hubble parameter, which describes the expansion rate of the universe.
As seen below, $H$ is an important bridge between the theory and the observation of cosmology.

One can read from the first Friedmann equation that, energy density $\rho$ makes the universe change its volume.
In principle, the change may be either expansion or contraction.
Since the astronomical observations tell us that our universe is expanding \cite{1929PNAS...15..168H},
we will mainly focus on the expansion direction unless otherwise mentioned.
In addition, from the second Friedmann equation, one can see that pressure $p$ affects the acceleration of cosmic expansion:
if $p > -\rho/3$, the universe will decelerate; if $p < -\rho/3$, the universe will accelerate.
It is easy to infer from these two Friedmann equations that,
if all the components in the universe were determined,
the expansion history of our universe can be completely understood.

\

\subsection{Dark Energy}
\label{sec:2.2}

As is well known, DE is a longstanding puzzle in modern cosmology.
In 1917, in order to maintain a static universe,
Einstein added a cosmological constant $\Lambda$ in his field equations of GR \cite{1917SPAW.......142E}.
Afterwards, he removed the cosmological constant term because of the discovery of the cosmic expansion,
and claimed that the introduction of $\Lambda$ is the biggest blunder of his life.
In 1967, Zel'dovich reintroduced the cosmological constant by taking the vacuum fluctuations into account \cite{1967JETPL...6..316Z}.
In 1989, Weinberg published a review article about the cosmological constant problems \cite{1989RvMP...61....1W},
and divided the theoretical attempts of solving these problems into five categories.
In 1998, Riess {\it et al.} and Perlmutter {\it et al.} discovered that
the universe is expanding at an increasing rate \cite{Riess:1998cb,Perlmutter:1998np}.
This great discovery declares the return of DE.

Now, it is widely believed that our universe mainly contains four components: baryon matter, DM, radiation, and DE.
Thus, the first Friedmann equation can be rewritten as
\be
  H(z)=H_{0}\sqrt{\Omega_{r0}(1+z)^{4}+\Omega_{b0}(1+z)^{3}+\Omega_{dm0}(1+z)^{3}+\Omega_{k0}(1+z)^{2}+\Omega_{de0}X(z)}.
\ee
Here $H_{0} = 100h (km \cdot s^{-1} \cdot Mpc^{-1})$ is the present-day value of the Hubble parameter $H$,
$h$ is the Hubble constant, $z = a^{-1}-1$ is the redshift,
$\Omega_{i0} \equiv \rho_{i0}/\rho_{c0}= \rho_{i0}/3H_{0}^{2}M_{p}^{2}$ denote the present fractional densities of various component,
where the subscript ``r'', ``b'', ``dm'', ``k'' and ``de'' represent
radiation, baryon matter, DM, spatial curvature and DE, respectively.
It is clear that $\sum_{i}\Omega_{i0}=1$.
In addition, the total fractional matter density $\Omega_{m}=\Omega_{b}+\Omega_{dm}$,
the effective energy density of spatial curvature $\rho_{k} \equiv -3M_{p}^{2}k/a^{2}$,
and the DE density function
\be
X \equiv \frac{\rho_{de}(z)}{\rho_{de0}} = \exp\left[3\int_{0}^{z}dz'{1+w(z')\over 1+z'}\right],
\ee
where $w \equiv p_{de}/\rho_{de}$ is the equation of state (EoS) of DE,
which is the most important quantity characterizing the properties of DE
\cite{2003PhRvL..90c1301H,2005PhRvD..71b3506H,Huang:2009rf,Wang:2010vj,2011JCAP...07..011L,Hu:2015ksa}.

The simplest DE model is the $\Lambda$CDM model, which has an EoS $w=-1$.
So far, this model still has the best performance in fitting the current observational data,
and thus has been viewed as the standard model of cosmology.
However, the standard model suffers from two cosmological constant problems \cite{Padmanabhan:2002si,Padmanabhan:2004qc}:
(a) Why $\rho_{\Lambda} \approx 0$, namely why it is so small? This is the so-called fine-tuning problem.
(b) Why $\rho_{\Lambda} \sim \rho_m$ now? This is the so-called coincidence problem.
To explore the problem of DE, thousands of papers have been written on this subject.
Unfortunately, although hundreds of DE models have been proposed in the past 18 year,
the nature of DE is still in the dark.

As mentioned above, in 1989 Weinberg divided the theoretical attempts on DE problem into five categories \cite{1989RvMP...61....1W}:
\begin{enumerate}
\item
Symmetry. There have been many attempts, for example, no-scale supersymmetry \cite{Ellis:1983sf} or complexification of coordinates \cite{tHooft:2006uhw}. However those proposals either still contain some fine tuning in the Lagrangian, or involving exotic symmetries which is not evident in nature.
\item
Anthropic principle. It is assumed that we live in a multiverse, where different energy density of DE can be realized \cite{Dicke:1961gz, Carter:1974zz}. We live with the observed DE density because it allows long enough time for galaxy formation and thus fits for observers to live. Later, the discovery of string landscape \cite{Bousso:2000xa, Susskind:2003kw} seems to support this idea. This explanation is different from a scientific explanation in a traditional way and the existence of multiverse is hard to verify.
\item
Tuning mechanisms. In this class of models a scalar field is introduced which can dynamically reduce the density of DE. Simple attempts of this kind results in vanishing Newton's constant and thus not desired \cite{Dolgov:1982gh}. Recent progress has been made in this direction by making use of a generalized class of scalar-tensor theories \cite{Charmousis:2011bf}.
\item
Modified gravity. Modified gravity can help for the DE problem in many ways. For example, in unimodular gravity \cite{vanderBij:1981ym, Unruh:1988in}, one requires $\det g = -1$ and this changes the value of DE as an integration constant. As another example, self-accelerating solution can be found in massive gravity \cite{deRham:2014zqa}, where at large scales, gravitational attraction becomes suppressed by a Yukawa-like gravitational force.
\item
Quantum gravity. For example, from the Hartle-Hawking wave function of the universe \cite{Hartle:1983ai}, exponentially small DE is predicted \cite{Hawking:1984hk}. However at the same time an empty universe is predicted which does not agree with observations.
\end{enumerate}
After the discovery of cosmic acceleration, some new ideas were proposed.
Now, one can add three new categories:
\begin{enumerate}[resume]
\item
Holographic principle. This will be the focus of this review.
\item
Back-reaction of gravity. As a nonlinear theory, in general relativity inhomogeneities can backreact on the FRW background \cite{Rasanen:2011ki}. However the amount of inhomogeneities of our universe does not seem to be enough to drive the observed acceleration of our universe.
\item
Phenomenological models. DE can be modeled with scalar fields with various potentials or kinetic terms \cite{Copeland:2006wr}. And in fact most of other dark energy models can be phenomenologically reconstructed by scalar fields.
\end{enumerate}
As introduced in \cite{Li:2011sd}, all the DE models belong to the above eight categories.
In this review, we focus on the sixth category: holographic principle.
We refer the reader to Ref. \cite{Li:2011sd} for a detailed description of all the eight categories.

\

\subsection{Cosmological Observations}
\label{sec:2.3}

Let us briefly introduce how to connect the theories and the observations of DE.
In physics, it is most common to connect theoretical models and the observational data through the $\chi^{2}$ statistic
\footnote{There are some alternatives to the $\chi^{2}$ statistic.
For more details, see \cite{Gott:2000mv,2011PhRvD..84h3521B,Ma:2016sio,Nielsen:2015pga}.}.
For a physical quantity $\xi$ with experimentally measured value $\xi_{\rm obs}$,
standard deviation $\sigma_{\xi}$, and theoretically predicted value $\xi_{\rm th}$,
the $\chi^2$ function is given by
\be
  \chi_{\xi}^2(\mathbf{p})={(\xi_{\rm obs}-\xi_{\rm th}(\mathbf{p}))^2\over\sigma_{\xi}^2},
\ee
where $\mathbf{p}$ denotes the model parameters.
If there are many different cosmological observations that give many different $\chi_{\xi_i}^2$,
the total $\chi^2$ can be expressed as the sum of all $\chi_{\xi_i}^2$s, i.e.
\be \label{eq:chi1}
  \chi^2(\mathbf{p})=\sum_{i}\chi_{\xi_i}^2(\mathbf{p}).
\ee
Note that Eq. (\ref{eq:chi1}) only holds true for the case where the measurements of $\xi_i$s are independent events.
If the measurements of $\xi_i$s are related to each other,
the $\chi^2$ function need to be generalized to the form
\be \label{eq:chi2}
  \chi^2(\mathbf{p})=\sum_{i,j}\Delta_{i}(Cov^{-1})_{i,j}\Delta_{j}.
\ee
Here $\Delta_{i} \equiv \xi_{\rm i, obs}-\xi_{\rm i, th}(\mathbf{p})$
is the vector consisting the difference between the observational values and the theoretical values of all the $\xi_i$s,
and $Cov$ is a covariance matrix characterizing the error information of the data.
Moreover, assuming the measurement errors are Gaussian, the likelihood function satisfies
\be \label{eq:likelihood}
{\mathcal{L}} \propto e^{-\chi^2/2}.
\ee
The best-fit model parameters correspond to a minimal $\chi^{2}$ and a maximal ${\mathcal{L}}$.
In practice, the $\chi^2$ statistics are often performed by using the Markov Chain Monte-Carlo (MCMC) technology \cite{Lewis:2002ah}.

In the following, we will introduce in detail
three kinds of most mainstream cosmological observations used in the numerical studies of DE
(including type Ia supernova, baryon acoustic oscillation and cosmic microwave background),
and describe how these observations are included into the $\chi^2$ statistics.
In addition, we will also give a qualitative overview for some other cosmological observations associated with DE.
We refer the reader to Ref. \cite{2013PhR...530...87W} for a more comprehensive and more detailed review on the observational probes of DE.

\

\subsubsection{Type Ia Supernova}
\label{sec:2.3.1}

Type Ia supernova (SNIa) is a sub-category of cataclysmic variable stars that result from the thermonuclear explosions of white dwarfs,
when these white dwarfs reach the Chandrasekhar limit and ignite carbon at their centers \cite{2000tias.conf...33L,Hillebrandt:2000ga}.
There are two kinds of mechanisms yielding SNIa:
the first is the so-called ``single degenerate'',
in which a white dwarf accreting from a binary companion is pushed over the Chandrasekhar limit;
the second is the so-called ``double degenerate'',
in which gravitational radiation causes an orbiting pair of white dwarfs to merge and exceed the Chandrasekhar limit.
SNIa can be used as standard candles to measure the luminosity distance $d_L(z)$ \cite{Leibundgut:1996qm,Perlmutter:1996ds,Leibundgut:2000xw},
and thus provides a most straightforward tool to measure the expansion history of the universe.

\begin{figure*}
\centering
\includegraphics[width=0.8\textwidth]{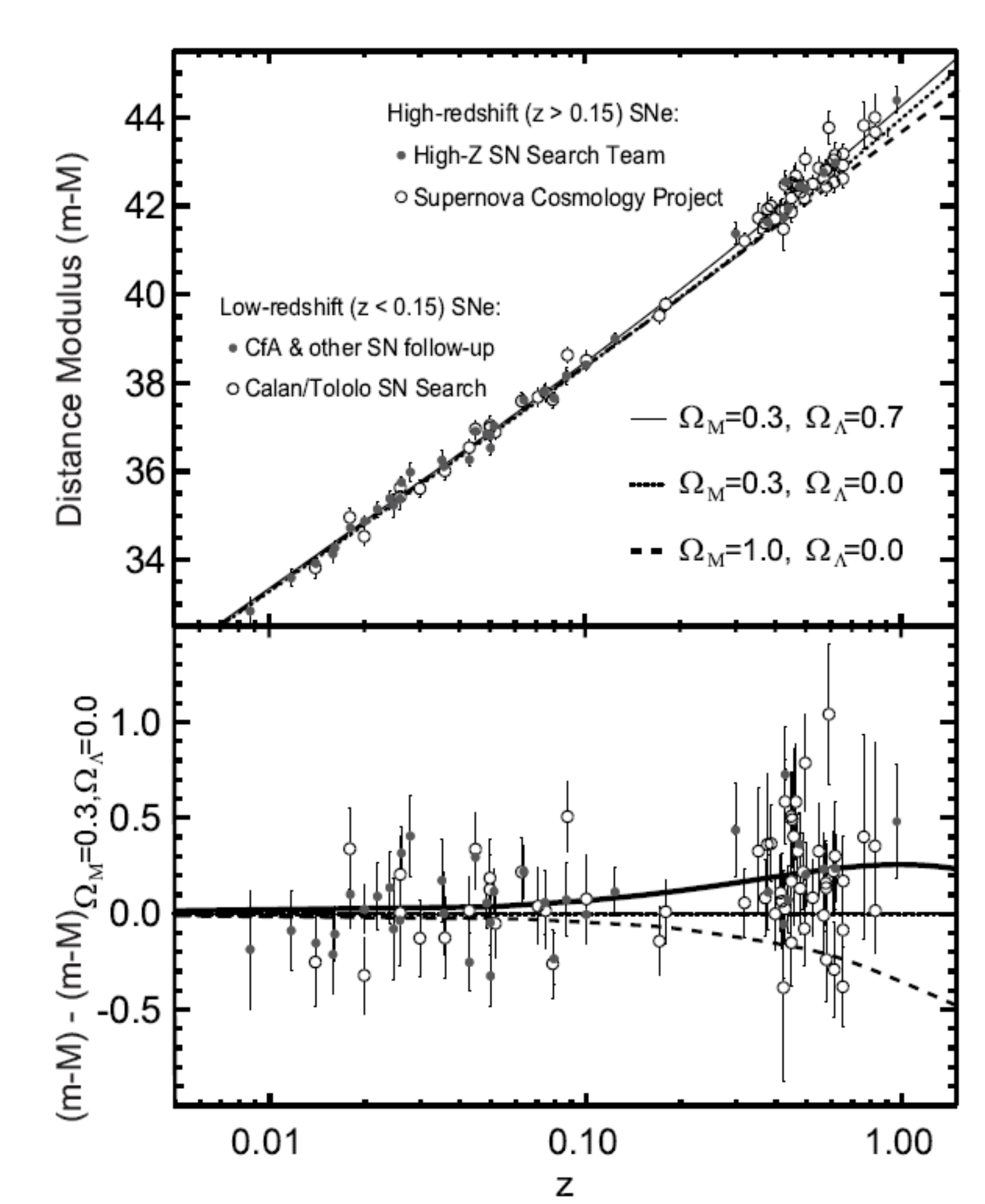}
\caption{Discovery data: Hubble diagram of SNIa measured by the Supernova Cosmology Project and the High-z Supernova Team.
Bottom panel shows residuals in distance modulus relative to an open universe with $\Omega_{m0}=0.3$ and $\Omega_{\Lambda 0} = 0$.
From \cite{Frieman:2008sn}, based on \cite{Riess:1998cb,Perlmutter:1998np}.}
\label{fig1}
\end{figure*}

In 1998, using 16 distant and 34 nearby supernovae from the Hubble space telescope (HST) observations,
Riess {\it et al.} first discovered the acceleration of expanding universe \cite{Riess:1998cb}.
Soon after, based on the analysis of 18 nearby supernovae from the Calan-Tololo sample and 42 high-redshift supernovae,
Perlmutter {\it et al.} confirmed the discovery of cosmic acceleration \cite{Perlmutter:1998np}.
The discovery of the universe's accelerating expansion (see Fig. \ref{fig1})
was another big surprise since Edwin Hubble discovered the cosmic expansion in 1929.
Because of this great discovery, Saul Perlmutter, Brian Schmidt, and Adam Riess won the Nobel prize in physics 2011.

In recent years, several high quality supernova (SN) datasets have been released,
such as ``gold04'' \cite{Riess:2004nr}, ``gold06'' \cite{Riess:2006fw}
``SNLS'' \cite{Astier:2005qq}, ``ESSENCE'' \cite{WoodVasey:2007jb},
``Union'' \cite{Kowalski:2008ez}, ``Constitution'' \cite{Hicken:2009dk},
``SDSS'' \cite{Kessler:2009ys}, ``Union2'' \cite{Amanullah:2010vv}, ``SNLS3'' \cite{Conley:2011ku} and ``Union2.1'' \cite{Suzuki:2011hu}.
The largest SN sample is ``Joint Light-curve Analysis'' (JLA) dataset \cite{Betoule:2014frx},
which consists of 740 supernovae. JLA data includes 118 supernovae at $0<z<0.1$ from several low-redshift samples,
374 supernovae at $0.03<z<0.4$ from the Sloan Digital Sky Survey (SDSS) SN search,
239 supernovae at $0.1<z<1.1$ from the Supernova Legacy Survey (SNLS) observation
and 9 supernovae at $0.8<z<1.3$ from HST measurement.
As an example, we will describe how to include the JLA SN data into the $\chi^2$ statistics.

In practice, SNIa's distance modulus are often used to construct the $\chi^2$ function of SN.
The theoretical value of distance modulus can be computed as
\be
  \mbox{\bf $\mu$}_{th} = 5 \log_{10}\left[\frac{d_L(z_{hel},z_{cmb})}{Mpc}\right] + 25,
\ee
where $z_{cmb}$ and $z_{hel}$ are the CMB restframe and heliocentric redshifts of SN.
The luminosity distance ${d}_L$ is given by
\be
  {d}_L(z_{hel},z_{cmb}) = (1+z_{hel}) r(z_{cmb}),
\ee
where
\be
\label{eq:rz}
r(z) = H_0^{-1}\, |\Omega_k|^{-1/2} {\rm sinn}\left[|\Omega_k|^{1/2}\, \int_0^z\frac{dz'}{E(z')}\right].
\ee
Here $E(z) \equiv H(z)/H_0$ is the reduced Hubble parameter,
${\rm sinn}(x)=\sin(x)$, $x$, $\sinh(x)$ for $\Omega_k<0$, $\Omega_k=0$, and $\Omega_k>0$, respectively.

On the other hand, the observed value of distance modulus can be expressed as
\be
  \mbox{\bf $\mu$}_{obs}= m_{B}^{\star} - M_B + \alpha \times X_1
  -\beta \times {\cal C},
\ee
where $m_B^{\star}$ is the observed peak magnitude in the rest-frame of the $B$ band,
$X_1$ describes the time stretching of light-curve, and ${\cal C}$ describes the SN color at maximum brightness.
As mentioned above, the JLA dataset consist of 740 SNIa;
for each SNIa, the observed values of $m_B^{\star}$, $X_1$ and ${\cal C}$ are given in Ref. \cite{Betoule:2014frx}.
In addition, $\alpha$ is the SN stretch-luminosity parameter, $\beta$ is the SN color-luminosity parameter,
and $M_B$ is the absolute B-band magnitude that depends on the host galaxy properties \cite{Schlafly:2010dz,Johansson:2012si}.
In the recipe of \cite{Betoule:2014frx}, $\alpha$ and $\beta$ are treated as free model parameters;
in contrast, $M_B$ is analytically marginalized in the process of numerical fitting.

The $\chi^2$ of JLA data can be calculated as
\be  \label{eq:chi2_SN}
  \chi^2_{SN} = \Delta \mbox{\bf $\mu$}^T \cdot \mbox{\bf Cov}^{-1} \cdot \Delta\mbox{\bf $\mu$},
\ee
where the data vector $\Delta \mbox{\bf $\mu$}\equiv \mbox{\bf $\mu$}_{obs}-\mbox{\bf $\mu$}_{th}$.
$\mbox{\bf Cov}$ is the total covariance matrix, which can be expressed as
\be
\mbox{\bf Cov}=\mbox{\bf D}_{\rm stat}+\mbox{\bf C}_{\rm stat}
+\mbox{\bf C}_{\rm sys}.
\ee
Here $\mbox{\bf D}_{\rm stat}$ is the diagonal part of the statistical
uncertainty, which is given by \cite{Betoule:2014frx},
\ba
\mbox{\bf D}_{\rm stat,ii}&=&\left[\frac{5}{z_i \ln 10}\right]^2 \sigma^2_{z,i}+
  \sigma^2_{\rm int} +\sigma^2_{\rm lensing} + \sigma^2_{m_B,i} \nonumber\\
&&   +\alpha^2 \sigma^2_{X_1,i}+\beta^2 \sigma^2_{{\cal C},i}
+ 2 \alpha C_{m_B X_1,i} - 2 \beta C_{m_B {\cal C},i} -2\alpha\beta C_{X_1 {\cal C},i}.
\ea
The first three terms account for the uncertainty in redshift due to peculiar velocities,
the intrinsic variation in SN magnitude, and the variation of magnitudes caused by gravitational lensing.
$\sigma^2_{m_B,i}$, $\sigma^2_{X_1,i}$, and $\sigma^2_{{\cal C},i}$
denote the uncertainties of $m_B$, $X_1$ and ${\cal C}$ for the $i$-th SN.
In addition, $C_{m_B X_1,i}$, $C_{m_B {\cal C},i}$ and $C_{X_1 {\cal C},i}$
are the covariances between $m_B$, $X_1$ and ${\cal C}$ for the $i$-th SN.
Moreover, $\mbox{\bf C}_{\rm stat}$ and $\mbox{\bf C}_{\rm sys}$
are the statistical and the systematic covariance matrices, given by
\be
\mbox{\bf C}_{\rm stat}+\mbox{\bf C}_{\rm sys}=V_0+\alpha^2 V_a + \beta^2 V_b +
2 \alpha V_{0a} -2 \beta V_{0b} - 2 \alpha\beta V_{ab},
\ee
where $V_0$, $V_{a}$, $V_{b}$, $V_{0a}$, $V_{0b}$ and $V_{ab}$ are six $740\times740$ matrices.
Notice that the values of all the physical quantities and matrices mentioned in this paragraph are given in Ref. \cite{Betoule:2014frx}.
The reader can refer to the original JLA paper \cite{Betoule:2014frx},
as well as their publicly released code,
for more details of calculating JLA data's $\chi^{2}$ function.

It must be mentioned that, as the rapid growth of the number of SNIa discovered in the astronomical observation,
the systematic errors in the SN observation have become the major factor that confines the ability to accurately probe the nature of DE.
The distinct sources of systematic uncertainties include calibration errors,
dust or host-galaxy extinction, and gravitational lensing \cite{2013PhR...530...87W}.
In addition, the studies on various SNIa datasets
(including SNLS3 \cite{Wang:2013yja}, Union2.1 \cite{Mohlabeng:2013gda}, Pan-STARRS1 \cite{Scolnic:2013efb} and JLA \cite{Shariff:2015yoa,Li:2016dqg})
all indicated that SN color-luminosity parameter $\beta$ should evolve along with redshift $z$;
while the redshift-evolution of $\beta$ will have significant effects on the parameter estimation of various cosmological models \citep{Wang:2013tic,Wang:2014oga,Wang:2014fqa}.
\footnote{The possible redshift-dependence of the intrinsic scatter $\sigma_{\rm int}$ \cite{Marriner:2011mf},
as well as different choice of SN light-curve fitters \cite{Bengochea:2014iha,Hu:2015opa},
may also cause the systematic uncertainties of SNIa.}
Therefore, the control of the systematic uncertainties of SNIa have become one of the biggest challenges in SN cosmology.

Some interesting analysis techniques are also proposed to reduce the systematic uncertainties of SNIa.
For instance, Wang proposed a data analysis technique, called flux-averaging (FA),
to reduce the systematic errors caused by the weak lensing effect of SNIa \cite{2000ApJ...536..531W};
it has been proved that using FA can also reduce the bias in distance estimate
induced by some other systematic effects \cite{Wang:2003gz,Wang:2004py,Wang:2009sn,Wang:2011sb,Wang:2013yja,2016arXiv160601779W}.
In addition, it was argued that, compared with the $\chi^{2}$ analysis,
applying Bayesian graphs to the SNIa data analysis has potential to reduce the systematic errors of SNIa
\cite{Ma:2016sio}.

\

\subsubsection{Baryon Acoustic Oscillation}
\label{sec:2.3.2}

Baryon acoustic oscillation (BAO) is the periodic fluctuation in the density of the visible baryonic matter of the universe
\cite{1968ApJ...151..459S,1970ApJ...162..815P,1970Ap&SS...7....3S}.
Different from the SNIa that provides a ``standard candle'' for astronomical observations,
the BAO provides a ``standard ruler'', which is the radius of the sound horizon at the drag epoch,
to explore the expansion history of the universe \cite{1998ApJ...504L..57E,2003ApJ...594..665B,Seo:2003pu}.
Today's BAO scale is mainly measured at low redshifts through the large-scale galaxy surveys \cite{Eisenstein:2005su,Percival:2009xn}
\footnote{Today's BAO scale can be measured at high redshifts through 21 cm emission from reionization \cite{2006PhR...433..181F}, too.}.
Moreover, an anisotropic BAO analysis that measures the BAO feature in the line-of-sight and transverse directions
can separately measure Hubble parameter $H(z)$ and the comoving angular diameter distance $D_M(z)$, which is defined as
\be
\label{eq:dm}
D_M(z) \equiv H_0^{-1}\, |\Omega_k|^{-1/2} {\rm sinn}\left[|\Omega_k|^{1/2}\, \int_0^z\frac{dz'}{E(z')}\right].
\ee
Therefore, BAO can provide an important complement to the SNIa data.

\begin{figure*}
\centering
\includegraphics[width=0.8\textwidth]{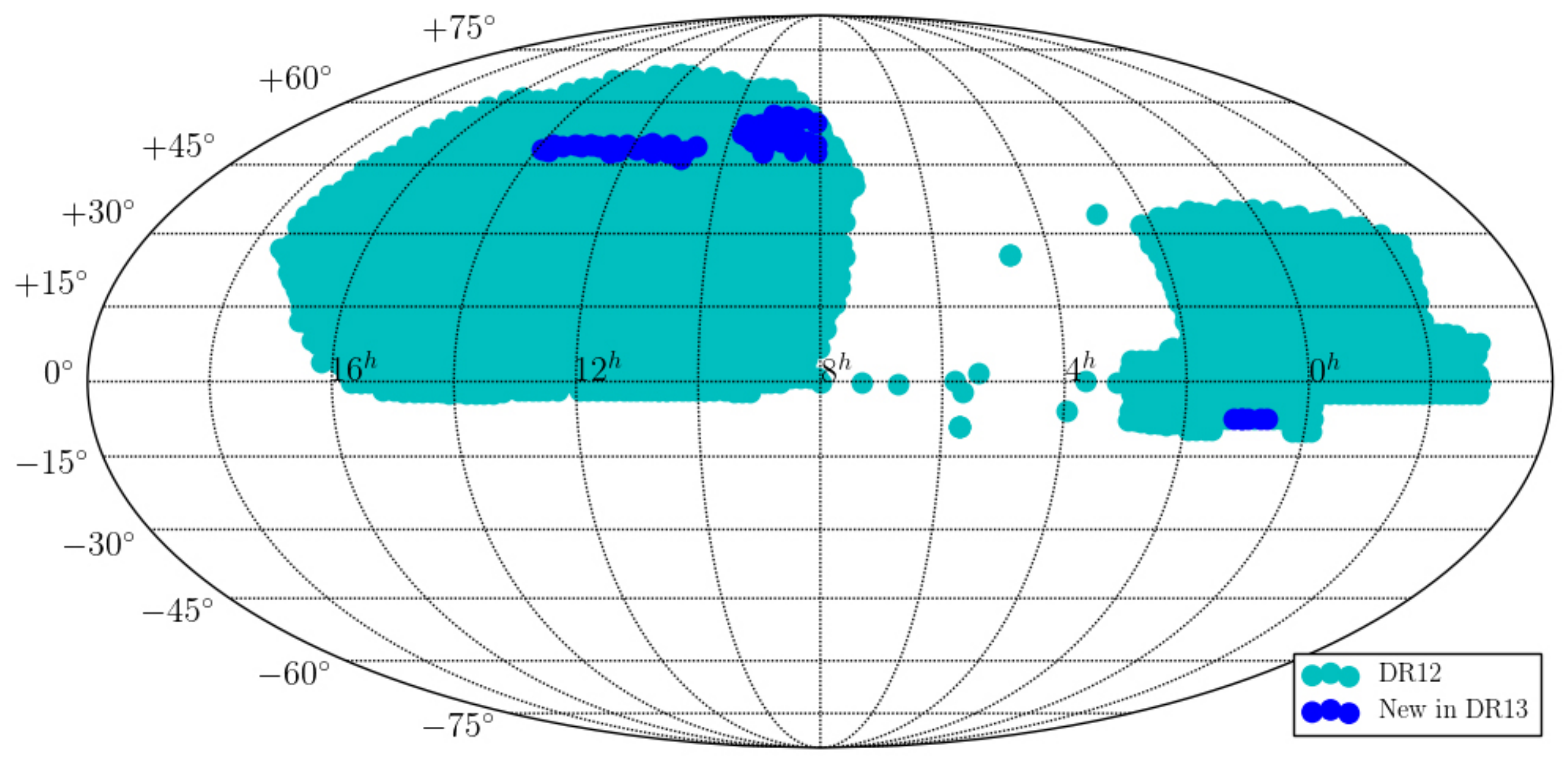}
\caption{Coverage of the first data release for SDSS-IV in equatorial coordinates.
The blue areas show the locations of the new plates released in DR13.
The green represents the area covered by SDSS-III in DR12.
From \cite{2016arXiv160802013S}.}
\label{fig2}
\end{figure*}

The most famous astronomical project of measuring today's BAO feature is the Sloan Digital Sky Survey (SDSS) \cite{York:2000gk}.
As one of the most successful surveys in the history of astronomy, SDSS was launched in 2000.
Now it has progressed through four phases (SDSS-I, SDSS-II, SDSS-III, and SDSS-IV),
and has created the most detailed three-dimensional maps of the universe with the spectra for more than three million astronomical objects.
So far, the latest SDSS dataset is the thirteenth Data Release (DR13) \cite{2016arXiv160802013S},
which is the first data release for SDSS-IV (see Fig. \ref{fig2}).
In the following, we will introduce how to use the SDSS BAO data to constrain DE.

In the literature, to analyse the BAO feature from the SDSS data,
a lot of characteristic quantities have been proposed in the last decade \cite{Eisenstein:2005su,Percival:2009xn,Chuang:2011fy}.
As an example, here we describe the usage of SDSS DR12 \cite{Alam:2016hwk}.
In \cite{Alam:2016hwk}, Alam et al. adopted two characteristic quantities,
$D_M(z)r_{s,fid}/r_s(z_d)$ and $H(z)r_s(z_d)/r_{s,fid}$, to constrain various DE models.
Here $r_s(z)$ is the comoving sound horizon, given by
\be \label{eq:rs}
  r_s(z)=\int_{z}^{\infty}\frac{c_s(z^\prime)}{H(z^\prime)}dz^\prime,
\ee
where $c_s(z)= 3^{-1/2}c[1+\frac{3}{4}\rho_{b}(z)/\rho_{r}(z)]^{-1/2}$ is the sound speed in the photon-baryon fluid.
$z_d$ is the redshift of the drag epoch, whose fitting formula is \cite{Eisenstein:1997ik}
\be
  z_d={1291(\Omega_{m0}h^2)^{0.251}\over 1+0.659(\Omega_{m0}h^2)^{0.828}}\left[1+b_1(\Omega_{b0}h^2)^{b_2}\right],
\ee
where
\be
  b_1=0.313(\Omega_{m0}h^2)^{-0.419}\left[1+0.607(\Omega_{m0}h^2)^{0.674}\right], \quad b_2=0.238(\Omega_{m0}h^2)^{0.223}.
\ee
In addition, $r_{s,fid} = 147.78$ Mpc is the fiducial value of $r_s(z_d)$.

Now, the $\chi^2$ function for the BAO data from the SDSS DR12 can be written as
\be \label{eq:chi2BAO}
\chi^2_{BAO}=\Delta p_i \left[ \mbox{Cov}^{-1}_{BAO}(p_i,p_j)\right]
\Delta p_j,
\hskip .2cm
\Delta p_i= p_i - p_i^{data}.
\ee
Ref. \cite{Alam:2016hwk} gave six BAO data points:
\ba
p_1=D_M(0.38)r_{s,fid}/r_s(z_d), \hskip .2cm p_1^{data}= 1518,\nonumber\\
p_2= H(0.38)r_s(z_d)/r_{s,fid}, \hskip .2cm p_2^{data}= 81.5,\nonumber\\
p_3=D_M(0.51)r_{s,fid}/r_s(z_d), \hskip .2cm p_3^{data}= 1977, \nonumber\\
p_4=H(0.51)r_s(z_d)/r_{s,fid}, \hskip .2cm p_4^{data}= 90.4,\nonumber\\
p_5=D_M(0.61)r_{s,fid}/r_s(z_d), \hskip .2cm p_5^{data}=2283,\nonumber\\
p_6=H(0.61)r_s(z_d)/r_{s,fid}, \hskip .2cm p_6^{data}= 97.3.
\ea
The covariance matrix $Cov_{BAO}$ is given by
\be \label{eq:CMB_cov}
\mbox{Cov}_{BAO}(p_i,p_j)=\sigma(p_i)\, \sigma(p_j) \,\mbox{NormCov}_{BAO}(p_i,p_j),
\ee
where $\sigma(p_i)$ is the 1$\sigma$ error of observed quantity $p_i$,
and $\mbox{NormCov}_{BAO}(p_i,p_j)$ is the corresponding normalized covariance matrix.
The values of $\sigma(p_i)$ and $\mbox{NormCov}_{BAO}(p_i,p_j)$ are listed in the table 8 of \cite{Alam:2016hwk}.

\

\subsubsection{Cosmic Microwave Background}
\label{sec:2.3.3}

\begin{figure*}
\centering
\includegraphics[width=0.8\textwidth]{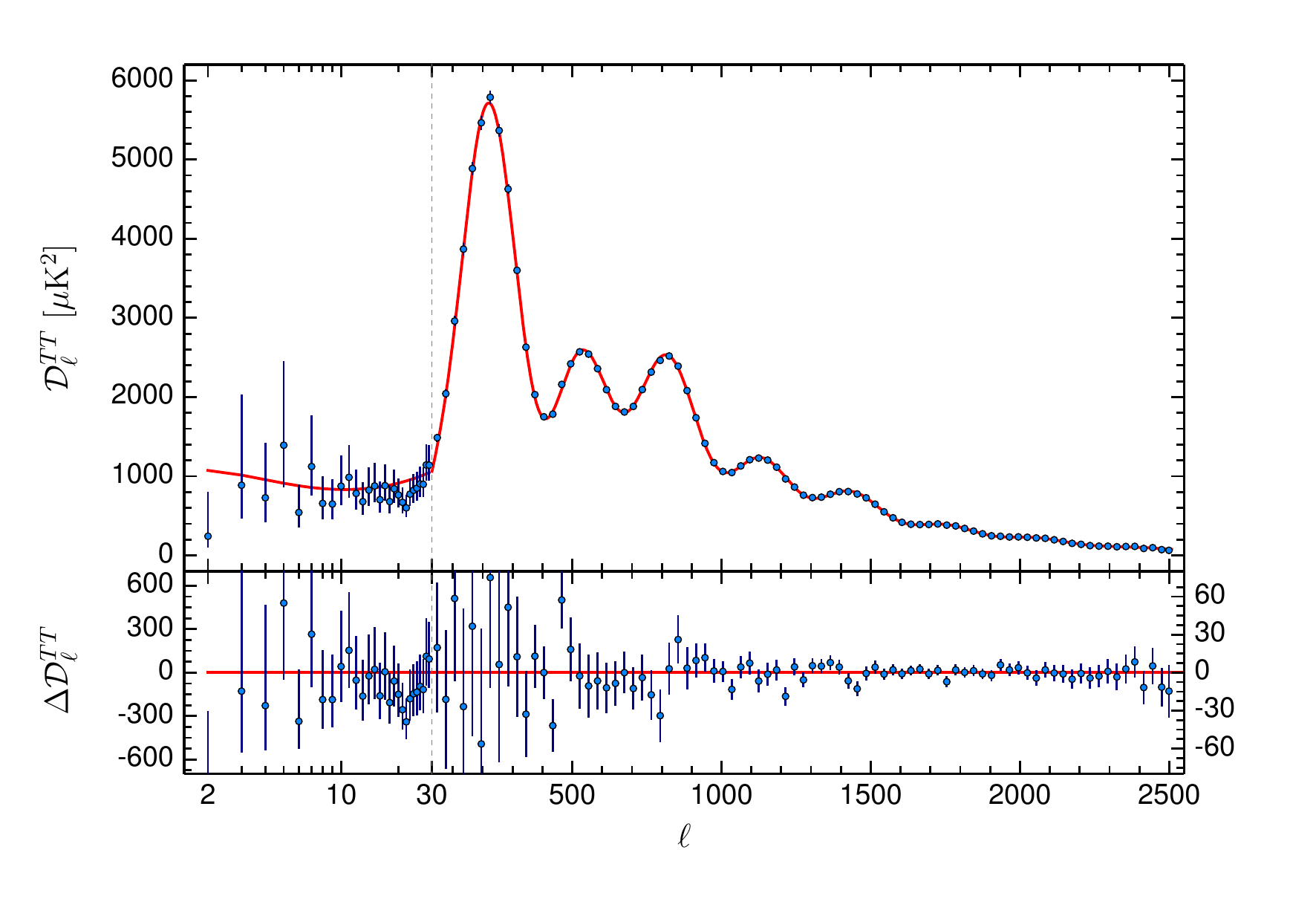}
\caption{The  Planck 2015 temperature power spectrum.
The upper panel shows the best-fit base $\Lambda$CDM theoretical spectrum fitted to the Planck TT+lowP likelihood,
while the lower panel show the residuals with respect to this model.
From \cite{2015arXiv150201589P}.}
\label{fig3}
\end{figure*}

Cosmic microwave background (CMB) is the legacy of the cosmic recombination epoch,
and it contains abundant information of the early universe.
In 1964, Penzias and Wilson detected the CMB for the first time \cite{1965ApJ...142..419P};
because of this great discovery, they won the Nobel Prize in Physics 1978.
The discovery of CMB provided strong evidence that supports the Big Bang theory of the universe \cite{1948PhRv...73..803A},
and opened a golden age of modern cosmology.
In 1989, the first generation of CMB satellite, the Cosmic Background Explorer (COBE), was launched.
It discovered the CMB anisotropy for the first time \cite{1992ApJ...396L...1S},
and thus opened the era of the precise cosmology.
Two of COBE's principal investigators, Mather and Smoot, received the Nobel Prize in Physics 2006.
In 2001, the second generation of CMB satellite, the Wilkinson
Microwave Anisotropy Probe (WMAP) \cite{Bennett:2003ca,Spergel:2003cb}, was launched.
It precisely measured the CMB spectrum
and probed various cosmological parameters with a higher accuracy \cite{Spergel:2006hy,Komatsu:2008hk,2011ApJS..192...18K}.
In 2009, the Planck satellite, as the successor to WMAP, was launched.
The latest scientific results of the Planck satellite were published in 2015 \cite{2015arXiv150201589P} (see Fig. \ref{fig3}).
In the following, we will introduce how to use the Planck 2015 data to perform cosmology-fits.

As is well known, CMB anisotropy data can provide the strongest constraints on cosmological parameters.
Therefore, the inclusion of CMB data will be very helpful to break the degeneracies among DE and cosmological parameters.
In the literature, people often make use of the ``distance priors'' data extracted from the CMB observations to constrain DE.
These distance priors include the ``shift parameter'' $R$ and the ``acoustic scale'' $l_A$.

The shift parameter $R$ is defined as \cite{Efstathiou:1998xx}
\be
R \equiv \sqrt{\Omega_{m0} H_0^2} \,r(z_*),
\ee
where $r(z)$ is the comoving distance given in Eq. (\ref{eq:rz}).
In addition, $z_*$ is the redshift of the photon decoupling epoch, whose fitting formula is given by \cite{1996ApJ...471..542H}
\be \label{eq:z*}
z_*=1048[1+0.00124(\Omega_{b0} h^2)^{-0.738}][1+g_1(\Omega_{m0} h^2)^{g_2}],
\ee
where
\be
g_1=\frac{0.0783(\Omega_{b0} h^2)^{-0.238}}{1+39.5(\Omega_{b0} h^2)^{0.763}}, \quad g_2=\frac{0.560}{1+21.1(\Omega_{b0} h^2)^{1.81}}.
\ee
In addition, the acoustic scale $l_A$ is defined as
\be \label{eq:la}
  l_A\equiv \pi r(z_*)/r_s(z_*),
\ee
where $r_s(z)$ is the comoving sound horizon given in Eq. (\ref{eq:rs}).
These two distance priors, together with the baryonic matter parameter $\omega_b \equiv \Omega_b h^{2}$,
provide an efficient summary of CMB data as far as DE constraints go \cite{2007PhRvD..76j3533W}.

Now, the $\chi^2$ function for the Planck 2015 distance prior data can be written as
\be \label{eq:chi2CMB}
\chi^2_{CMB}=\Delta q_i \left[ \mbox{Cov}^{-1}_{CMB}(q_i,q_j)\right]\Delta q_j,
\hskip .2cm
\Delta q_i= q_i - q_i^{data}.
\ee
Ref. \cite{2015arXiv150201590P} gave three CMB data points:
\ba
q_1=R(z_*), \hskip .2cm q_1^{data}= 1.7382,\nonumber\\
q_2=l_A(z_*), \hskip .2cm q_2^{data}= 301.63,\nonumber\\
q_3=\omega_b, \hskip .2cm q_3^{data}= 0.02262.
\ea
The covariance matrix for $(q_1, q_2, q_3)$ is given by
\be \label{eq:CMB_cov2}
\mbox{Cov}_{CMB}(q_i,q_j)=\sigma(q_i)\, \sigma(q_j) \,\mbox{NormCov}_{CMB}(q_i,q_j),
\ee
where $\sigma(q_i)$ is the 1$\sigma$ error of observed quantity $q_i$,
and $\mbox{NormCov}_{CMB}(q_i,q_j)$ is the corresponding normalized covariance matrix.
The values of $\sigma(q_i)$ and $\mbox{NormCov}_{BAO}(q_i,q_j)$ are listed in the table 4 of \cite{2015arXiv150201590P}.
\footnote{For some other CMB distance priors data, e.g. see \citep{Wang:2013mha,Huang:2015vpa} and references therein.}

In addition to the CMB distance priors, one can also use the full CMB power spectrum to constrain DE
by applying the Markov Chain Monte Carlo (MCMC) global fit technique \cite{2015arXiv150201589P,2015arXiv150201590P}.
For simplicity, we will not introduce how to use the MCMC global fit technique to perform cosmology-fits.
For the details of global fit technique, the reader can refer to \cite{Li:2008cj} and references therein.

\

\subsubsection{Other Cosmic Probes}
\label{sec:2.3.4}

The target of this subsection is to give a qualitative overview for some other cosmic probes,
including weak lensing, galaxy clusters, redshift-space distortion, Alcock-Paczynski effect, standard sirens, redshift drift and cosmic age test.
Different from SNIa, BAO and CMB, these probes are seldom used to constrain DE models in the literature,
because of the lack of actual data or the existence of significant systematic errors.
Moreover, up to now there are still many debates about how to make use of these probes to constrain DE models.
So in this work, we do not describe how to calculate the $\chi^2$ functions of these probes.
We refer the reader to Ref. \cite{2013PhR...530...87W} for a more detailed description about these DE probes.

\

\begin{itemize}
 \item Weak Lensing
\end{itemize}
Weak lensing (WL) is the slight distortion of distant galaxies' images,
due to the gravitational bending of light by structures in the Universe.
Typically, the distortions of galaxies' size and shape are of the order of $1\%$.
WL can provide a direct measure of the distribution of matter (independent of any assumptions about galaxy biasing),
and thus provide a useful tool to probe DE
through its influence on the growth of structure \cite{Huterer:2001yu,Weinberg:2002rd,Abazajian:2002ck,Linder:2003dr,He:2017alg}.
In principle, the effect of WL on the distant sources can represent on the distortions in the shapes, sizes and brightness.
In practice, the shape distortions (called ``cosmic shear'') have been used much more widely.
A variety of statistical approaches have been used to extract information from cosmic shear,
including cosmic shear power spectrum \cite{Takada:2003ef}, cross-correlation tomography \cite{Jain:2003tba},
galaxy-galaxy lensing \cite{Hu:2003pt}, cosmography \cite{Bernstein:2003es}, and so on.
For more details about the WL observation and its applications on the DE probe,
see \cite{2013PhR...530...87W} and references therein.

\

\begin{itemize}
 \item Galaxy Clusters
\end{itemize}
Galaxy clusters (GC) are the largest gravitationally bound objects in the universe.
They typically contain 50 to 1000 galaxies and have a diameter from 2 to 10 Mpc.
The studies of GC are crucial
in helping to establish the standard model of cosmology \cite{1933AcHPh...6..110Z,1980ApJ...241..486H,1989ApJ...347..563P}.
Today, GC are still capable to test cosmology in a variety of ways.
For example, cluster abundances provide a important tool for constraining the growth of structure in the matter distribution.
By comparing the predicted space density of massive halos to the observed space density of clusters,
one can get the cosmological constraints on
the present fractional matter density $\Omega_{m0}$ and the amplitude of the matter power-spectrum $\sigma_8$ \cite{1993MNRAS.262.1023W}.
In addition, one can also obtain the constraints on DE by using the X-ray cluster gas mass fraction \cite{Allen:2002sr,Allen:2007ue}.
For more details about the GC observation and its applications on the DE probe,
see \cite{2011ARA&A..49..409A} and references therein.

\

\begin{itemize}
 \item Redshift-Space Distortion
\end{itemize}
Redshift-space distortion (RSD) is an observational phenomenon of anisotropic galaxy distribution in redshift space,
which is due to the peculiar velocities of the galaxies causing a Doppler shift
in addition to the redshift caused by the Hubble flow \cite{1987MNRAS.227....1K}.
Observations of RSD in spectroscopic galaxy surveys offer an attractive method for observing the build-up of cosmological structure,
which depends both on the expansion rate of the Universe and the theory of gravity.
By modeling the full redshift-space galaxy power spectrum,
one can extract the parameter combination $f(z)\sigma_8(z)$,
the product of the growth rate and the matter clustering amplitude \cite{2009MNRAS.393..297P}.
Now this quantity is often used to constrain DE together with the BAO information \cite{2012MNRAS.420.2102S,2012MNRAS.426.2719R}.
Therefore, RSD has become an important cosmic probe for DE \cite{2009MNRAS.397.1348W,Koyama:2009gd,Jennings:2010uv}.

\

\begin{itemize}
 \item Alcock-Paczynski Effect
\end{itemize}
The Alcock-Paczynski (AP) effect refers to the geometric distortion
when an incorrect cosmological model (with an incorrect value of the product $H(z)D_A(z)$)
is assumed for transforming redshift to comoving distance,
induced by the fact that measured distances along and perpendicular to the line of sight are fundamentally different \cite{1979Natur.281..358A}.
The AP effect can be measured through
the statistical study of galaxies clustering \cite{1996MNRAS.282..877B,1996ApJ...470L...1M,2004MNRAS.348..745O},
the symmetry properties of galaxy pairs \cite{2010Natur.468..539M,2012MNRAS.420.1079J,2012PhRvD..86b3530B},
and the cosmic voids \cite{1995ApJ...452...25R,2012ApJ...754..109L,2014MNRAS.443.2983S}.
In addition, it is argued that measuring the redshift dependence of AP effect
may also derive useful cosmological constraints on DE \cite{Li:2014ttl,Li:2015jra,2016arXiv160905476L}.

\

\begin{itemize}
 \item Standard Sirens
\end{itemize}
Since the great breakthrough of the direct gravitational waves (GW) detection of the Advanced LIGO \cite{Abbott:2016blz,Abbott:2016nmj},
GW astronomy has become the most popular and most active research area in astrophysics.
The observation of GW has great potential to make interesting contributions to the studies of DE,
because it can open an entirely different route to distance measurement.
In 1986, Schutz found that the luminosity distance of the binary neutron stars or binary black holes
can be independently determined by observing the GW generated by these systems \cite{1986Natur.323..310S}.
If their redshifts can be determined via other method,
then they could be used to probe DE through the Hubble diagram \cite{Holz:2005df,Dalal:2006qt,Arun:2007hu,Deffayet:2007kf,Linder:2007ge}.
Because of the analogy between GW observations and BAO measurements,
this approach is often referred to as ``standard siren'',
which has drawn more and more attentions \cite{Calabrese:2016bnu,Oguri:2016dgk,Giudice:2016zpa,2016JCAP...10..006C}
\footnote{DE can also leave characteristic features
on the spectrum of primordial gravitational waves \cite{Zhang:2005nw,Wang:2008fg,Wang:2008vp},
which may be detected via the measurements of CMB B-mode polarization \cite{1975JETP...40..409G,Kamionkowski:1996zd,Kesden:2002ku,Smith:2006nka}.}.

\

\begin{itemize}
 \item Redshift Drift
\end{itemize}
Along with the expansion of the universe, the redshift of a comoving cosmological source will change over time.
In 1962, Sandage was the first to propose that
measuring this ``redshift drift'' can provide a useful tool to test cosmology \cite{1962ApJ...136..319S}.
In 1998, Loeb repopularized the idea, noting that high-resolution spectrographs on large telescopes
may measure the effect in absorption-line spectra of high-redshift quasars \cite{1998ApJ...499L.111L}.
In other words, the variations of redshifts can be obtained by direct measurements of the quasar Lyman-$\alpha$ absorption lines
at sufficiently separated epochs (e.g., $10-30$ yrs).
Then, it can be used to directly measures the expansion of the universe.
The redshift drift (also called Sandage-Loeb test) is unique in its coverage of the ``redshift desert'' at $2\leq z\leq 5$,
where other DE probes are unable to provide useful information about this redshift region \cite{Corasaniti:2007bg}.
Therefore, redshift drift is expected to be an important complementary to other DE probes \cite{Balbi:2007fx,Uzan:2008qp,Jain:2009bm}.

\

\begin{itemize}
 \item Cosmic Age Test
\end{itemize}
The cosmic age problem is a longstanding issue in cosmology \cite{Chaboyer:1998fu}.
The conflict between the ages of some old globular clusters and the age of a decelerating universe
was one of the significant early arguments for cosmic acceleration \cite{1996Natur.381..581D,1997ApJ...484..581S,Hansen:2002ij}.
The return of cosmological constant $\Lambda$ has greatly alleviated the cosmic age problem \cite{Alcaniz:1999kr}.
However, the cosmic age puzzle remains in the standard cosmology.
For example, the existence of an old quasar APM 08279+5255 at $z=3.91$ \cite{Hasinger:2002wg} is still a mystery,
because it is even older than the cosmic age
given by almost all the mainstream cosmological models \cite{Friaca:2005ba,Jain:2005gu,Wei:2007ig}.
In other words, to accommodate this anomalous object,
some more complicated cosmological model should be taken into account \cite{Wang:2006qw,Wang:2008te,Wang:2010su}.
In addition, the cosmic age test can also be used to
distinguish DE models from inhomogeneous universe models \cite{Lan:2010ky,Yan:2014eca}.

\

\section{Holographic Dark Energy}
\label{sec:3}

In this section, we introduce the key idea of the HDE model;
in addition, we also introduce the theoretical explorations and the observational constraints of this model.

\

\subsection{The HDE Model}
\label{sec:3.1}

\

\subsubsection{Applying HP to The DE Problem}
\label{sec:3.1.1}

As pointed out in the introduction,
HP it is the most important cornerstone of quantum gravity,
and has great potential to solve many long-standing issues of various physical fields.
Now we apply the HP to the DE problem.
Let us consider a universe with a characteristic length scale $L$.
The HP tell us that all the physical quantities inside the universe,
including the energy density of DE $\rho_{de}$, can be described by some quantities on the boundary of the universe.
It is clear that only two physical quantities,
the reduced Planck mass $M_p$ and the cosmological length scale $L$,
can be used to construct the expression of $\rho_{de}$.
Based on the dimensional analysis, we have
\be \label{eq:expansion_DE}
\rho_{de} = C_1 M_p^4 + C_2 M_p^2L^{-2} + C_3 L^{-4} + \ldots
\ee
where $C_1$, $C_2$, $C_3$ are constant parameters\footnote{Note that in general, $C_1$, $C_2$ and $C_3$ may be time dependent. The time dependence of $C_2$ and $C_3$ can be absorbed into the redefinition of the IR cutoff $L$. On the other hand, $C_1$ only depend on the UV physics, and thus more likely a constant because of the time translation symmetry of the fundamental theory (although there can be exceptions such as theories with time varying $G_N$).}.
The first term is strongly disfavored by naturalness ($10^{120}$ times larger than the observational value!) \cite{Peebles:2002gy};
this is the famous fine-turning problem (also called the old cosmological constant problem) \cite{Li:2011sd}.

Cohen, Kaplan and Nelson \cite{1999PhRvL..82.4971C} noted that the $C_1$ term is not compatible with HP. Hinted by HP, local quantum field theory should not be a good description for a black hole, or states at the scale of its Schwarzschild radius. Especially, the traditional estimate $\rho_{de} \sim C_1M_p^4$ from local quantum field theory should not be present from this argument. Rather, the local quantum field theory acquires a non-trivial UV cutoff $\Lambda$. To see this, note that the energy within a Schwarzschild radius $L$ is $L^3\Lambda^4$. By requiring this energy to be less than the mass of a corresponding black hole, we have $L^3\Lambda^4 < L M_p^2$. As a result, the vacuum fluctuation estimated from this UV-cut-off quantum field theory is $\rho_{de} \sim \Lambda^4 \lesssim M_p^2 L^{-2}$. Thus the $C_1$ term is not present, and the expansion in (\ref{eq:expansion_DE}) should start from the second term.

Moreover, compared with the second term, the third and the other terms are negligible.
Therefore, the expression of $\rho_{de}$ can be rewritten as
\be \label{eq:hdedensity}
\rho_{de} = 3C^2 M_p^2L^{-2},
\ee
where $C$ is another constant parameter

It must be stressed that, this expression of $\rho_{de}$ is obtained by combining the HP and the dimensional analysis,
instead of adding a DE term into the Lagrangian.
Due to this unique feature, HDE remarkably differs from any other theory of DE.

\

\subsubsection{Future Event Horizon as The Characteristic Length Scale}
\label{sec:3.1.2}

As a next step, it is crucial to choose the specific expression of characteristic length scale $L$.
The simplest choice is the Hubble scale $L=1/H$,
giving a energy density that is comparable to the present-day DE \cite{Horava:2000tb,2002PhRvL..89h1301T}.
However, Hsu had proved that this choice does not work because it will yield a wrong EoS of DE \cite{2004PhLB..594...13H}.
In addition, choosing the particle horizon does not work either
because it is impossible to obtain an accelerated expansion on this basis.

In 2004, one of the present authors (Miao Li) suggested that $L$ can be chosen as the future event horizon \cite{Li:2004rb}
\be \label{eq:future event horizon}
 L=a\int_{t}^{\infty}\frac{dt'}{a}=a\int_{a}^{\infty}\frac{da'}{Ha'^{2}}.
\ee
Note that this horizon is the boundary of the volume a fixed observer may eventually observe.
As seen below, this choice will give a very competitive DE model.

Now consider the universe dominated by the HDE and the pressureless matter.
For this case, the Friedmann equation can be written as
\be \label{eq:simpleFriedmann}
  3M_p^2H^2=\rho_{de}+\rho_{m},
\ee
or equivalently,
\be \label{eq:simpleFriedmann2}
  E(z)\equiv \frac{H(z)}{H_0} =\left(\frac{\Omega_{m0}(1+z)^3}{1-\Omega_{de}(z)}\right)^{1/2}.
\ee
Here \label{eq:hdeOm}
\be
  \Omega_{de} \equiv {\rho_{de} \over \rho_{c}}={C^2 \over L^2H^2},
\ee
where $\rho_c \equiv 3M_{P}^2H^2$ is the critical density of the universe.
Taking derivative of $\Omega_{de}$ with respect to $\ln a$,
and using Eq. \ref{eq:future event horizon},
one can obtain a differential equation for $\Omega_{de}$
\be \label{eq:hdeOm1}
  \Omega'_{de}=2 \Omega_{de} \left(-{H' \over H}-1+{\sqrt{\Omega_{de}}\over C}\right),
\ee
where the prime denotes derivative with respect to $\ln a$.
From Eq. \ref{eq:simpleFriedmann}, one can get
\be \label{eq:hdeH}
  -{H' \over H}={3 \over 2}-{\Omega_{de} \over 2}-{\Omega_{de}^{3/2} \over C}.
\ee
Combining Eq. \ref{eq:hdeOm1} with Eq. \ref{eq:hdeH},
one can obtain the following equation governing the dynamical evolution of the HDE model
\be \label{eq:hdeOmega}
  {d \Omega_{de}\over dz}=-{\Omega_{de}(1-\Omega_{de})\over 1+z} \left(1+{2\sqrt{\Omega_{de}}\over C}\right).
\ee
Since $0<\Omega_{de}<1$, $d \Omega_{de}/dz$ is always negative, namely the fraction density of HDE always increases along with redshift $z \rightarrow -1$.
This means that the expansion of the universe will never have a turning point, so that the universe will not re-collapse in the future.

\begin{figure*}
\centering
\includegraphics[width=0.8\textwidth]{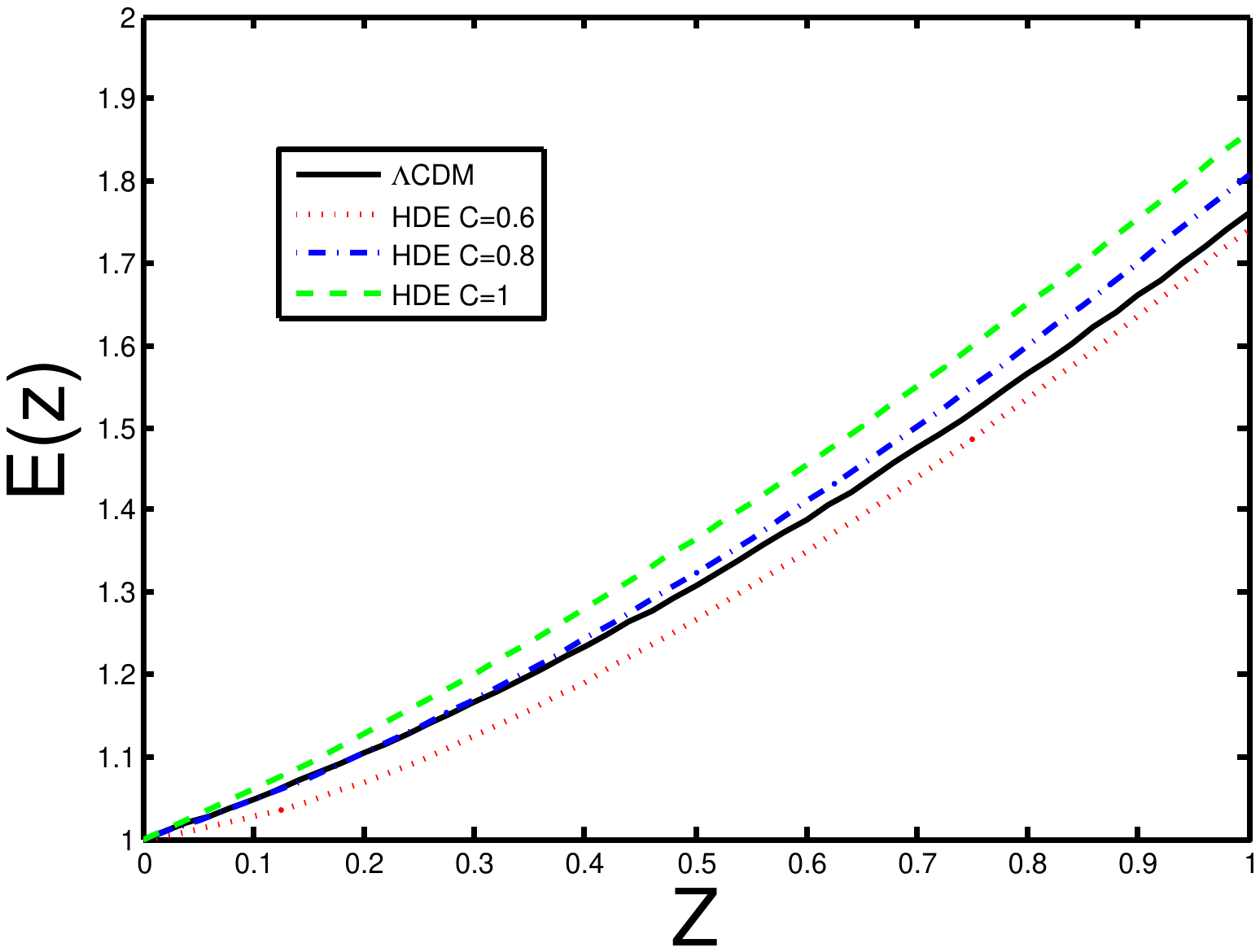}
\caption{The redshift evolution of Hubble parameter $H(z)$ of the HDE model. In this figure $\Omega_{m0}=0.3$ is always adopted.
The red dotted line, the blue dash-dotted line and the green dashed line correspond to the cases of $C=0.6$, $C=0.8$ and $C=1$, respectively.
To make a comparison, we also plot the result of the $\Lambda$CDM model.}
\label{fig4}
\end{figure*}

Solving numerically Eq. \ref{eq:hdeOmega} and substituting the corresponding results into Eq. \ref{eq:simpleFriedmann2},
the redshift evolution of Hubble parameter $H(z)$ of the HDE model can be obtained (As examples, see Fig. \ref{fig4}).

\

\subsubsection{Some Important Properties of HDE}
\label{sec:3.1.3}

\

\begin{itemize}
 \item EoS of HDE
\end{itemize}

\

From energy conservation,
\be \label{eq:matter conservation}
  \rho'_{m}+3\rho_{m} = 0.
\ee
\be \label{eq:de conservation}
  \rho'_{de}+3(1+w)\rho_{de} = 0.
\ee
Taking derivative of Eq. \ref{eq:hdedensity} with respect to $\ln a$,
and making use of Eq. \ref{eq:de conservation},
one can get the EoS of HDE
\be \label{eq:hdew}
  w = -{1\over 3}-{2\sqrt{\Omega_{de}}\over 3C}.
\ee

When the HDE is sub-dominant (i.e., at early universe with $\Omega_{de} \ll 1$), $w \simeq -1/3$ thus $\Omega_{de} \sim a^{-2}$.
When the HDE is dominant (i.e., at late universe with $\Omega_{de}\simeq 1$),
$w \simeq -1/3-2/3C$ thus the universe experiences accelerating expansion as long as $C>0$;
in other words, choosing future event horizon as the characteristic length scale $L$ indeed yields a kind of energy component behaving as DE.
Moreover, if $C=1$, $w=-1$ thus HDE will be similar to the cosmological constant $\Lambda$;
if $C>1$, $w>-1$ thus HDE will be a quintessence DE \cite{1999PhRvL..82..896Z}, corresponding to an eternal cosmic expansion;
if $C<1$, $w<-1$ thus HDE will be a phantom DE in the far future \cite{2002PhLB..545...23C},
corresponding to a cosmic doomsday called big rip \cite{Caldwell:2003vq,Li:2012via}
\footnote{In addition to quintessence and phantom, a scalar field DE may also be a quintom DE,
whose EoS can evolve across the cosmological constant boundary \cite{Feng:2004ad}.
See \cite{Cai:2009zp} for a review of quintom DE.}.
This means that $C$ is the key parameter that determines the property of HDE.

It should be mentioned that the value of $C$ cannot be derived from the theoretical framework of the HDE model,
and it can only be obtained by fitting the observational data.

\

\begin{itemize}
 \item Explanation for The Coincidence Problem
\end{itemize}

The coincidence problem can be reinterpreted as a problem of
why the ratio between the DE density and the radiation density is very tiny
at the onset of the radiation dominated epoch \cite{1999PhRvD..59l3504S}.
We assume that in the inflation epoch there are only two energy components: the HDE and the inflation energy;
the latter is almost constant during the inflation epoch, and decayed into radiation after the inflation.

Choosing the inflation energy scale as $10^{14}$ Gev,
A rough estimate shows that the ratio between $\rho_{de}$ and $\rho_{r}$ is about $10^{-52}$ \cite{Li:2004rb}.
Since during inflation epoch the HDE is diluted as $\Omega_{de} \sim a^{-2}$,
this is equal to $\exp (-2N)$ with $N=60$, namely the minimal number of e-folds in the inflation scenario.
In other words, HDE provides an explanation of the coincidence problem,
as long as inflation only last for about 60 e-folds.
A similar result was obtained in \cite{Kim:2007kw}.

\

\subsection{Theoretical Explorations for The HDE Model}
\label{sec:3.2}

In addition to the dimensional analysis mentioned above,
there are also a number of other theoretical motivations leading to the form of HDE.
We shall briefly review some of the motivations in this subsection.

\

\subsubsection{Entanglement Entropy from Quantum Information Theory}
\label{sec:3.2.1}

It is suggested that vacuum entanglement energy associated with the entanglement entropy of the universe is the origin of DE \cite{Lee:2007zq}.
The entanglement entropy of the quantum field theory vacuum with a horizon can be generically written as
\be \label{eq:HDEentangEntropy}
  S_{\rm Ent} = {\varrho R_h^2 \over l^2},
\ee
where $\varrho$ is constant parameter that depends on the nature of the field,
$R_h=a\int_{t}^{\infty}\frac{dt'}{a}$ is the future event horizon, and $l$ is the ultraviolet cutoff from quantum gravity.
The entanglement energy is conjectured to satisfy
\be \label{eq:HDEentangEnergy}
  dE_{\rm Ent} = T_{\rm Ent}dS_{\rm Ent},
\ee
where $T_{\rm Ent}=1/(2\pi R_h)$ is the Gibbons-Hawking temperature.
Integrating Eq. \ref{eq:HDEentangEnergy}, one gets
\be \label{eq:HDEentangEnergyCalc}
  E_{\rm Ent} = {\varrho N_{\rm dof}R_h \over \pi l^2},
\ee
where $N_{\rm dof}$ is the number of light fields present in the vacuum.
Thus the energy density is
\be \label{eq:HDEentanEntropyy}
  \rho_{de} = 3C^2 M_p^2 R_h^{-2},
\ee
where $C={\sqrt{\beta N_{\rm dof}}\over 2\pi l M_p}$
is in principle calculable in the quantum information theory. It is clear that this energy density has the same form with HDE. Here $\beta\sim 0.3$ from lattice simulation \cite{Muller:1995mz}, $N_\mathrm{dof}\sim 10^2$ and $l\sim 1/M_p$. Thus, $C$ is naturally of order one. However due to theoretical uncertainties it is difficult to predict the precise value of $C$.

\

\subsubsection{Holographic Gas as Dark Energy}
\label{sec:3.2.2}

So far, the nature of a general strongly correlated gravitational system has not been well understood.
The studies of condensed matter physics show that sometimes a system which appears nonperturbative
can be described by weakly interacting quasi-particle excitations.
In \cite{Li:2008qh} it is suggested that the quasi-particle excitations of such a system may be described by a gas of holographic particles,
with modified degeneracy
\be \label{eq:HDEhologasW}
  w = w_0 k^A V^{B}M_p^{3B-A},
\ee
where $V$ is the volume of the system, both $w_0$, $A$, and $B$ are dimensionless constants.
Inspired by holography, when taking $T\propto V^{-1/3}$ and $S\propto V^{2/3}$,
one can obtain $B=(A+2)/3$,
and the corresponding energy density can be written as
\be \label{eq:HDEhologasRho}
\rho = {A+3 \over A+4}{ST \over V},
\ee
where $S$ and $T$ are the entropy and temperature of the system.

Then Eq. \ref{eq:HDEhologasRho} can be applied to the cosmology.
After adopting the Gibbons-Hawking entropy $S=8\pi^2M_p^2R^2$ and temperature $T=1/(2\pi R)$ ($R$ is the radius of the universe),
one can obtain
\be \label{eq:HDEhologasRhoDE}
  \rho = 3{A+3 \over A+4}M_p^2 R^{-2}.
\ee
This has the same form as HDE with
\be \label{eq:HDEhologasC}
  C^2 = {A+3 \over A+4}.
\ee
It is clear that the holographic gas model always satisfies $C<1$,
implying that the fate of our universe is phantom like.

\

\subsubsection{Casimir Energy in de Sitter Space}
\label{sec:3.2.3}

The Casimir energy is one of the important predictions in quantum field theory \cite{1948PhRv...73..360C}. The Casimir effect in de Sitter space is systematically studied in \cite{Fischetti:1979ue, Hartle:1979uf, Hartle:1980nn}.
It is suggested that the Casimir energy of electromagnetic field in static de Sitter space
can be taken as a possible origin of DE \cite{Li:2009pm,Li:2009zy}.
It can be written as
\be \label{eq:HDECasimir}
  E_{\rm Casimir} = {1\over 2}\sum_\omega |\omega|,
\ee
where the absolute value of $\omega$ is the energy with respect of time $t$ of the static patch.
$E_{\rm Casimir}$ can be calculated using heat kernel method with $\zeta$ function regularization.
The final result is
\be \label{eq:HDECasimirRes}
  E_{\rm Casimir} = {3\over 8\pi}\left(\ln\mu^2-\gamma-\Gamma'(-1/2)/\Gamma(-1/2)\right)
  \left({L\over l_p^2}-{1\over L}\ln\left({2L\over l_p^2}\right)\right) +{\cal O}(1/L),
\ee
where $L$ is the de Sitter radius, $\gamma$ is the Euler constant and $\Gamma'(-1/2)\simeq -3.48$.
Here a cutoff at stretched horizon is imposed, which has a distance $l_p$ away from the classical horizon.
Note that the dominate term scales as $E_{\rm Casimir}\sim L/l_p^2$.
Thus the energy density scales as $\rho_{\rm Casimir}\sim M_p^2L^{-2}$, which is the form of HDE.

\

\subsubsection{Dark Energy from Entropic Force}
\label{sec:3.2.4}

Verlinde conjectured that gravity may be an entropic force, instead of a fundamental force of nature \cite{Verlinde:2010hp}.
\cite{Li:2010cj} investigated the implication of the conjecture for DE.
It is suggested that the entropy change of the future event horizon should be considered
together with the entropy change of the test holographic screen.
Consider a test particle with physical radial coordinate $R$,
which is the distance between the particle and the ``center'' of the universe where the observer is located.
The energy associated with the future event horizon $R_h$, using Verlinde's proposal, can be estimated as
\be \label{eq:HDEentropicE}
  E_h \sim N_h T_h \sim R_h/G,
\ee
where $N_h\sim R_h^2/G$ is number of degrees of freedom on the horizon,
and $T_h\sim 1/R_h$ is the Gibbons-Hawking temperature.
Following Verlinde's argument (instead of Newtonian mechanics),
the energy of the horizon induces a force to a test particle of order $F_h \sim GE_hm/R^2$,
which can be integrated to obtain a potential
\be \label{eq:HDEentropicV}
  V_h\sim -{R_h m \over R} = -C^2m/2,
\ee
where after the integration one can take the limit $R\rightarrow R_h$,
and $C$ is a constant reflecting the order one arbitrarily.
Using standard argument leading to Newtonian cosmology,
this potential term for a test particle will show up in the Friedmann equation
as a DE component $\rho_{de} = 3c^2M_p^2R_h^{-2}$.
Again it is the form of HDE.

\

\subsubsection{HDE from Action Principle}
\label{sec:3.2.5}

Most DE models are from the action principle.
It is argued that the form of HDE can also be derived from the action principle \cite{2012arXiv1210.0966L}.

The FLRW metric can be rewritten as
\be \label{eq:newmetric}
ds^2=-N^2(t)dt^2+a^2(t)[\frac{dr^2}{1-k r^2}+r^2d\Omega_2^2].
\ee
Now consider the action
\be \label{eq:newaction}
S=\frac{1}{16\pi G}\int dt[\sqrt{-g}(R-\frac{2C}{a^2(t) L^2(t)})-\lambda(t)(\dot{L}(t)+\frac{N(t)}{a(t)})]+S_{m},
\ee
where $R$ is the Ricci scalar, $\sqrt{-g}=N a^3$, and $S_{m}$ denotes the action of matter.
Note that the first two terms in the action are just the Einstein-Hilbert action plus the energy density of HDE.
As for the last term, $\lambda(t)$ is just a Lagrange multiplier, which forces the cut-off in the energy density is given by the event horizon.
Please note that $\dot{L}(t)+N(t)/a(t))=0$ is a local variant of the definition of event horizon.
By taking the variations of $N, a, \lambda, L$, and redefining $Ndt$ as $dt$,
one can obtain the corresponding equations of motion
\ba
(\frac{\dot{a}}{a})^2+\frac{k}{a^2}=\frac{C}{3a^2L^2}+\frac{\lambda}{6a^4}+\frac{8\pi}{3}\rho_{m},\nonumber \\
\frac{2\ddot{a}a+\dot{a}^2+k}{a^2}=\frac{C}{3a^2L^2}-\frac{\lambda}{6a^4}-8\pi p_{m},
\ea
and
\ba
\dot{L}&=&-\frac{1}{a},\ \ \ \ \ L=\int_t^{\infty}\frac{dt'}{a(t')}+L(a=\infty)\nonumber, \\
\dot{\lambda}&=&-\frac{4ac}{L^3},\ \
\lambda=-\int_0^{t}dt'\frac{4a(t')C}{L^3(t')}+\lambda(a=0).
\ea
In \cite{2012arXiv1210.0966L} the authors proved that $L(a \rightarrow \infty)=0$,
so $aL$ is exactly the future event horizon.
It is remarkable that the above equations of motion make equivalent the local and the global definition of event horizon.
This is an elegant property of HDE.
Moreover, based on the formulas above, one can obtain the energy density of DE
\be \label{Eq:newrhohde}
\rho_{de}=\frac{1}{8\pi G}\left(\frac{C}{a^2L^2}+\frac{\lambda}{2a^4}\right),
\ee
which is characterized by the future event horizon $aL$, and a new term $\frac{\lambda}{2a^4}$.
In this term, the $\lambda(a=0)$ component evolves in the same way as radiation,
thus can be naturally interpreted as dark radiation \cite{Hamann:2010bk}.

\

\subsection{Observational Constraints on The HDE Model}
\label{sec:3.3}

Now, we introduce the observational constraints on the HDE model.

\

\subsubsection{Parameter Estimation for The HDE Model}
\label{sec:3.3.1}

As mentioned in subsection \ref{sec:3.1.3},
the parameter $C$ plays an essential role in determining the evolution of the HDE.
If $C=1$, the EOS of HDE will be asymptotic to that of a cosmological constant,
and the universe will enter the de Sitter phase in the future;
if $C>1$, the EOS of DE will always be greater than -1, and the HDE will behave as quintessence DE;
if $C<1$, the EOS of HDE will eventually cross the phantom boundary $w=-1$,
leading to a phantom universe with big rip as its ultimate fate.
Since the result of $C$ cannot be derived from the theoretical framework of the HDE model,
it is crucial to determine the value of $C$ by using the cosmological observations.

\begin{table*}
\centering
\caption{Some observational constraints on the HDE model obtained in recent ten years.
Both the best-fit values and 1$\sigma$ errors of model parameters (including $\Omega_{m0}$ and $C$) are listed.}
\label{tab:3}
\centering
\begin{tabular}{cccc}
\hline\hline
References  & Observational Data & $\Omega_{m0}$ & $C$ \\ \hline\hline

\cite{Huang:2004wt}  & gold04  & $0.46^{+0.08}_{-0.13}$ & $0.21^{+0.45}_{-0.14}$\\

\hline

\cite{Zhang:2005hs}  & gold04+WMAP3+SDSS(parameter $A$) & $0.28^{+0.03}_{-0.03}$ & $0.81^{+0.23}_{-0.16}$\\

\hline

\cite{Chang:2005ph}  & Chandra(X-ray gas mass fraction) & $0.24^{+0.06}_{-0.05}$ & $0.61^{+0.45}_{-0.21}$\\

\hline

\cite{Zhang:2007sh}  & gold06+WMAP3+SDSS & $0.29^{+0.03}_{-0.03}$ & $0.91^{+0.26}_{-0.18}$\\

\hline

\cite{Ma:2007pd}   & gold06+WMAP3+SDSS+Chandra & $0.276^{+0.017}_{-0.016}$  &  $0.748^{+0.108}_{-0.109}$\\

\hline

\cite{Xu:2012aw}  &  Union2+WMAP7+BAO(SDSS DR7) & $0.273^{+0.017}_{-0.017}$ & $0.696^{+0.074}_{-0.074}$\\

\hline

\cite{Xu:2013mic}  &  Union2.1+WMAP7+BAO+RSD & $0.283^{+0.017}_{-0.017}$ & $0.750^{+0.098}_{-0.100}$\\

\hline

\cite{Li:2013dha}  &  Planck2013+WP+lensing & $0.248^{+0.079}_{-0.079}$ & $0.508^{+0.207}_{-0.207}$\\

\hline

\cite{Wang:2013zca}  &  SNLS3(linear $\beta$)+Planck2013+BAO & $0.288^{+0.015}_{-0.013}$ & $0.768^{+0.112}_{-0.068}$\\

\hline

\end{tabular}
\end{table*}

A large number of research works had been done to constrain the parameter spaces of the HDE model by using various observational data.
In table \ref{tab:3}, we list some observational constraints on the HDE model obtained in recent ten years.
A most distinct feature of this table is that all the combinations of observational data mildly favor the case of $C<1$,
which corresponds to a phantom universe with big rip.
Similar results were obtained in \cite{Shen:2004ck,Kao:2005xp,Yi:2006bw,2012MPLA...2750115Z,Cui:2015oda,Wang:2016fem}.
Moreover, along with the rapid increase of the number of observed data points in recent years,
the error bar of parameter $C$ has become smaller and smaller,
thus this feature has become more and more obvious. Phantom-like universe and big rip has profound implication for the fate of our universe. We shall discuss more about the fate of the universe and possible ways to avoid the big rip singularity in Section \ref{sec:4.7}.

\begin{figure*}
\centering
\includegraphics[width=0.8\textwidth]{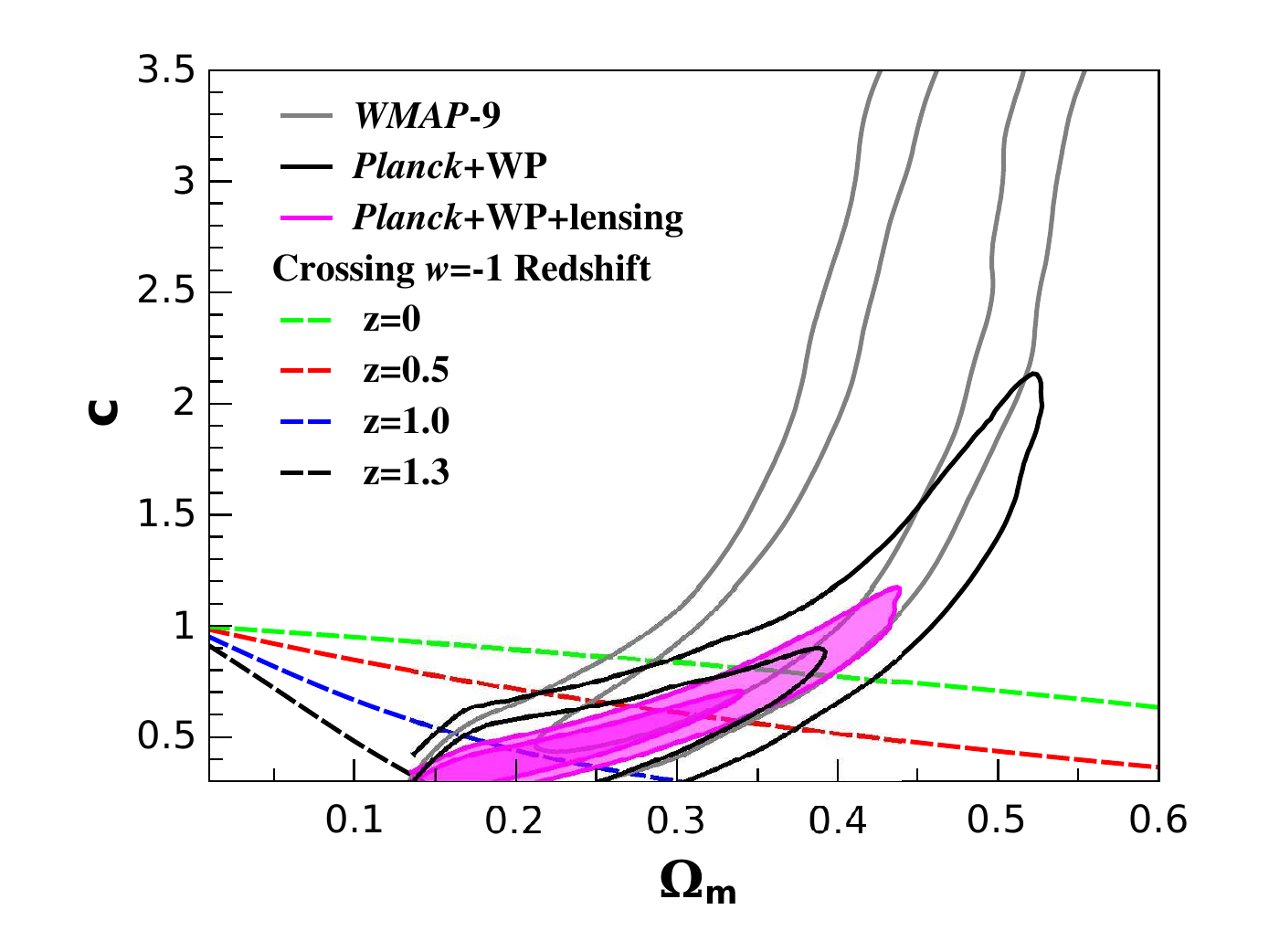}
\caption{The effects of different CMB data on the cosmological constraints of the HDE model.
Both the marginalized 1$\sigma$ and 2$\sigma$ CL contours are plotted in the $\Omega_{m0} - C$ plane.
Dashed lines mark the $w=-1$ crossing at $z = 0, 0.5, 1.0, 1.3$.
From \cite{Li:2013dha}.}
\label{fig5}
\end{figure*}

Some works studied the effects of adopting different observational data on the fitting results of the HDE model.
For example, \cite{Li:2013dha} discussed the effects of different CMB data on the cosmological constraints of the HDE model (see Fig. \ref{fig5}).
It can be seen that the WMAP-9 data alone do not lead to any effective constraint on parameter $C$,
while the Planck+WP results show the preference for $C < 1$ at the 1$\sigma$ confidence level (CL).
Adding the lensing data tightens the constraint,
and the present phantom behavior of HDE is preferred at the more than 1$\sigma$ CL.
Besides, It if found that in the HDE model $\Omega_{m0}$ is constrained to be $0.26 - 0.28$ at 1$\sigma$ CL;
in contrast, using WMAP-9 data alone cannot lead to an effective constraint on $\Omega_{m0}$ in the HDE model.
These results imply that the Planck data can give much better constrains on the HDE model than the WMAP data.

\begin{figure*}
\centering
\includegraphics[width=0.8\textwidth]{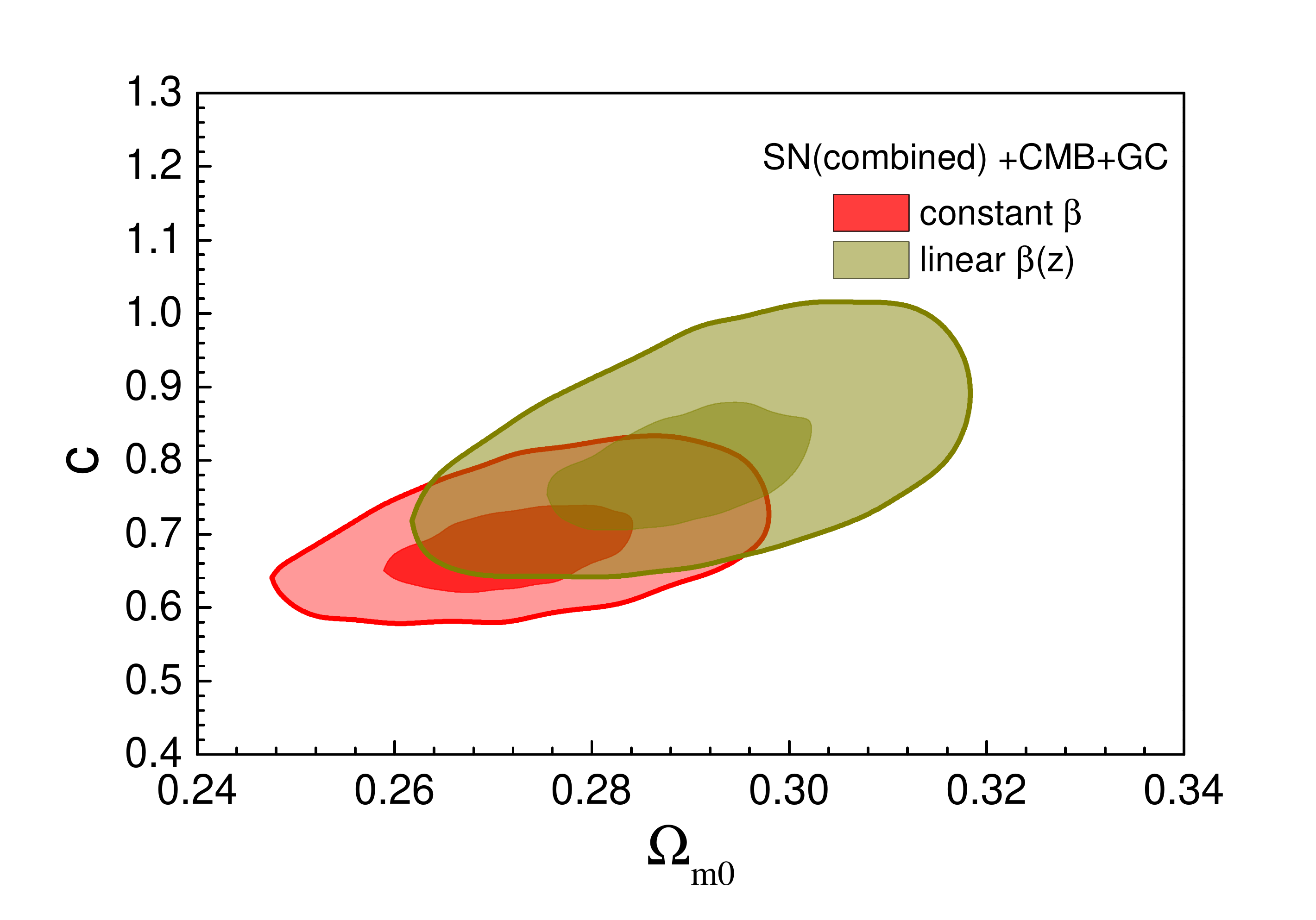}
\caption{The effects of the redshift-dependence of SN color-luminosity parameter $\beta$ on the cosmological constraints of the HDE model.
The marginalized 1$\sigma$ and 2$\sigma$ CL contours are plotted in the $\Omega_{m0} - C$ plane.
Both the results of constant $\beta$ and linear $\beta(z)$ are presented.
From \cite{Wang:2013zca}.}
\label{fig6}
\end{figure*}

In addition, \cite{Wang:2013zca} discussed the effects of the redshift-dependence of SN color-luminosity parameter $\beta$
on the cosmological constraints of the HDE model (see Fig. \ref{fig6}).
It is found that for the constant $\beta$ case, the best-fit result is $\Omega_{m0}=0.274$ and $C=0.687$;
for the linear $\beta(z)$ case, the best-fit result is $\Omega_{m0}=0.288$ and $C=0.768$.
This means that considering the $\beta$'s evolution will enlarge the values of $\Omega_{m0}$ and $C$.
Moreover, for these two cases, the 2$\sigma$ CL ranges of parameter space are quite different.
These results reveal that ignoring the evolution of $\beta$ may cause systematic bias on parameter estimation.
Therefore, the possible evolution of the supernova population with redshift should be taken into account seriously.

\

\subsubsection{More Numerical Studies on The HDE Model}
\label{sec:3.3.2}

\

\begin{itemize}
 \item Statefinder Diagnostic for The HDE Model
\end{itemize}

The scale factor of the Universe $a$ can be Taylor expanded around today's cosmic age $t_0$ as follows:
\be
a(t)=1+\sum\limits_{\emph{n}=1}^{\infty}\frac{A_{\emph{n}}}{n!}[H_0(t-t_0)]^n,
\ee
where
\be \label{eq:An}
A_{\emph{n}}=\frac{a(t)^{(n)}}{a(t)H^n},~~n\in N,
\ee
with $a(t)^{(n)}=d^na(t)/dt^n$.
The Hubble parameter $H(z)$ contains the information of the first derivative of $a(t)$.
The deceleration parameter $q$ is given by
\be
q=-A_2=-\frac{\ddot{a}}{aH^{2}},
\ee
which contains the information of the second derivatives of $a(t)$.
Moreover, different letters of the alphabet have been used to describe higher derivatives of $a(t)$.
For examples, $A_3$ corresponds to the jerk $j$, $A_4$ corresponds to the snap $s$,
and $A_5$ corresponds to the lerk $l$ (See \cite{Visser:2003vq,Capozziello:2008qc,Dunajski:2008tg} and references therein).
These quantities are called geometrical diagnostic in the sense that
they only depend upon the scale factor $a(t)$ and hence upon the metric describing space-time.

A most famous geometrical diagnostic is the so-called ``statefinder'' pair $\{r,s\}$ \cite{Sahni:2002fz}, defined as
\be \label{eq:statefinder}
r\equiv \frac{\stackrel{...}{a}}{aH^3},~~~~s\equiv\frac{r-1}{3(q-1/2)}.
\ee
Note that the statefinder probes the expansion of the universe through the third derivatives of $a(t)$,
and can also be expressed as
\be \label{eq:statefinder r}
  r=1+\frac{9}{2}w(1+w)\Omega_{de}-\frac{9}{2}w'\Omega_{de},
\ee
\be \label{eq:statefinder s}
  s=1+w-\frac{1}{3}\frac{w'}{w}.
\ee
Here the prime denotes derivative with respect to $\ln a$.
It is clear that different cosmological model will yield different evolution trajectories in the $s-r$ plane.
The spatially flat $\Lambda$CDM model corresponds to a fixed point
\be \label{eq:statefinderlcdm}
\{s,r\}\bigg\vert_{\rm \Lambda CDM} = \{ 0,1\}.
\ee
Departure of a given DE model from this fixed point
provides a good way of establishing the ``distance'' between this model and the $\Lambda$CDM model.
As demonstrated in Refs. \cite{Alam:2003sc,Zhang:2005rj,Yi:2007gu,Wang:2008fx},
the Statefinder diagnostic can effectively differentiate between a wide variety of DE models.

\begin{figure*}
\centering
\includegraphics[width=0.8\textwidth]{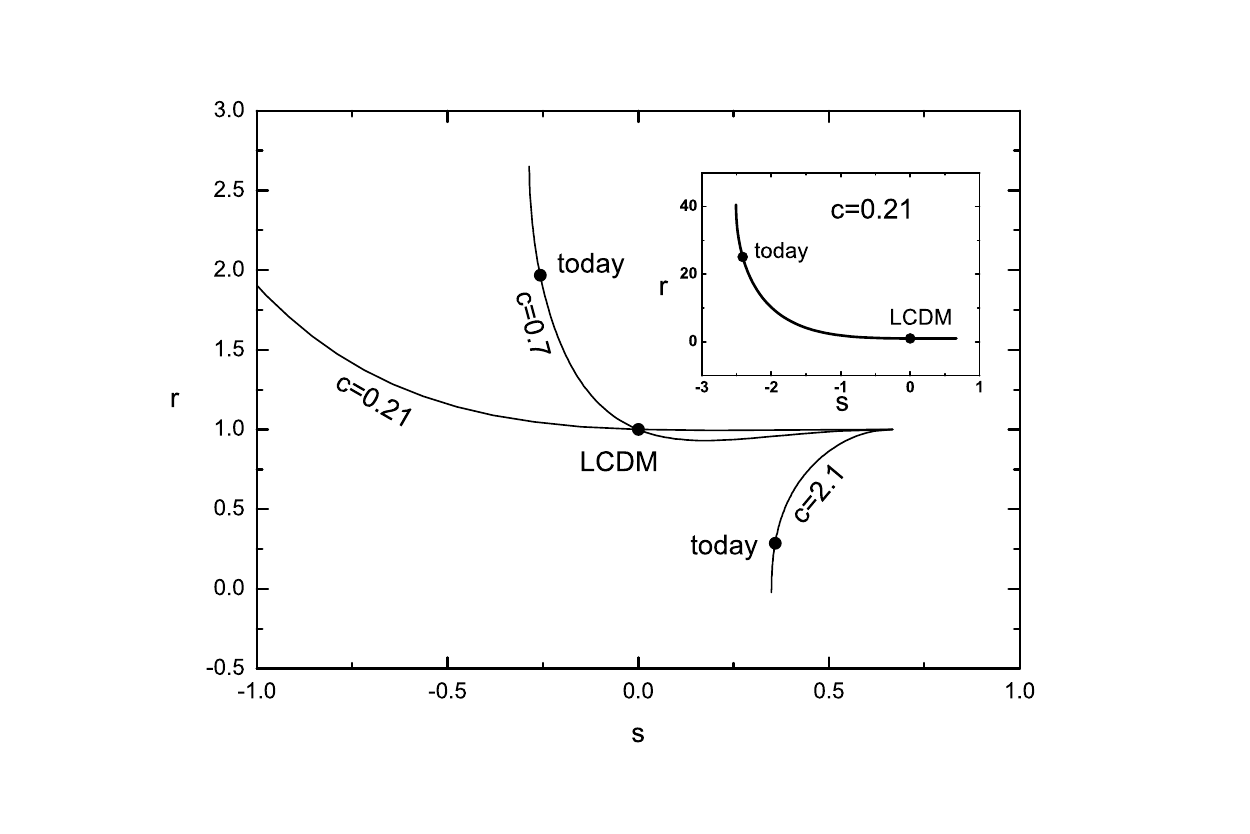}
\caption{The statefinder diagrams $r(s)$ for the HDE model in the cases of $C = 0.21, 0.7, 2.1$, respectively.
The inset shows the complete curve for the case of $C = 0.21$.
The round dots represent the values of today for these cases.
The $\Lambda$CDM model corresponds to a fixed point $\{ 0,1\}$.
From \cite{Zhang:2005yz}.}
\label{fig7}
\end{figure*}

\cite{Zhang:2005yz} studied the HDE model from the viewpoint of statefinder.
The corresponding results are shown in Fig. \ref{fig7}.
It can be seen that the evolutionary trends of the HDE model for the cases of $C < 1$ and $C > 1$ are upward and downward, respectively.
Moreover, for the situations of $C < 1$, the trajectories pass through the $\Lambda$CDM fixed point;
while for the $C > 1$ cases, the tracks never reach the $\Lambda$CDM fixed point.
In addition, for the $C = 0.21$ case, $r$ can arrive at a very large value ($\sim 40$).
So making use of the statefinder diagnostic can easily differentiate the HDE models with different $C$.
In other word, the parameter $C$ plays a crucial role in determining the properties of HDE,
as well as the ultimate fate of the universe.

\

\begin{itemize}
 \item Future Redshift Drift Constraints on The HDE Model
\end{itemize}

As introduced in subsection \ref{sec:2.3.4},
the ``redshift drift'' technique can directly measure the expansion rate of the universe in the redshift desert $2\leq z \leq5\leq$
by detecting the redshift variation in the absorption-line spectra of Lyman-$\alpha$ forest of distant quasars.
The redshift variation is defined as \cite{1998ApJ...499L.111L},
\be \label{eq:SL}
 \Delta v \equiv \frac{\Delta z}{1+z}=H_0\Delta t_o\bigg[1-\frac{E(z)}{1+z}\bigg],
\ee
where $\Delta t_o$ is the time interval of observation, which is often set as 10, 20 or 30 years.
According to the Monte Carlo simulations, the uncertainty of $\Delta v$ expected by the cosmic dynamics experiment
can be expressed as \cite{2008MNRAS.386.1192L}
\be \label{eq:SLerror}
\sigma_{\Delta v}=1.35 \bigg(\frac{S/N}{2370}\bigg)^{-1}\bigg(\frac{N_{\mathrm{QSO}}}{30}\bigg)^{-1/2}
\bigg(\frac{1+z_{\mathrm{QSO}}}{5}\bigg)^{x}~\mathrm{cm}/\mathrm{s},
\ee
where $S/N$ is the signal-to-noise ratio,
$N_{\mathrm{QSO}}$ is the number of observed quasars, $z_{\mathrm{QSO}}$ represents their redshift,
and the last exponent $x=-1.7$ for $2<z<4$ and $x=-0.9$ for $z>4$.
\footnote{A previous expression of $\sigma_{\Delta v}$ was given in \cite{2005Msngr.122...10P}.}
By simulating the Sandage-Loeb (SL) test data uniformly distributed over the redshift bin of $z_{\mathrm{QSO}} \in [2, 5]$,
one can obtain the future redshift drift constraints
on various cosmological models \cite{Zhang:2007zga,2012PhRvD..86l3001M,2013PhRvD..88b3003L,2014JCAP...07..006G,2014JCAP...12..018G}.

\begin{figure*}
\centering
\includegraphics[width=0.8\textwidth]{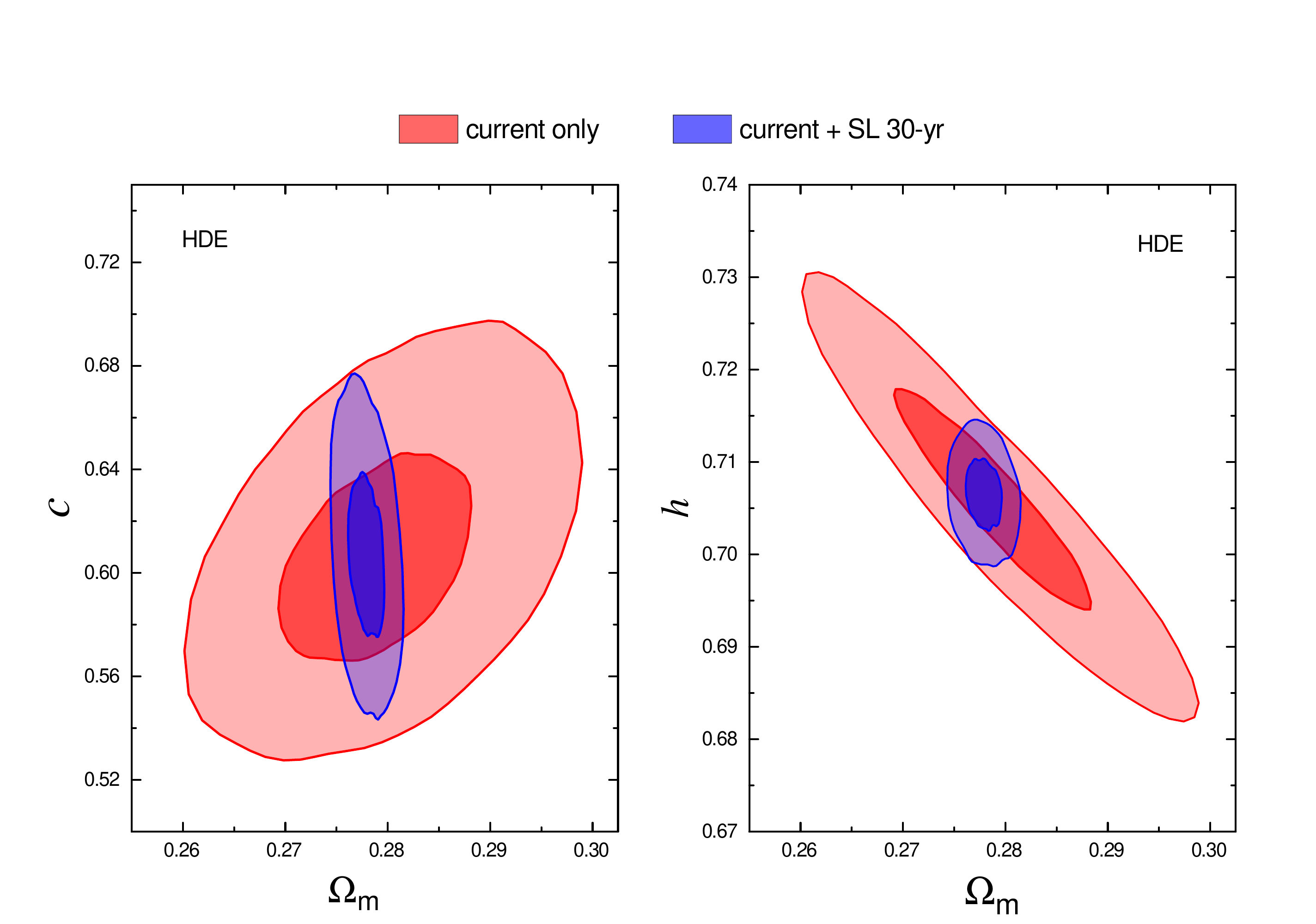}
\caption{The 1$\sigma$ and 2$\sigma$ CL contours
in the $\Omega_{m0}-C$ plane (left panel) and $\Omega_{m0}-h$ plane (right panel) for the HDE model.
For comparison, both the results given by current only and current+SL 30-year data are plotted in this figure.
From \cite{2016arXiv160705643H}.}
\label{fig8}
\end{figure*}

This method had also been used to study the HDE model.
For example, \cite{Zhang:2007zga} simulated the SL 10-year data of 240 quasars,
and used them to forecast the future redshift drift constraints on HDE.
It is found that SL test can provide a extremely strong bound on $\Omega_{m0}$,
while its constraint on $C$ is rather weak.
In addition, \cite{2016arXiv160705643H} explored the impact of SL test on the precision of cosmological constraints for the HDE models,
by adding 30 simulated SL 30-year data to the current observational data (A combination of SNIa+BAO+CMB+$H_0$ data).
The corresponding results are shown in Fig. \ref{fig8}.
It can be seen that adding the SL test data effectively breaks the existing strong degeneracy between various model parameters,
and thus gives much tighter limits on these parameters.
Therefore, future redshift drift measurements have great potential to significantly improve the observational constraints on the HDE model.

\

\section{More Topics in HDE Cosmology}
\label{sec:4}

The application of HP to DE has profound implications.
Compared to a featureless cosmological constant,
much more questions can be asked within the context of HDE
\footnote{For more examples, see \cite{Naderi:2014awa,Mehrabi:2015kta,vanPutten:2015wma}.}.
In this section, we will explore some of those questions.

\

\subsection{Spatial Curvature}
\label{sec:4.1}

As already noted in \cite{Li:2004rb}, when HDE is a sub-dominant component, the energy density of HDE scales as spatial curvature: $\rho_{de}\propto a^{-2}$. This is conceptually different from the cosmological constant, whose energy density does not change no matter dominant or not.

In the conventional picture of $\Lambda$CDM + inflation, our spatial curvature is very likely to be negligible. This is because there is no reason that inflation happens for 60 e-folds sharp. Once inflation lasts much longer than the last 60 observable e-folds, which is typical for inflation model construction, the spatial curvature is diluted much more. As result, today we should observe negligible spatial curvature.

However, the framework of HDE predicts 60 e-folds of inflation. Thus the spatial curvature of our present universe may not be many orders of magnitudes less than order one. In this subsection we review the consequences of the spatial curvature in our present universe. Issues more related to inflation will be reviewed in Subsection \ref{sec:4.5}.

In \cite{Huang:2004ai}, HDE in a closed universe is studied. Besides the motivations mentioned above, it is also noted that for $C=1$, the future of our universe with only HDE is de Sitter space. And de Sitter space can be sliced into homogeneous and isotropic sections with positive spatial curvature as well. (One can also slice de Sitter into spatial sections with negative spatial curvature, for example the static patch metric of de Sitter. However, this is not a homogeneous and isotropic solution and thus not describing cosmology). Thus a consistent description of HDE should include the study of HDE in closed universe. We shall here adopt the convention that $a_0=1$ for our present universe.

With spatial curvature, the defining equation of HDE still applies:
\be
    \rho_{de} = 3 C^2 M_p^2 L^{-2}~.
\ee
To find $L$, one note that in non-flat space
\be
  \label{eq:ehc}
  \int_0^{r(t)} \frac{dr}{\sqrt{1-kr^2}}  = \int_t^\infty \frac{dt}{a} = \frac{R_h}{a}~.
\ee
One can then solve $r(t)$ from Eq. \ref{eq:ehc}:
\be
  r(t) = \frac{1}{\sqrt{k}}  \sin \left ( \frac{\sqrt{k} R_h}{a} \right )~.
\ee
It is natural to choose the IR cutoff as
\be
  \label{eq:decL1}
  L = a r(t)~.
\ee

Similarly to the case without spatial curvature,
it is convenient to study the EoS $w$ and the evolution of DE using variable $\Omega_{de} \equiv \rho_{de}/\rho_{c}$,
where $\rho_c \equiv 3M_{P}^2H^2$ is the critical density of the universe.
In a closed universe, we still have
\be
  HL = \frac{C}{\sqrt{\Omega_{de}}} ~.
\ee
Inserting the IR cutoff $L$ in Eq. \ref{eq:decL1}, one obtains
\be
  \dot L = HL + a \dot r
  = \frac{c}{\sqrt{\Omega_{de}}} - \sqrt{1-kr^2} = \frac{C}{\sqrt{\Omega_{de}}} - \cos \left ( \frac{\sqrt{k} R_h}{a}  \right )~.
\ee
Comparing this result with the definition of $w$ (where $\dot\rho_{de}$ can be calculated using $\dot L$), one can read off the equation of state
\be
  w = - \frac{1}{3} \left [ 1 + \frac{C}{2} \sqrt{\Omega_{de}} \cos \left ( \frac{\sqrt{k} R_h}{a}  \right ) \right ]~.
\ee

The time evolution of HDE with spatial curvature can be solved by the master equation
\be
  \frac{\Omega'_{de}}{\Omega^2_{de}} = ( 1- \Omega_{de}) \left [ \frac{2}{C} \frac{1}{\sqrt{\Omega_{de}}} \cos \left ( \frac{\sqrt{k} R_h}{a} \right ) + \frac{1}{1 + a\Omega_{k0}/\Omega_{de0}} \frac{1}{\Omega_{de}} \right ]~.
\ee
which can be derived similarly as the flat space case. (We here adopt the convention that $\Omega_{k} + \Omega_{m} + \Omega_{r} + \Omega_{de}=1$. In \cite{Li:2004rb} the convention of $\Omega_{k}$ differs by a sign.)

In such a universe, the fate of the universe is similar to the case of HDE in flat space. When $C=1$, the energy density of DE will approach to a constant. When $C>1$, the energy density will eventually be diluted due to cosmological expansion.

In \cite{Gong:2004cb}, the formalism is generalized to include open universe cosmology. It is noted that by defining a function
${\rm sinn}(x) \equiv \sin(x)$, $x$, $\sinh(x)$ for $\Omega_k<0$, $\Omega_k=0$, and $\Omega_k>0$,
and replacing $\sqrt{k}$ with $\sqrt{|k|}$, the formalism also covers open universes. The model has been tested against observations. Using the SN data, a closed universe is favored at $\Omega_{k0} = -0.35^{+0.38}_{-0.17}$ and $C=1.0^{+0}_{-0.17}$
\footnote{where the upper bound is set as a prior as $C\leq 1$. We here adopt the convention that $\Omega_{k} + \Omega_{m} + \Omega_{r} + \Omega_{de}=1$. In \cite{Gong:2004cb} the convention of $\Omega_{k}$ differs by a sign.}.
When using both SN and CMB data, the best fit value favors a flatter universe at $\Omega_{k0} = -0.02 \pm 0.10$ and $C=0.84_{-0.03}^{+0.16}$.
In both cases a flat universe is consistent with observations.

Another approach for adding spatial curvature to HDE is proposed by \cite{Zhang:2014ija}. In this approach, the future event horizon $R_h$ is directly used as the IR cutoff, instead of using $a r(t)$. Following similar calculations, the equation of state of HDE in this approach is
\be
  w = - \frac{1}{3} - \frac{2}{3C} \sqrt{\Omega_{de}}~,
\ee
which is identical to the equation of state of HDE in flat universe. But note that $\Omega_{de}$ evolves differently in flat universe and non-flat universes.

\cite{Zhang:2014ija} also provided updated observational constraints on the HDE parameters, by using the combined SN+BAO+CMB+$H_0$ data.
For the original model of HDE with curvature \cite{Li:2004rb, Gong:2004cb},
$C=0.644^{+0.057}_{-0.043}$ and $\Omega_{k0} = 1.582^{+2.401}_{-3.045}\times 10^{-3}$.
For the case of $L=R_h$, $C=0.654^{+0.052}_{-0.051}$ and $\Omega_{k0} = 4.902^{+3.024}_{-2.705}\times 10^{-3}$.
Here although a slightly open universe is favored,
one should note that using the same data, $\Lambda$CDM favors an even opener universe with $\Omega_{k0} = 7.636^{+5.821}_{-5.284}\times 10^{-3}$.
\footnote{There is an important difference between the data fitting process
of the Planck 2015 paper \cite{2015arXiv150201589P} and the work of Zhang et al. \cite{Zhang:2014ija}.
In \cite{2015arXiv150201589P}, the full CMB power spectrum are used to constrain $\Lambda$CDM model;
while in \cite{Zhang:2014ija}, only the CMB distance prior data are used to constrain $\Lambda$CDM model.
Therefore, although the fitting results of these two papers are slightly different, both of them are reliable.}
Further, \cite{Zhang:2014ija} considered the time evolution of color-luminosity parameter $\beta$,
and in that case, a open universe is favored at about $2\sigma$.

\

\subsection{Neutrino}
\label{sec:4.2}

Neutrino physics has become an increasingly important part of modern cosmology (see, for example, \cite{Halzen:1998mb, deGouvea:2004gd} and references therein). Currently the masses of the neutrinos are not known. Only the mass hierarchy is known to be
\be
\Delta m_{21}^2=(7.65\pm 0.65)\times 10^{-5}\mathrm{eV}^2~,
\ee
and
\be
\Delta m_{32}^2=(2.40\pm 0.35)\times 10^{-3}\mathrm{eV}^2.
\ee
Thus at least one neutrino has mass larger than $0.04$eV. The mass hierarchy of the neutrinos may be distributed in two ways, namely the normal hierarchy, where the 3rd generation neutrino has greater mass than the first two generations; and the inverted hierarchy, where the 3rd generation neutrino has less mass than the first two generations. The normal hierarchy predicts that the total mass of the three generations has lower bound
\be
  \sum m_\nu \geq 0.05 \mathrm{eV}~,
\ee
and the inverted hierarchy predicts that
\be
  \sum m_\nu \geq 0.1 \mathrm{eV}~.
\ee

In cosmology, the neutrinos are ultra-relativistic (radiation) in the early universe and have become non-relativistic (matter) now. It happens that the mass scale of neutrinos is close to the energy scale of recombination. As a result, CMB physics is sensitive to the mass of neutrinos. At the background level the matter radiation equality time is shifted and at the perturbation level free streaming smooths the small scale power spectrum \cite{Bashinsky:2003tk}. As a result, the CMB physics gives the strongest to date upper bound on the total mass of neutrinos, $\sum m_\nu < 0.17$eV (95\% CL, Planck15 TT, TE, EE + lowP + BAO) \cite{2015arXiv150201589P}. This greatly narrows down the parameter space of $\sum m_\nu$ (compared to the SN neutrino mass bound $\sum m_\nu<1$eV and the beta decay bound $\sum m_\nu<6$eV), and is not far from the neutrino mass lower bound of inverted hierarchy.

\begin{figure*}
\centering
\includegraphics[width=0.8\textwidth]{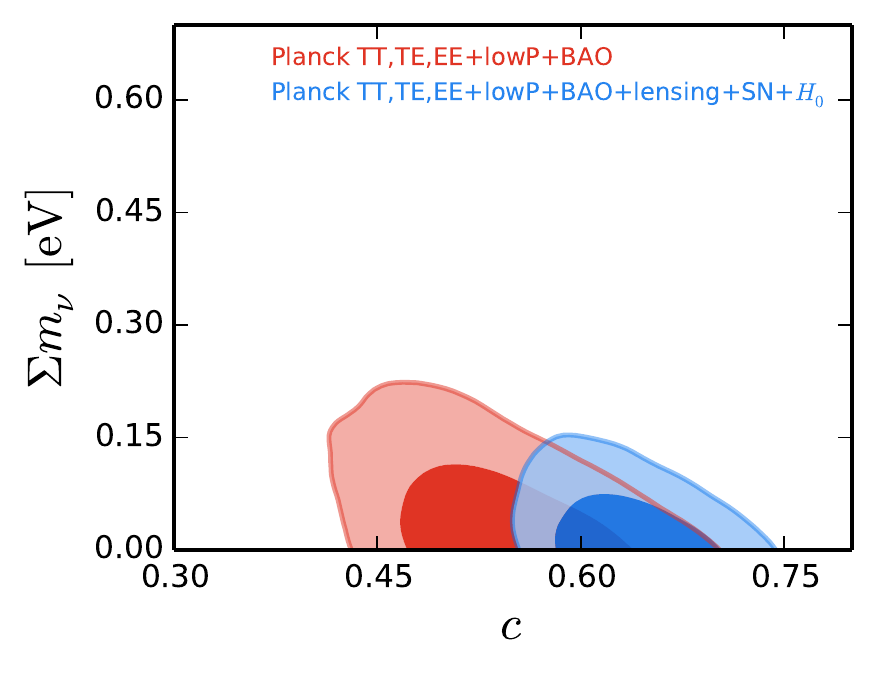}
\caption{marginalized constraints ($1\sigma$ and $2\sigma$ CL) in the $\sum m_\nu$--$C$ plane for the HDE model,
from the Planck TT,TE,EE+lowP+BAO (red) and Planck TT,TE,EE+lowP+BAO+lensing+SN+$H_{0}$ (blue) data combinations.
From \cite{Zhang:2015uhk}.}
\label{fig9}
\end{figure*}

The impact of neutrinos in the context of HDE is studied in details.
A full Markov Chain Monte Carlo exploration of HDE with spatial curvature and massive neutrinos is presented in \cite{Wang:2012uf}.
The study of neutrino in HDE in light of Planck 2015 data is presented in \cite{Zhang:2015uhk} and \cite{2016arXiv160800672W}.
To address the DE perturbations (as impacted by neutrino),
the PPF framework \cite{Fang:2008sn} is used where the HDE perturbations are modeled as scalar field perturbations.
Interestingly, the degeneracy between $\sum m_\nu$ and $H_0$ is different between $\Lambda$CDM and HDE \cite{Zhang:2015uhk}.
In $\Lambda$CDM, these two parameters has anti-correlation but in HDE these two parameters are positively correlated.
As a result, much tighter bound of neutrino mass can be obtained in HDE compared with $\Lambda$CDM.
At $2\sigma$ CL, the Planck 2015 TT, TE, EE + lowP + BAO + lensing + SN + $H_0$ bound is (See Fig.~\ref{fig9})
\be
  \sum m_\nu < 0.197 \quad (\Lambda\mathrm{CDM})~,
\ee
and
\be
  \sum m_\nu < 0.113 \quad (HDE)~.
\ee
Thus in HDE, the total neutrino mass is approaching the bound of the inverted hierarchy.
In \cite{2016arXiv160800672W} the normal hierarchy and inverted hierarchy are considered separately with HDE.
It is found that the minimal $\chi^2$ is smaller in the case of normal hierarchy,
though the difference $\Delta \chi^2$ (about 4 in a combined analysis) is not yet enough to distinguish normal hierarchy
and inverted hierarchy given the observational data to-date.

\

\subsection{Instability of Perturbation}
\label{sec:4.3}

A cosmological constant $\Lambda$ does not have fluctuations. Once dynamical DE is considered, fluctuations has to be considered, which affects evolution of the gravitational potential and growth of structure. For scalar field DE models, the fluctuation of the scalar field can be studied using field theory and is implemented in cosmological codes such as the PPF \cite{Fang:2008sn}. Indeed, in the literature the HDE is sometimes modeled by PPF, but as HDE is by nature a cosmological constant with nontrivial UV-IR relations, theoretical study of the HDE perturbations is deserved, including the stability of HDE, and the nature of its fluctuations.

The stability of HDE is first studied in \cite{Myung:2007pn}. Unfortunately, the analysis is based on sound speed, which is within the framework of fluid and does not note the unique feature of HDE. In \cite{Li:2008zq}, the analysis of HDE is performed based on the defining nature of HDE.

As the nature of HDE is not totally clear, assumptions have to be made for the study of stability issue. In \cite{Li:2008zq}, two assumptions are made, namely, the fluctuation of HDE completely comes from the fluctuation of the size of future event horizon (as evident in the original definition of HDE); and spherical symmetric type of fluctuations (for technical simplicity).
It is convenient to use the Newtonian gauge. The perturbed metric is
\be\label{eq:Newtonian-gauge}
  ds^2 = - [1+2\Phi(r,t)] dt^2 + a^2(t) [1-2\Phi(r,t)] d \mathbf{x}^2~.
\ee
With the scalar type fluctuations, the future event horizon becomes
\be
  R_h(0,t) = \int_0^{r_h(t)} a(t)[1-\Phi(r,t)]dr~,
\ee
where $r_h(t)$ is the the coordinate distance to the future event horizon, which can be written as
\be
  r_h \equiv r_{h0} + \delta r_h~, \quad
  \delta r_h = \int_t^\infty \frac{2\Phi(r_{h0}(t'), t')}{a(t')}~.
\ee
Thus the fluctuation of the future event horizon can be written as
\be
  \delta R_h(0,t) \equiv R_h(0,t) - R_{h0}
  = a(t) \left [
    \int_t^\infty \frac{2\Phi(r_{h0}(t'),t')}{a(t')}dt'
    - \int_0^{r_{h0}} \Phi(r,t) dr
  \right ]~,
\ee
and the HDE energy density has fluctuation
\be
  \delta \rho_{de} = -2 \rho_{de} \frac{\delta R_h}{R_h}~.
\ee

\begin{figure}[htbp]
  \centering
  \includegraphics[width=0.4\textwidth]{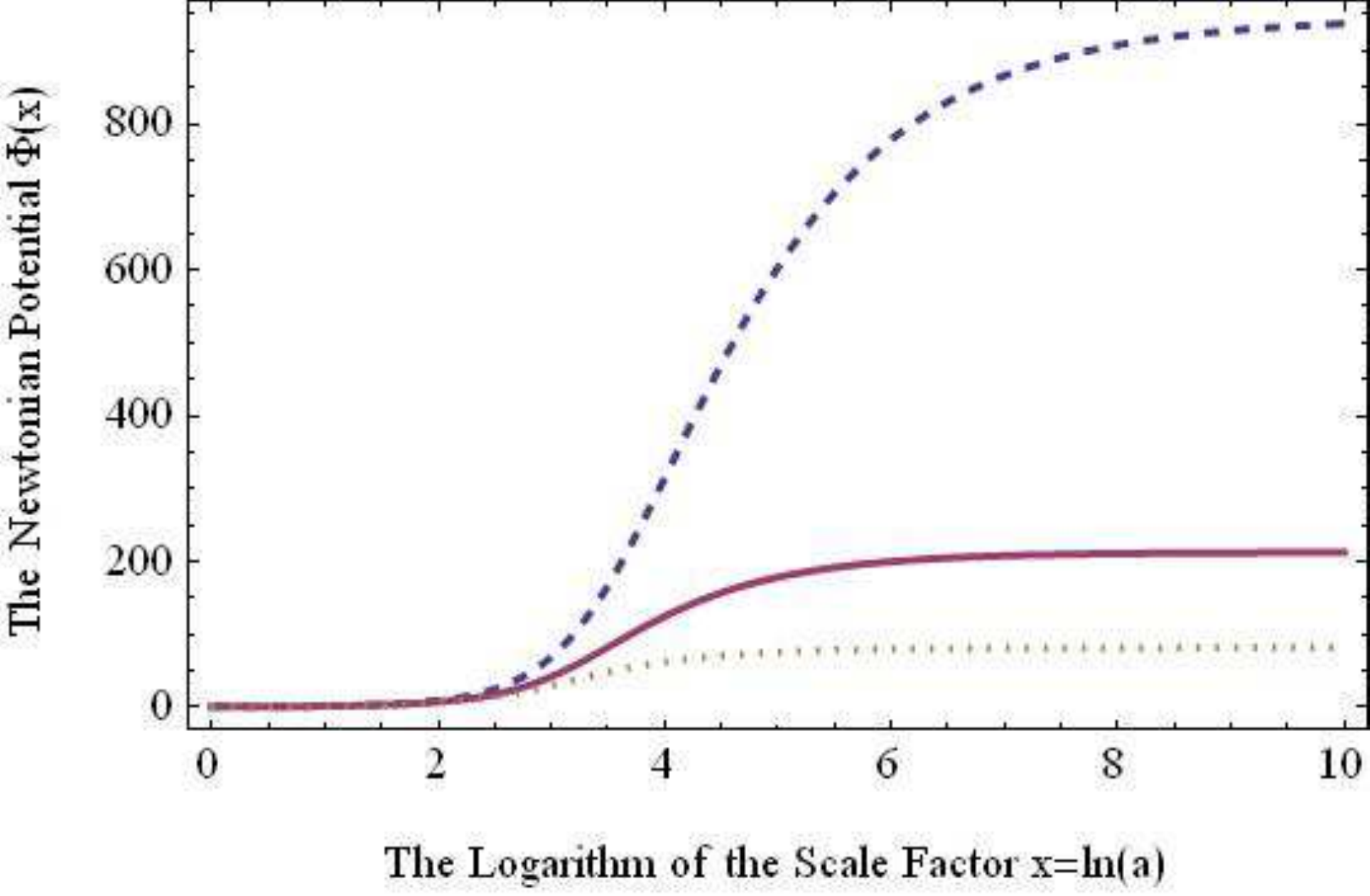}\hspace{0.1\textwidth}
  \includegraphics[width=0.4\textwidth]{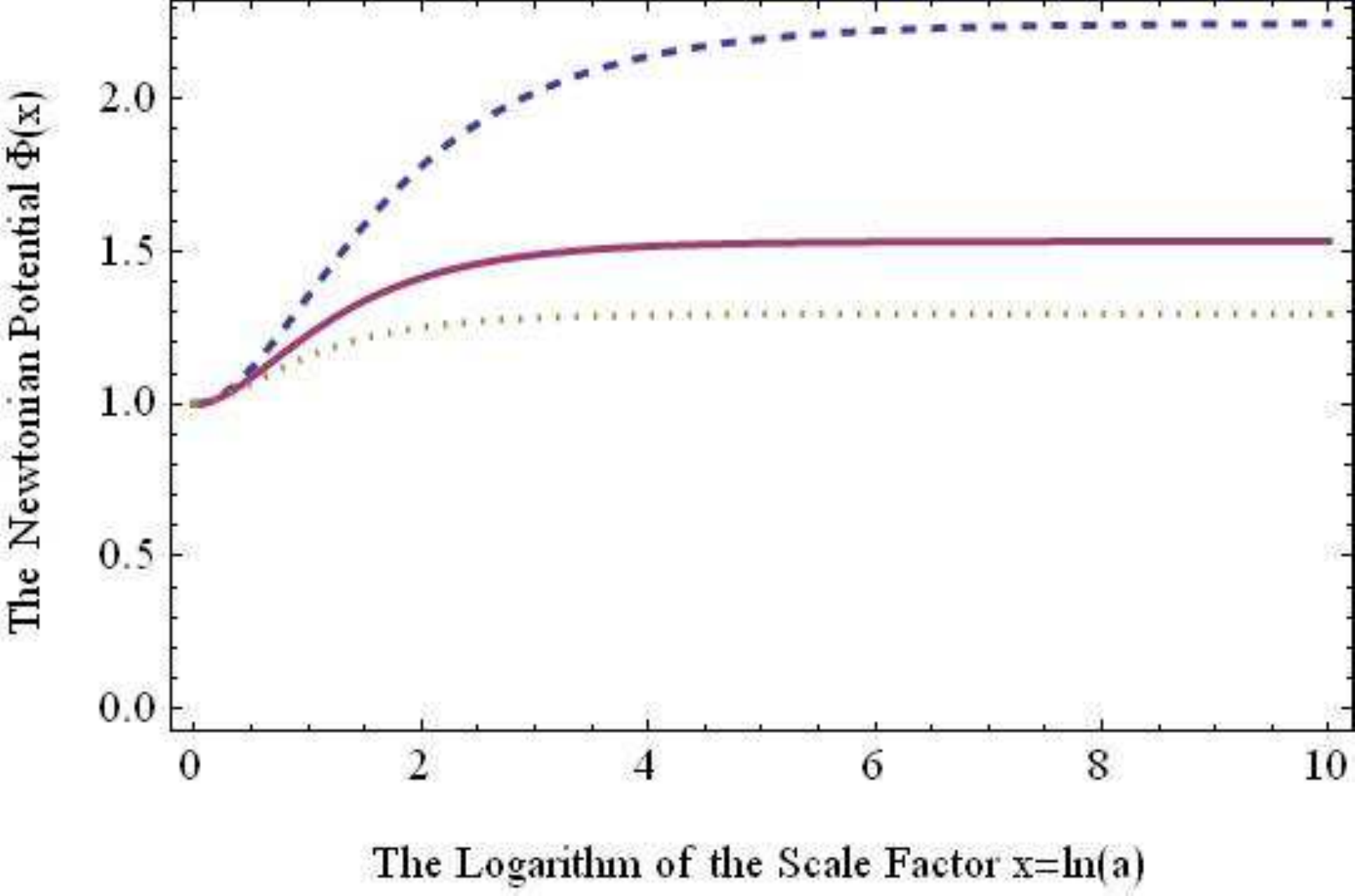}
  \caption{\label{fig10} The perturbation of HDE in matter dominate era with $\Omega_{de}=0.01$ (left panel) and DE dominated era with $\Omega_{de}=0.72$ (right panel). The blue dashed, red solid and yellow dotted lines are for $C = 0.8$, $C = 1.0$ and $C = 1.2$, respectively. One observes that in all cases, the perturbations are stable at late times. From \cite{Li:2008zq}.}
\end{figure}

To study the implication of such perturbations, one can insert the fluctuations into the Einstein equations. The resulting integral-differential equation is difficult to solve precisely. However, fortunately in the sub-Hubble limit and in the super-Hubble limit, the behavior can be studied by analytical method. It can be shown that the sub-Hubble fluctuations are decaying modes; while the super-Hubble fluctuations approaches to a constant. During the process of Hubble-crossing, the fluctuation can grow. But the Hubble-crossing happens for a short period of time and the growth of fluctuation is bounded. For example, when $C=0.8$, in the mater dominated era the HDE fluctuation can grow by $O(100)$ at horizon crossing, and in the HDE domination about 2 times. For larger $C$ the growth is less significant. The evolution of HDE in matter dominated and DE dominated eras are plotted in Fig.~\ref{fig10}.

Thus one gets the conclusion that the HDE fluctuations are stable (there is no unbounded growth). If the quantum initial fluctuation of HDE is small enough, then it is consistent to study HDE in the late universe without fluctuations.

In \cite{Li:2008zq} only classical fluctuations are studied. It remains interesting to study the quantum initial fluctuations of the HDE. More input from quantum gravity and HP may be needed to complete such a study.

\

\subsection{Time-Varying Gravitational Constant}
\label{sec:4.4}

As a non-renormalizable theory, gravity is believed to be sensitive to the UV physics. In the framework of HDE, the UV cutoff is determined by the IR scales under consideration. Thus it is possible that the gravity theory is modified in the framework of HDE. Here we review a simply possibility where the Newton's gravity constant is modified with time dependence \cite{Jamil:2009sq, Lu:2009iv, Alavirad:2013nfy}. More complicated modified gravity models with HDE will be reviewed later at Section \ref{sec:6}.

Theoretically, the attempts to study time-varying gravitational constant dates back to \cite{Dirac:1938mt} in the framework of Kaluza-Klein compactification with a varying volume of extra dimension. Since then in numerable models of modified gravity arise with varying gravitational constant. Putting the time variation of gravitational constant together with HDE, one can derive the new master equation for HDE
\be
  \Omega'_{de} = \Omega_{de} (1-\Omega_{de})
  \left [ 1 + \frac{2\sqrt{\Omega_{de}}}{C} \right ]
  - \Omega_{de} (1-\Omega_{de}) \frac{G'}{G} ~.
\ee
The solution of this equation of course depends on the explicit time dependence of $G$. Analytical solution can be obtained in simple cases, for example in the case where $\Delta_G \equiv G'/G = \dot G / (HG)$ is a constant.

Observationally, the time variation of the Newton's constant G is constrained from astrophysics to be
\be \label{eq:varg1}
  \left| \frac{\dot G}{G}  \right| < 4.10\times 10^{-11}\mathrm{yr}^{-1}
\ee
from the Hulse-Taylor binary pulsar, helio-seismological data, SNIa and pulsating white dwarf star G117-B15A (see \cite{Lu:2009iv} and the references therein). The Big Bang nuclei-synthesis (BBN) gives a much tighter bound \cite{Copi:2003xd}
\be \label{eq:varg2}
  -3.0\times 10^{-13} \leq \frac{\dot G}{G} \leq 4.0 \times 10^{-13} \mathrm{yr}^{-1}~.
\ee
However, this tight bound is based on two data points at the time of BBN and now. It would still be possible that the gravitational constant has some significant change in between in a non-monotonic way. Thus whether to use constraint in Eq. \ref{eq:varg1} or Eq. \ref{eq:varg2} is a model dependent choice.

This HDE scenario with time-varying $G$ is tested against data. In a flat universe, based on SN, BAO, CMB and Hubble data, it is shown \cite{Lu:2009iv} that, a time-independent value of Newton's constant is consistent with observations, with $\Delta_G = -0.0016^{+0.0049}_{-0.0049}$ at 1$\sigma$ confidence level. The corresponding constraint on $C$ is $0.80^{+0.16}_{-0.13}$. Allowing spatial curvature, similar analysis obtains $\Delta_G = -0.0025^{+0.0080}_{-0.0050}$ at 1$\sigma$ confidence level, with $C=0.80^{+0.19}_{-0.14}$, which is again consistent with the null hypothesis.

\

\subsection{Inflation}
\label{sec:4.5}

HDE originates from a UV/IR relation. It is interesting to see that in terms of the physics phenomena, HDE also relates UV and IR in a special way. There have been two known periods of cosmological accelerations in our universe. One is inflation happening at an extremely UV scale (and is the highest energy scale that we have so far probed); the other is DE, related to the extreme IR scale. In HDE, those two epochs of physics are connected by the cosmological coincidence problem. The cosmic coincidence problem is solved by that we have about 60-e-folds of inflation\footnote{In fact, the number 60 is approximate and depend on the detail of reheating. But the key point is that, at the start of observable inflation (where the scale corresponding to the current size of observable universe exits the Hubble horizon), HDE energy density should be comparable with the inflaton energy density. This is because when HDE is subdominant, HDE energy density scales as curvature. The conventional argument about the importance of curvature at the beginning of inflation applies here.}. On the other hand, HDE become non-negligible again at the beginning of inflation\footnote{Note that the inflationary Hubble radius is not a future event horizon (as inflation eventually ends). Thus if one choose future event horizon as the IR cutoff to define HDE, HDE does not have to dominate throughout inflation.}.

The relation between HDE and inflation was already noticed in the original work of HDE \cite{Li:2004rb}. In \cite{Chen:2006qy}, the impact of HDE for inflation is studied in more details.

It is assumed that HDE does not have density fluctuations. Although the quantum fluctuation of HDE during inflation is not yet known, it is noticed that the sub-Hubble HDE fluctuations decay \cite{Li:2008zq}. Thus the vanishing HDE fluctuation is the most natural choice for inflation.

For simplicity, HDE is considered together with the minimal single field slow roll inflation. The Friedmann equation during inflation takes the form
\be
  3M_p^2 H^2 = \frac{1}{2} \dot\varphi^2 + V(\varphi) + 3 C^2 M_p^2 R_h^{-2}~.
\ee
Also, as the simplest possibility, it is assumed that HDE does not couple to the inflaton. Thus the evolution equation for $\varphi$ is still the conventional continuity equation
\be
  \ddot \varphi + 3H \dot\varphi + V_\varphi = 0~.
\ee
They are the master equations for the homogeneous and isotropic background evolution. To derive the time evolution of HDE, we note that the trick of using $\Omega_{de}$ to solve the HDE sector is still valid. One gets
\be
  \Omega'_{de} = -2 \Omega_{de} (1-\Omega_{de}) \left ( 1- \frac{\sqrt{\Omega_{de}}}{c}  \right )~,
\ee
where prime is with respect to $\ln a$. This equation can be solved (for example, using Mathematica). As expected, HDE starts from a dominate value, and after a few e-folds of inflation, HDE is diluted exponentially (similar to curvature).

\begin{figure}[htbp]
  \centering
  \includegraphics[width=0.8\textwidth]{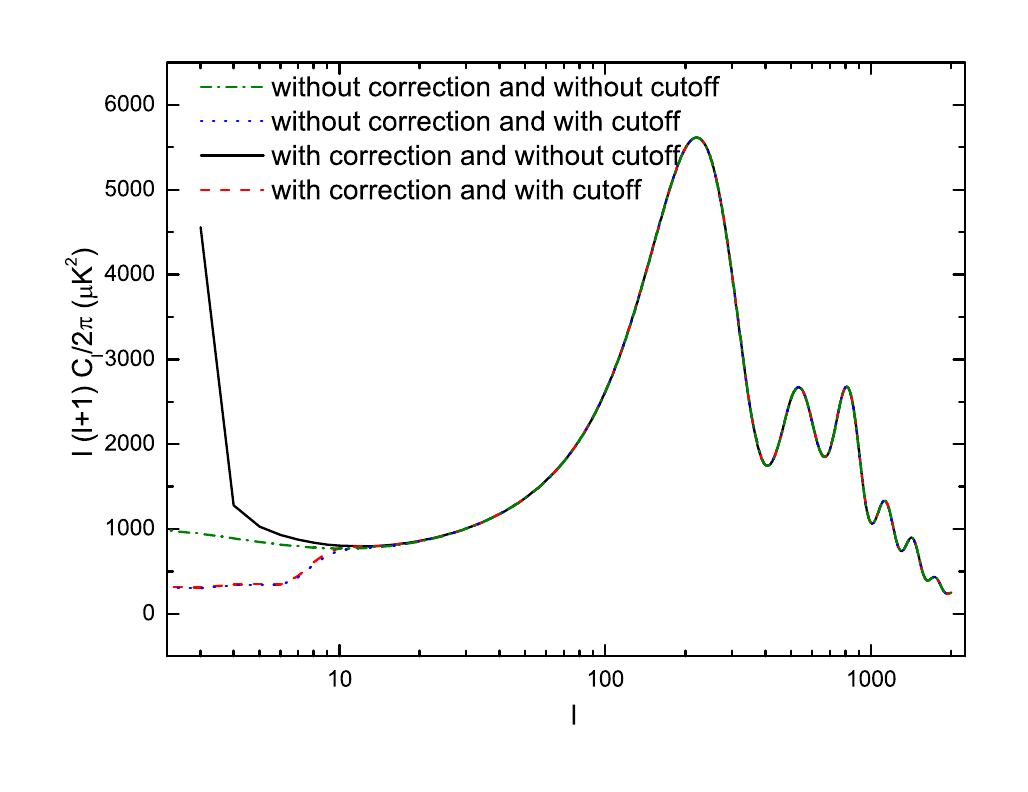}
  \caption{\label{fig11} The HDE impact on the CMB temperature power spectrum. The situations with and without HDE correction \ref{eq:HDEinflation}, and the cutoff proposed in \cite{Enqvist:2004xv} are plotted. From \cite{Chen:2006qy}.}
\end{figure}

To study the perturbation theory with HDE, one follows the assumption that HDE does not introduce additional fluctuations to the theory. As a result, the fluctuations of the gravitational potential satisfies the same evolution equation
\be
  \ddot \Phi + \left ( H - \frac{2\ddot\varphi}{\dot\varphi}\right ) \dot \Phi
   + \left ( 4 \dot H - H \frac{2\ddot\varphi}{\dot\varphi} + \frac{\dot\varphi^2}{M_p^2}   \right ) \Phi
   - \frac{\nabla^2}{a^2}\Phi = 0~.
\ee
Here $\Phi$ is the gravitational potential in the Newtonian gauge defined in Eq. \ref{eq:Newtonian-gauge}.
The equation can be solved by the standard procedure. The conserved curvature perturbation \cite{Chen:2006wn} can be calculated as
\be \label{eq:HDEinflation}
  P_\zeta = \frac{H^4}{4\pi^2\dot\varphi^2}
  \exp \left [
    4c^2
    \int_t^{t_{LS}} \frac{dt}{R_h^2 H} \left ( 1-\frac{1}{R_hH}  \right )
  \right ]~.
\ee
The spectral index thus receives a correction
\be
  \delta n_s = - \frac{10c^2}{R_h^2 H^2} \left ( 1 - \frac{1}{R_hH}  \right )~.
\ee
This correction can becoem of order one when the future event horizon is close to Hubble size (when HDE dominates the energy density of the early stage of inflation). This correction does not fit the observations well (although cosmic variance dominates the uncertainty of the largest scales of CMB, and thus the possibility is not totally ruled out). The resulting CMB power spectrum is plotted in Fig.~\ref{fig11}.

One should also note that the UV/IR relation may also affect the initial fluctuation of the inflaton \cite{Enqvist:2004xv}. This is because, near the start of inflation, the size of the future event horizon is not much larger than the Hubble radius (this is how HDE can contribute significantly to the energy density of the universe at the start of inflation). As a result, one should cut off the super-Hubble fluctuations. With this observation, HDE can also help solving the low $\ell$ suppression problem of the CMB power spectrum.

\

\subsection{Black Hole}
\label{sec:4.6}

We recall that black holes are the key of understanding the HP, both the original HP, and the non-trivial UV/IR relation which eventually leads to HDE. There have not been very successful attempts so far. In \cite{Myung:2007tx}, a toy model relating HDE and a gas of black holes is studied.

Clearly, a gas of black holes does not behave like HDE. This is because the classical black holes are pressureless and thus $w_\mathrm{BH}=0$. This is different from HDE. In AdS/CFT, the CFT side has equation of state $w_\mathrm{CFT}=1/3$ because the black hole spacetime (in AdS) is dual to relativistic radiation. This is even farther away from a component of DE like HDE.

However, it is noted in \cite{2007AIPC..903..560B} that there is a duality between black hole with mass energy $E=M$, entropy $S=A/4$ and temperature $T$, and a quantum system of weakly interacting gas with internal energy $E'$, entropy $S'$ and temperature $T'$. The dictionary is
\be
  S' \rightarrow E=M~, \quad
  E' \rightarrow S = A/4 = \pi M^2~,\quad
  T' \rightarrow 1/T = 8\pi M~.
\ee
Such a nontrivial relation shows hints for that the quantum pressure of black holes may be different from the classical intuition. One notes that the energy surrounded by a spherical horizon can be integrated to be $E=2TS$. Compared to the law of thermodynamics $E = TS - pV$, one obtains
\be
  w_\mathrm{QG}= -\frac{1}{2}~.
\ee
This is still too large to fit for observations. Yet it is a component which can drive cosmological accelerations. It is argued that the physical origin of such a quantum equation of state may be because of quantum Casimir energy of the black hole gas. It remains interesting to see whether work along this line would result in more realistic HDE scenarios.

\

\subsection{Big Rip Singularity}
\label{sec:4.7}

To close this section, we finally review the study of the fate of our universe under the framework of HDE.
Using the up-to-date observation, in the simplest HDE model,
$C$ is or order 0.8 and $C\geq 1$ is rejected by more than 5$\sigma$.
In such models, the future event horizon of the universe is shrinking. Such a universe may end up with a big rip singularity.

However, HDE is intrinsically quantum gravitational. There have been many efforts to study the evolution of HDE near the big rip and it is noticed that effects from quantum gravity can rescue the universe from a big rip.

\cite{Elizalde:2005ju} first studied the generality of the big rip singularity in HDE. It is noted that for the case of $C<1$, classically it corresponds to a universe with finite time span, and thus the finite life time of the universe $t_s$ also behaves as an IR cutoff. For example, the following IR cutoff is considered in the paper
\be
  \frac{L_\Lambda}{C} = \frac{2t_s \left ( \frac{L_p+L_f}{\pi t_s}  \right )^2}
  { \left [ 1 + \left ( \frac{L_p+L_f}{\pi t_s}  \right )^2 \right ]^2 } ~,
\ee
where $L_\Lambda$ is the IR cutoff, and $L_p$ and $L_f$ are the particle horizon and the future event horizon, respectively. This simple example is chosen, because it leads to a simple solution
\be
  H = \frac{1}{2} \left ( \frac{1}{t} + \frac{1}{t_s - t}  \right )~,
\ee
and thus
\be
  a = a_0 \sqrt{\frac{t}{t_s - t} }~.
\ee
And there is indeed a big rip singularity similarly to the original HDE scenario. It is noticed that there are additional IR scales such as $dL_p/dt$ and/or $dL_f/dt$.

Near the big rip singularity, the energy scale of the universe becomes so high that quantum effects of gravity may play a key role. To resolve the big rip singularity, \cite{Elizalde:2005ju} uses conformal anomaly to model the quantum gravitational back-reaction near the big rip. The conformal anomaly $T_A$ can be computed to be
\be \label{eq:tacalc}
  T_A \equiv -\rho_A + 3p_A = b \left ( F + \frac{2}{3} \nabla^2 R \right ) + b'G + b''\nabla^2 R~,
\ee
where $G$ is the Gauss-Bonnet combination
\be
  G \equiv R^2 - 4 R_{ \mu\nu } R^{ \mu\nu } + R_{ \mu\nu\rho\lambda } R^{ \mu\nu\rho\lambda }~,
\ee
and $F$ is the Weyl tensor squared
\be
  F \equiv \frac{1}{3} R^2 - 2 R_{\mu\nu} R^{\mu\nu} + R_{ \mu\nu\rho\lambda } R^{ \mu\nu\rho\lambda }~.
\ee
The coefficients $b$ and $b'$ are related to the field content in the effective action
\be
  b = \frac{N + 6N_{1/2}+12N_1+611N_2-8N_{HD}}{120(4\pi)^2}~,
\ee
\be
  b' = -\frac{N + 11N_{1/2}+62N_1+1411N_2-28N_{HD}}{360(4\pi)^2}~,
\ee
where $N$, $N_{1/2}$, $N_1$, $N_2$, $N_{HD}$ are the numbers of the scalar, spinor, vector, graviton, and higher derivative conformal scalars, respectively. If the higher derivative conformal scalars do not contribute significantly to the theory, then $b>0$ and $b'<0$. It is assumed in \cite{Elizalde:2005ju} that we indeed have $b'<0$.
Note that $b''$ can take any value because it can be renormalized by local counter term $R^2$.

Making use of the nature of conformal anomaly $T_A = -\rho_A + 3p_A$ and the continuity equation $\dot\rho_A + 3H(\rho_A + p_A)=0$, one can thus solve $p_A$ to get
\be
  T_A = -4\rho_A - \dot\rho_A/H~.
\ee
Thus $\rho_A$ can be solved as
\be
  \rho_A = - \frac{1}{a^4} \int_{t_0}^t dt a^4 H T_A
\ee
Inserting Eq. \ref{eq:tacalc}, one can write $\rho_A$ as
\be
  \rho_A = - \frac{1}{a^4} \int_{t_0}^t dt a^4 H
  \big [
    -12 b \dot H^2 + 24b' (-\dot H^2 + H^2\dot H + H^4)
    -(4b+6b'')(\stackrel{...}{H} + 7H\ddot H + 4 \dot H^2 + 12 H^2 \dot H)
  \big ]~.
\ee
Inserting this solution to the FRW equation, one finds
\be
  \frac{3H^2(1-C^2)}{8\pi G} = - 6b'H^4~.
\ee
Interestingly, for $C<1$, this equation has two solutions. Namely
\be
  H^2=0~, \qquad H^2 = \frac{1-C^2}{-16\pi Gb'}~.
\ee
Recall that $b'<0$ for ordinary matter. Note that it is not likely for the universe to become flat because then the energy scale of the universe drops. As a result, for $C<1$, the universe approaches to de Sitter space with finite Hubble parameter. The de Sitter radius of the universe is quantum gravitational size $\ell_\mathrm{dS} \sim \ell_\mathrm{Planck}$.

Another approach to avoid the cosmic singularity in HDE is considered by \cite{Zhang:2009xj}. It is noticed that in the brane world scenario, the Friedmann equation in our 3+1 dimensional brane is modified into
\be
  3M_p^2 H^2 = \rho \left ( 1+ \frac{\rho}{\rho_c}  \right )~,
\ee
where $\rho_c = 2\sigma$, and $\sigma$ is the brane tension
\be
  \sigma = \frac{6 (8\pi)^2 M_*^6}{M_p^2}~,
\ee
where $M_*$ is the fundamental Planck mass. Here a simple brane world model with one large extra dimension is assumed.

\begin{figure}[htbp]
  \centering
  \includegraphics[width=0.8\textwidth]{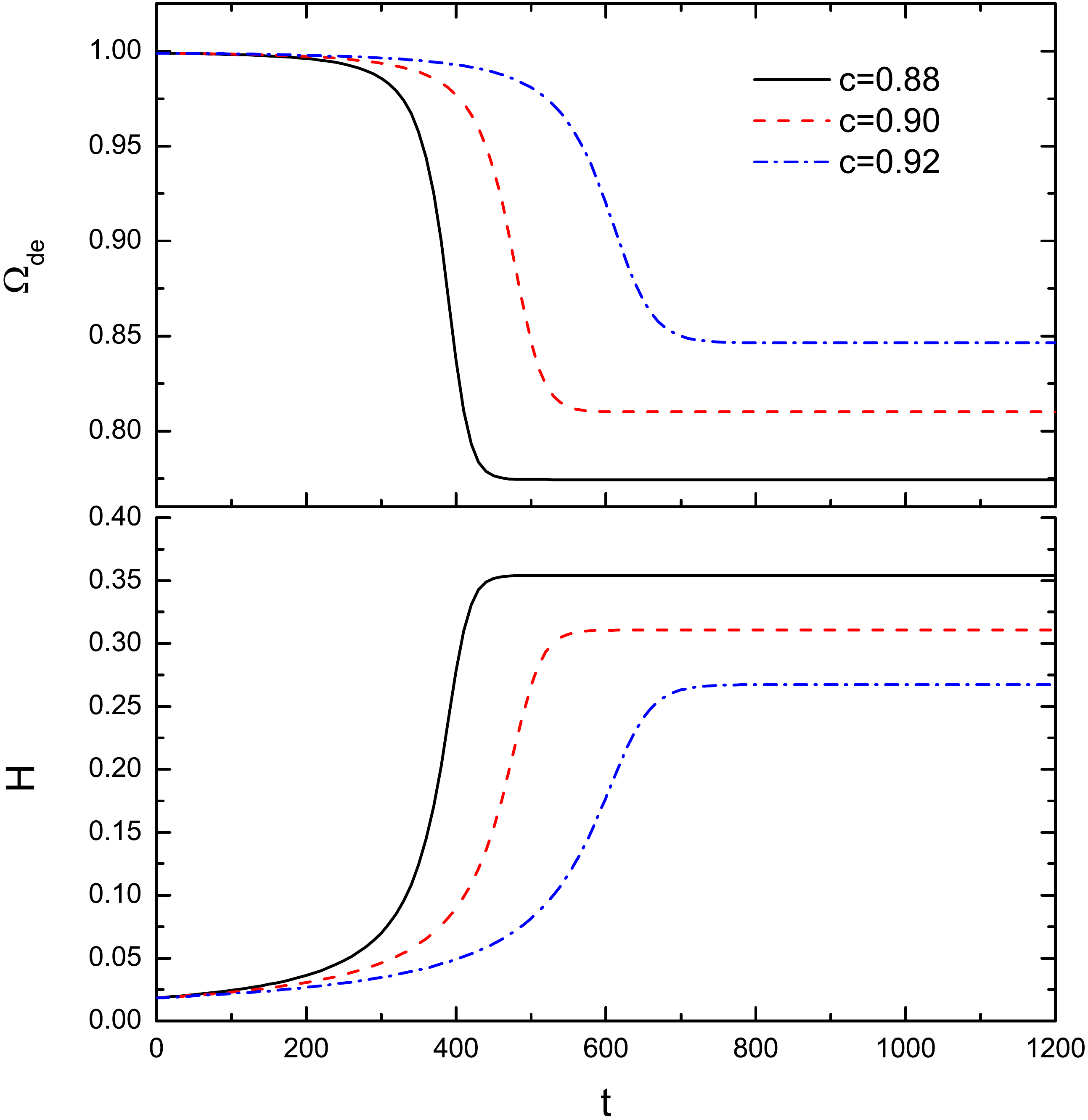}
  \caption{\label{fig12} The fate of our universe in the brane world scenario. The fractional DE density and the Hubble parameter at late time of the universe are plotted. From \cite{Zhang:2009xj}.}
\end{figure}

The presence of the large extra dimension and the brane world scenario correspond to an effective HDE parameter $C$,
\be
  C_\mathrm{eff} = C \sqrt{1+3C^2M_p^2R_h^{-2}\rho_c^{-1}}~.
\ee
At early times, the difference between $C_\mathrm{eff}(t)$ and $C$ is small. This is because the size of the future event horizon $R_h$ is large compared to the brane tension. And thus it introduces a very small correction. But once the future event horizon $R_h$ become comparable with the brane tension, brane world effect increases $C_\mathrm{eff}(t)$. As a result, one cannot get to $R_h \rightarrow 0$. The final de Sitter attractor can be solved as
\be
  R^\mathrm{min}_h = \frac{\sqrt{3}C^2 M_p}{\sqrt{(1-C^2)\rho_c}}~.
\ee
The big rip singularity is thus resolved, and the future event horizon of the universe become eventually compared with charasteric scales of the extra dimension. The numerical solution towards the attractor is plotted in Fig.~\ref{fig12}.

It is also noted that the big rip singularity can be resolved in the framework of interacting dark energy (IDE) if HDE decays to DM.
This will be introduced in detailed in the next section.

\

\section{Interacting Holographic Dark Energy}
\label{sec:5}

There is a theoretical possibility that DM and DE do not evolve separately but interact with each other.
In this section, we will introduce the research works of exploring the DE/DM interaction
in the framework of the interacting holographic dark energy (IHDE) scenario, from both the theoretical and the observational aspects.

\

\subsection{Theoretical Studies for The IHDE Model}
\label{sec:5.1}

The DE/DM interactions were first introduced to justify the currently small value of the cosmological constant
\cite{1988NuPhB.302..668W,1995A&A...301..321W};
afterwards they were found to be very useful to alleviate the coincidence problem
\cite{Amendola:2000uh,2001PhLB..521..133Z,2004GReGr..36.1483Z,Cai:2004dk}.
In addition, it has been proved that modified gravity (MG) models
can be expressed in terms of the DE/DM interaction in the Einstein frame
\cite{DeFelice:2010aj,He:2011qn,Zumalacarregui:2012us,Kofinas:2016fcp,Cai:2015emx}.
This equivalence implies that if we can determine the specific interaction term,
we will extend the gravitational theory beyond the scope of GR.
In the literature, these interacting models are widely studied
by using the parameterized post-Friedmann framework \cite{Hu:2007pj,Fang:2008sn,Skordis:2008vt,Baker:2011jy,Baker:2012zs,Pourtsidou:2013nha,Skordis:2015yra}.
In recent years, a lot of attention have been paid to study the DE/DM interaction in the IHDE scenario
\cite{Sadjadi:2007ts,MohseniSadjadi:2008na,Setare:2008bb,Setare:2007we,Chimento:2011dw}.
We refer the reader to Refs. \cite{Bolotin:2013jpa,Wang:2016lxa}
for more comprehensive and more detailed reviews on the topic of the DM/DE interaction.

Recently, the validity of DM/DE interaction is challenged from the consideration of radiative stability, if DE is made of quintesence \cite{DAmico:2016jbm,Marsh:2016ynw}. As the nature of HDE is holographic vacuum energy, such a local effective field theory analysis does not apply to HDE. It remains interesting to investigate whether radiative stability may be a concern for the holographic properties of spacetime, once interaction with DM is introduced.

\

\subsubsection{Dynamical Evolution of The IHDE Model in A Non-flat Universe}
\label{sec:5.1.1}

Now let us consider the IHDE model in a non-flat universe.
For this case, the first Friedmann equation can be written as
\be
  3M_{p}^{2}H^{2}=\rho_{dm}+\rho_{b}+\rho_{r}+\rho_{k}+\rho_{de}.
\ee
After taking into account the interaction between dark sectors,
the total energy density of all the dark sectors is still conserved,
but the energy density of DM and HDE evolve as
\be \label{eq:conservation1}
  \dot{\rho_{dm}}+3H\rho_{dm} = Q,
\ee
\be \label{eq:conservation2}
  \dot{\rho_{de}}+3H(1+w)\rho_{de} = -Q,
\ee
where $Q$ phenomenologically describes the interaction.

Owing to the lack of a fundamental theory of DM and DE,
the quantity $Q$ cannot be derived from the first principle.
In the literatures, the most common choice of $Q$ is
\be
 Q=H(\Gamma_{1}\rho_{dm}+\Gamma_{2}\rho_{de}),
\ee
where the coefficients $\Gamma_{1},\Gamma_{2}$ are constants that need to be determined by observational data.
It is convenient to use a single parameter instead of two,
so three choices are often made in the literatures: $\Gamma_{2}=0$, $\Gamma_{1}=0$ and $\Gamma_{1}=\Gamma_{2}=\Gamma_{3}$.
This leads to three most widely used interaction form
\be \label{eq:kernels}
Q_1=H\Gamma_{1}\rho_{dm};\quad
Q_2=H\Gamma_{2}\rho_{de};\quad
Q_3=H\Gamma_{3}(\rho_{dm}+\rho_{de}).
\ee
There are some other phenomenological interaction forms were proposed,
such as $Q=H \Gamma \rho_{dm} \rho_{de} / (\rho_{dm}+\rho_{de})$ \cite{delCampo:2015vha},
$Q=H \Gamma \rho_{dm}^{\xi_1} \rho_{de}^{\xi_2} / \rho_{c}^{\xi_1+\xi_2-1}$ \cite{Ma:2009uw},
$Q=\Gamma (\dot{\rho_{dm}}+\dot{\rho_{de}})$ \cite{Shahalam:2015sja}, and so on.

As shown in \cite{Zhang:2012uu},
making use of the energy conservation equations for all the components in the universe,
one can obtain the form of $p_{de}$,
\be \label{eq:pde}
p_{de}=-\frac{2}{3}\frac{\dot H}{H^2}\rho_c-\rho_c-{1\over3}\rho_r+{1\over3}\rho_k.
\ee
Substituting $p_{de}$ into Eq. (\ref{eq:conservation2}),
one can get a derivative equation of $\dot H$ and $\dot \Omega_{de}$
\be \label{eq:OH2}
2(\Omega_{de}-1){\dot H\over H}+\dot\Omega_{de}+H(3\Omega_{de}-3+\Omega_k-\Omega_r)=-H\Omega_I.
\ee
Here the effective dimensionless quantity for interaction is defined as
\be
\Omega_I\equiv\frac{Q}{H(z)\rho_c}.
\ee

In addition, as shown in the subsection \ref{sec:4.1}, in a non-flat universe the IR cut-off length scale $L$ takes the form
\be \label{eq:L1}
  L=ar(t),
\ee
where $r(t)$ satisfies
\be \label{eq:r(t)1}
  \int_0^{r(t)} {dr \over \sqrt{1-kr^2}}=\int_t^{+\infty}{dt\over a(t)}.
\ee
Eq. \ref{eq:L1} leads to another derivative equation of $\dot H$ and $\dot \Omega_{de}$
\be \label{eq:OL1.3}
  {\dot\Omega_{de}\over2\Omega_{de}}+H+{\dot H\over H}=\sqrt{{\Omega_{de}H^2\over c^2}-{k\over a^2}}.
\ee

Combining Eq. \ref{eq:OH2} with Eq. \ref{eq:OL1.3},
one can eventually obtain the following two equations
governing the dynamical evolution of the IHDE model in a non-flat universe,
\be \label{eq:OH3}
{1\over E(z)}{dE(z) \over dz}
=-{\Omega_{de}\over 1+z}\left({\Omega_k-\Omega_r-3+\Omega_I\over2\Omega_{de}}+{1\over2}+\sqrt{{\Omega_{de}\over
C^2}+\Omega_k} \right),
\ee
\be \label{eq:OH4}
{d\Omega_{de}\over dz}=
-{2\Omega_{de}(1-\Omega_{de})\over 1+z}\left(\sqrt{{\Omega_{de}\over
C^2}+\Omega_k}+{1\over2}-{\Omega_k-\Omega_r+\Omega_I\over 2(1-\Omega_{de})}\right).
\ee

\

\subsubsection{EoS of The IHDE Model}
\label{sec:5.1.2}

Then, we focus on the EoS $w$ of the IHDE model.
For simplicity, we just consider a flat universe dominated by HDE and the pressureless matter.
After taking into account the interaction between matter and HDE, we have
\be \label{eq:simpleconservation1}
  \dot{\rho_{m}}+3H\rho_{m} = Q,
\ee
\be \label{eq:simpleconservation2}
  \dot{\rho_{de}}+3H(1+w)\rho_{de} = -Q.
\ee
It is convenient to take the ratio of energy densities as \cite{Wang:2005jx}
\be
  r\equiv{\rho_{m} \over \rho_{de}}.
\ee
From Eqs. \ref{eq:simpleconservation1} and \ref{eq:simpleconservation2} we can obtain a differential equation for $r$
\be \label{eq:evolution of r}
  \dot{r}=3Hrw+\frac{(1+r)Q}{\rho_{de}}.
\ee
Notice that
\be \label{eq:constraint r}
r=\frac{1-\Omega_{de}}{\Omega_{de}};\quad
\dot{r}=-\frac{\dot{\Omega_{de}}}{\Omega_{de}^{2}},
\ee
then we can get
\be \label{eq:constraint w}
  w=-\frac{\Omega_{de}'}{3\Omega_{de}(1-\Omega_{de})}-\frac{Q}{3H(1-\Omega_{de})\rho_{de}}.
\ee
This formula holds true for all the IDE models.
In addition, as DE decays into pressureless matter (i.e. $Q>0$),
it will give rise to a more negative $w$.

For the original HDE model, the future event horizon is chosen as the IR cutoff.
From Eq. \ref{eq:OH4} we can get
\be \label{eq:IHDEom1}
  \Omega_{de}'=2\Omega_{de}(1-\Omega_{de})\left({\sqrt{\Omega_{de}} \over C}+{1\over2}-{Q \over 2H(1-\Omega_{de})\rho_c}\right).
\ee
Substituting Eq. \ref{eq:IHDEom1} into Eq. \ref{eq:constraint w}, we can obtain
\be \label{eq:IHDEw}
  w=-\frac{1}{3}-\frac{2\sqrt{\Omega_{de}}}{3C}-\frac{Q}{3H\rho_{de}}.
\ee

\begin{figure*}
\centering
\includegraphics[width=0.8\textwidth]{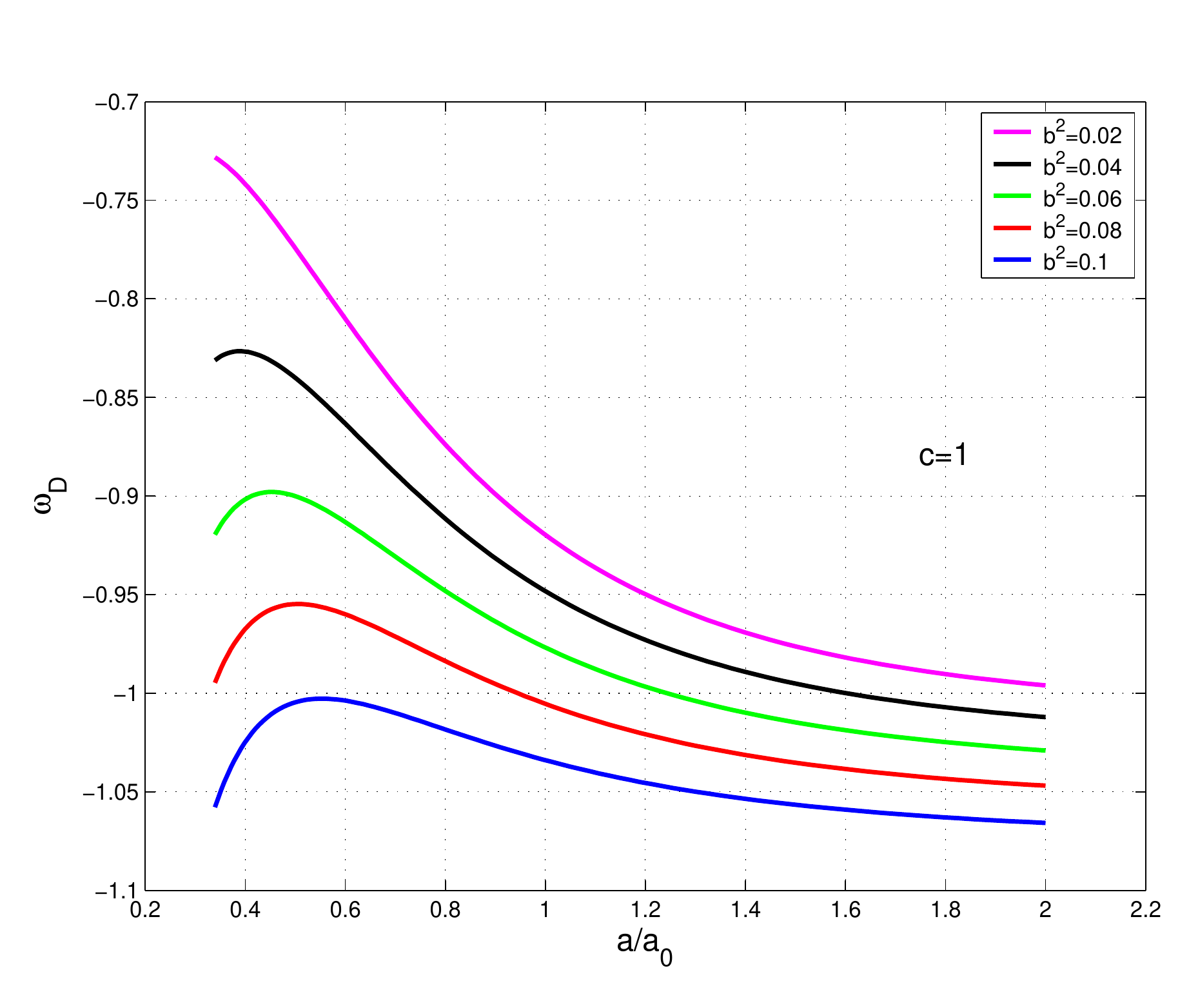}
\caption{The evolution behavior of the DE EoS for different coupling strength $b^2$.
A fixed $C=1$ is adopted in the analysis.
From \cite{Wang:2005jx}.}
\label{fig13}
\end{figure*}

In Ref. \cite{Wang:2005jx}, the authors considered a specific interaction form $Q=3b^2H\rho_c$, where $b$ is a dimensionless parameter.
Thus, the EoS of the IHDE model can be written as
\be \label{eq:IHDEw2}
  w=-\frac{1}{3}-\frac{2\sqrt{\Omega_{de}}}{3C}-\frac{b^2}{\Omega_{de}}.
\ee
The evolution behavior of the DE EoS for different interaction strength $b^2$ are plotted in Fig. \ref{fig13}.

As shown in \cite{Wang:2005jx}, to allow $w<-1$ at the present stage,
we only need
\be \label{eq:IHDEcon1}
  \frac{2\Omega_{de0}}{3}\left(1-\frac{\sqrt{\Omega_{de0}}}{C}\right)<b^2<\frac{8C^2}{81},
\ee
and
\be \label{eq:IHDEcon2}
  \sqrt{\Omega_{de0}}<C<\frac{2\sqrt{\Omega_{de0}}}{3\Omega_{de0}-1}.
\ee
In other words, the IHDE model can accommodate a transition of the DE from a normal state $w>-1$ to a phantom regime $w<-1$.
This conclusion had been extended to a universe with spatial curvature \cite{Wang:2005ph}.
\footnote{However, it was argued that the inclusion of the DM/DE interaction cannot lead to the phantom regime in the framework of HDE,
if the effective EoS $w^{eff}$, rather than the normal EoS $w$, is used in the analysis \cite{Kim:2005at}.
This conclusion holds true for the case of non-flat universe \cite{Setare:2006wh,Setare:2006sv}.}

\

\subsubsection{Alleviation of Coincidence Problem in The IHDE Scenario}
\label{sec:5.1.3}

Next, let us turn to the coincidence problem.
For the case without the DM/DE interaction, Eq. \ref{eq:evolution of r} can be reduced to
\be
  \frac{d\ln r}{dx} = 3w,
\ee
where $x \equiv \ln a$. For a constant $w$, we have
\be
  r = r_0a^{3w}.
\ee
It is clear that $r\sim O(1)$ only when $t$ is around $t_0$,
this is why the coincidence problem arises in the standard cosmology.

It is seen that the inclusion of the DM/DE interaction will greatly change the dynamics of $r$.
A special case was given in \cite{Pavon:2005yx}.
Choosing $Q=\Gamma \rho_{de}$, Eq. \ref{eq:evolution of r} can be reduced to
\be \label{eq:special r}
  \dot{r}=3Hr\left(w+\frac{1+r}{r}\frac{\Gamma}{3H}\right).
\ee
Moreover, choosing the Hubble scale $1/H$ as the characteristic length scale $L$,
one can also obtain the expression of DE EoS \cite{Pavon:2005yx}
\be \label{eq:special w}
  w=-\frac{1+r}{r}\frac{\Gamma}{3H}.
\ee
From Eqs. \ref{eq:special r} and \ref{eq:special w}, one can see that
\be
  \dot{r}=0.
\ee
This means that appropriately choosing the interaction term $Q$ and the characteristic length scale $L$ can lead to a constant $r$,
and thus completely solve the coincidence problem.

For the HDE model, $L$ has been chosen as the future event horizon;
for this case, it is impossible to get a constant $r$.
In other words, in the framework of the original HDE model,
the coincidence problem cannot be completely removed by adding the DM/DE interaction alone \cite{Li:2008zq}.
But in \cite{Hu:2006ar}, the authors demonstrated that
as long as the interacting term and the characteristic size of holographic bound are appropriately specified,
setting $\dot{r}=0$ will yield a positive solution of $r$, whose value is of $O(1)$.
In other words, the ratio $r$ has a stable constant solution at the late time,
and its value is not far from the current measured value.
This implies that certain amount of DM/DE interaction can make $r$ varies slowly with time,
and thus alleviate the coincidence problem significantly.
For more discussions about the alleviation of the coincidence problem in the IHDE scenario,
see Refs. \cite{Berger:2006db,Li:2006ci,Karwan:2008ig}.

\

\subsubsection{Generalized Second Law of Thermodynamics in The IHDE Scenario}
\label{sec:5.1.4}

As is well known, there is a deep connection between GR and thermodynamics \cite{Jacobson:1995ab}.
Therefore, the thermodynamics corresponding to an accelerated Universe has drawn a lot of attention
\cite{Brustein:1999ua,Cai:2005ra,Wang:2005pk,Izquierdo:2005ku,Setare:2006vz}.
Here we just discuss the validity of generalized second law of thermodynamics in The IHDE Scenario.

Now consider a IHDE mode with the interaction term $Q = \Gamma \rho_{de}$.
After defining the effective EoS \cite{Kim:2005at}
\be
w_{de}^{eff}=w+\frac{\Gamma}{3H}, \quad w_{m}^{eff}=-\frac{1}{r}\frac{\Gamma}{3H},
\ee
the continuity equations can be rewritten in their standard form
\be \label{eq:effconservation1}
  \dot{\rho_{m}}+3H(1+w_{m}^{eff})\rho_{m} = 0,
\ee
\be \label{eq:effconservation2}
  \dot{\rho_{de}}+3H(1+w_{de}^{eff})\rho_{de} = 0.
\ee
The entropy of the universe inside the future event horizon can be related to its energy and pressure in the horizon
through Gibbons equation \cite{Wang:2005pk,Izquierdo:2005ku}
\be \label{eq:effentropy1}
  d S_{m}=\frac{1}{T}(dE_{m}+p_{m}dV),
\ee
\be \label{eq:effentropy2}
  d S_{de}=\frac{1}{T}(dE_{de}+p_{de}dV).
\ee
Here $T=\frac{1}{2\pi L}$ is the temperature of the future event horizon,
$V=\frac{4\pi L^3}{3}$ is the volume containing all the matter and DE,
\be \label{eq:effenergy1}
  E_{m}=\frac{4\pi L^3}{3}\rho_{m}, \quad p_{m}=w_{m}^{eff}\rho_{m},
\ee
\be \label{eq:effenergy2}
  E_{de}=\frac{4\pi L^3}{3}\rho_{de}, \quad p_{de}=w_{de}^{eff}\rho_{de}.
\ee
In addition, the entropy of horizon is $S_{L} = \pi L^2$, so
\be \label{eq:effentropy3}
  d S_{L}=2\pi L \cdot dL.
\ee

Making use of Eqs. \ref{eq:effentropy1}, \ref{eq:effentropy2} and \ref{eq:effentropy3},
one can test the validity of generalized second law of thermodynamics in The IHDE Scenario.
Setare studied this topic in a closed universe;
by adopting the parameters $\Omega_{de0}=0.73$, $\Omega_{k0}=0.01$, $C=0.1$ and $b^2=0.2$,
he found \cite{Setare:2007at}
\be
  \frac{d}{dx}(S_{m}+S_{de}+S_{L})=\frac{M_p^2}{H^2}\left(-10.88+\frac{1482.88-167.42q}{H^2}\right)+\frac{1.33}{H^2},
\ee
where $q$ is the deceleration parameter.
If $q\leq8.85-H^2/15.4$, then $\frac{d}{dx}(S_{m}+S_{de}+S_{L})\geq0$.
In other words, the generalized second law of thermodynamics is respected for the special range of the deceleration parameter $q$.

\

\subsection{Observational Constraints on The IHDE Model}
\label{sec:5.2}

We have introduced the theoretical studies for the IHDE model.
Now, we turn to the observational constraints on this model.

\

\subsubsection{Parameter Estimation for The IHDE Model}
\label{sec:5.2.1}

For the IHDE model, it is crucial to determine the interaction strength from the cosmological observations.

In \cite{Wu:2007fs}, making use of the Glod04 and the ESSENCE SNIa samples,
the $A$ parameter of BAO measurement from the SDSS and the shift parameter $R$ from the WMAP3,
Wu et al. constrained the parameter space of HDE models with and without the interaction term $Q=9b^2M_p^2H^2$.
The fitting results are listed in table \ref{tab:4}.
From this table, one can see that all the combinations of data favor $b^2=0$ at 1$\sigma$ CL,
which means that a non-interacting HDE is favored by the cosmological observations.

\begin{table*}
\begin{center}
\caption{Fitting results for HDE models with and without the interaction. From \cite{Wu:2007fs}.}
\label{tab:4}
\begin{tabular}{|c|c|c|c|c|c|}
\hline
Model&Results &Gold04   &  Gold04+$A$+$R$ &ESSENCE &ESSENCE+$A$+$R$ \\
\hline
  &$\chi^2$ &158.27    & 158.66      &195.34        & 196.16  \\
With  &$\Omega_{m0}$  & $0.32^{+0.29}_{-0.13}$  & $0.29\pm 0.04$  &$0.27^{+0.23}_{-0.15}$ & $0.27^{+0.04}_{-0.03}$ \\
Interaction  &$b^2$  &$0^{+0.2}_{-0}$  & $0^{+0.01}_{-0}$  &$0.02^{+0.09}_{-0.02}$   & $0.002^{+0.01}_{-0.002}$  \\
&$C$  &$0.82^{+0.48}_{-0.18}$  & $0.88^{+0.40}_{-0.07}$ &$0.85^{+0.45}_{-0.18}$  & $0.85^{+0.18}_{-0.02}$  \\
\hline
&$\chi^2$& 158.27 &158.66 & 195.75 & 196.29 \\
$b^2=0$&$\Omega_{m0}$& $0.31^{+0.07}_{-0.1}$ & $0.29\pm 0.03$ & $0.27^{+0.03}_{-0.14}$ & $0.27^{+0.03}_{-0.02}$ \\
&$C$& $0.82^{+0.48}_{-0.04}$ & $c=0.88^{+0.24}_{-0.06}$ & $c=0.85^{+0.45}_{-0.02}$ & $0.85^{+0.1}_{-0.02}$ \\
\hline
\end{tabular}
\end{center}
\end{table*}

In \cite{Feng:2007wn}, by using the Glod04 SNIa sample, the shift parameter of the CMB from the WMAP3,
the BAO measurement from the SDSS, the $H(z)$ measurement and the lookback time data,
Feng et al. performed a statistical joint analysis of the IHDE model with the interaction term $Q=3b^2H(\rho_{m}+\rho_{de})$.
The corresponding results are summarized in table \ref{tab:5}.
Again, one can see that all the combinations of data favor a non-interacting HDE.

\begin{table}
\scriptsize \caption{Fitting results for the IHDE model with the interaction $Q=3b^2H(\rho_{m}+\rho_{de})$. From \cite{Feng:2007wn}.}
\label{tab:5}
\begin{center}
{\begin{tabular}{|c|ccc|c|}
\hline
Data&C&$\Omega_{de0}$&$b^2$&$\chi_{min}^2$\\
\hline
$\mathrm{SNIa+BAO}$&$0.53_{-0.22}^{+0.61}$&$0.72_{-0.04}^{+0.05}$&$-0.10_{-0.125}^{+0.131}$&156.24\\
\hline
$\mathrm{SNIa+BAO+CMB}$&$0.84_{-0.25}^{+0.46}$&$0.70_{-0.04}^{+0.04}$&$-0.004_{-0.012}^{+0.012}$&158.45\\
\hline
$\mathrm{SNIa+BAO+H(z)}$&$0.82_{-0.31}^{+0.89}$&$0.71_{-0.04}^{+0.05}$&$-0.005_{-0.075}^{+0.075}$&167.74\\
\hline
$\mathrm{SNIa+BAO+Lookback time}$&$0.62_{-0.28}^{+1.22}$&$0.72_{-0.05}^{+0.05}$&$-0.059_{-0.126}^{+0.148}$&159.48\\
\hline
$\mathrm{SNIa+BAO+CMB+H(z)}$&$0.84_{-0.25}^{+0.40}$&$0.71_{-0.04}^{+0.04}$&$-0.003_{-0.012}^{+0.010}$&167.75\\
\hline
$\mathrm{SN Ia+BAO+CMB+Lookback time}$&$0.83_{-0.25}^{+0.43}$&$0.71_{-0.04}^{+0.04}$&$-0.003_{-0.013}^{+0.012}$&160.08\\
\hline
\end{tabular}}
\end{center}
\end{table}

\begin{figure*}
\centering
\includegraphics[width=0.8\textwidth]{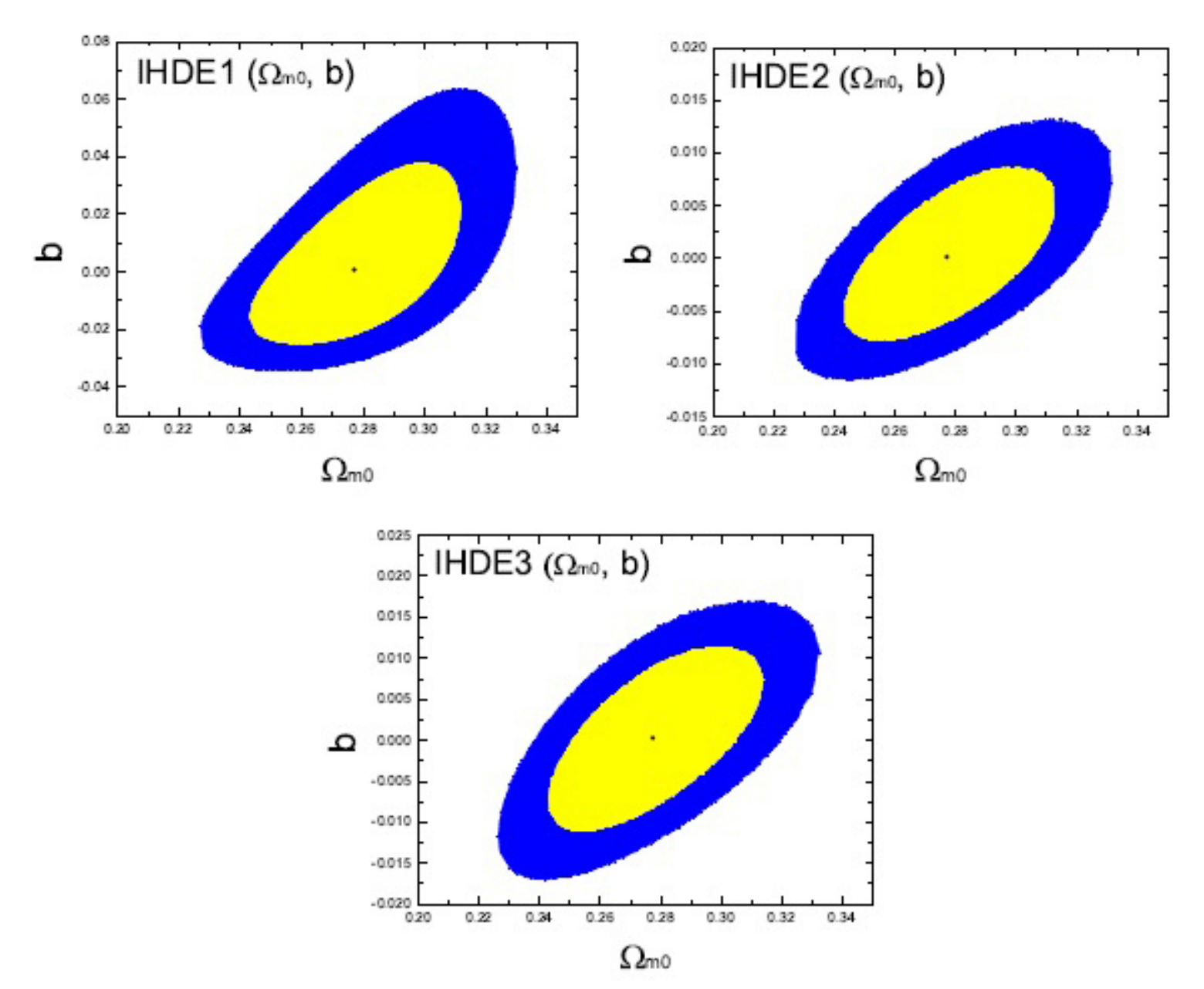}
\caption{Probability contours at $1\sigma$ and $2\sigma$ CL in the $\Omega_{m0}-b$ plane, for the three IHDE models.
From \cite{Li:2009zs}.}
\label{fig14}
\end{figure*}

Moreover, in \cite{Li:2009zs}, based on the Constitution SNIa sample,
the shift parameter of CMB given by the WMAP5 and the BAO measurement from the SDSS,
Li et al. placed the observational constraints on the HDE models with spatial curvature and three kinds of interaction
\be \label{eq:Likernels}
Q_1=-3bH\rho_{de};\quad
Q_2=-3bH(\rho_{de}+\rho_{m});\quad
Q_3=-3bH\rho_{m}.
\ee
The corresponding results are shown in table \ref{tab:6} (See also Fig. \ref{fig14}).
Once more, one can see that adding the spatial curvature and the interaction terms cannot effectively reduce the value of $\chi^2_{min}$,
which implies that there is no need to introduce the spatial curvature and the DM/DE interaction in the HDE cosmology.

\begin{table}
\caption{Fitting results for the HDE models with spatial curvature and three kinds of interaction. From \cite{Li:2009zs}.}
\begin{center}
\label{tab:6}
\begin{tabular}{cccccc}
  \hline\hline
  Model  &  $\Omega_{m0}$  &  $c$ &  $\Omega_{k0}$  &  $ b $   &  $\chi^2_{min}$  \\
  \hline
  HDE    & $0.277^{+0.022}_{-0.021}$ & $0.818^{+0.113}_{-0.097}$  &    &    & 465.912  \\
  \hline
  KHDE   & $0.278^{+0.037}_{-0.035}$ & $0.815^{+0.179}_{-0.139}$ &  $(7.7\times10^{-4})^{+0.018}_{-0.019}$ &  &  465.906 \\
  \hline
  IHDE1   & $0.277^{+0.035}_{-0.034}$ & $0.818^{+0.197}_{-0.257}$ &   &  $(6.1\times10^{-5})^{+0.036}_{-0.025}$ & 465.911 \\
  \hline
  IHDE2   & $0.277^{+0.034}_{-0.036}$ & $0.816^{+0.170}_{-0.223}$ &  & $(1.6\times10^{-4})^{+0.009}_{-0.008}$ & 465.910 \\
  \hline
  IHDE3   & $0.277^{+0.034}_{-0.036}$ & $0.815^{+0.164}_{-0.209}$ &  & $(3.0\times10^{-4})^{+0.011}_{-0.011}$ & 465.909  \\
  \hline
  KIHDE1  & $0.281^{+0.047}_{-0.043}$ & $0.977^{+0.563}_{-0.551}$ &  $0.030^{+0.066}_{-0.127}$  & $-0.046^{+0.243}_{-0.102}$ & 465.697 \\
  \hline
  KIHDE2  & $0.281^{+0.047}_{-0.044}$ & $0.974^{+0.559}_{-0.475}$  &  $0.030^{+0.070}_{-0.100}$ & $-0.042^{+0.191}_{-0.073}$ & 465.700 \\
  \hline
  KIHDE3  & $0.280^{+0.045}_{-0.042}$ & $0.961^{+0.231}_{-0.499}$  &  $0.061^{+0.038}_{-0.210}$ & $-0.048^{+0.113}_{-0.042}$ & 465.719 \\
  \hline\hline
\end{tabular}
\end{center}
\end{table}

These research works demonstrate that the current cosmological observations do not support the existence of DM/DE interaction in the HDE cosmology.
This conclusion is insensitive to the observational data or the specific interaction form used in the analysis.
These results are in agreement with some other numerical studies on the IHDE model \cite{Ma:2009uw,Zhang:2012uu,Feng:2016djj}.

\

\subsubsection{Other Cosmic Tests on The IHDE Model}
\label{sec:5.2.2}

\begin{itemize}
 \item Statefinder Diagnostic For The IHDE Model
\end{itemize}

As mentioned in the subsection \ref{sec:3.3.2},
The statefinder pair $\{r,s\}$ is a very useful tool to diagnose the DE models.
This diagnostic tool had been used to diagnose the IHDE model in \cite{Zhang:2007uh}.
Adopting an interaction term $Q=3b^2H(\rho_{m}+\rho_{de})$,
one can obtain
\ba \label{eq:rsIHDE}
r&=&1-\frac{3}{2}\Omega_{\rm de}w'+3\Omega_{\rm de}w\left(1-\frac{1}{C}\sqrt{\Omega_{\rm de}}\right),\\
s&=&1+w-\frac{w'}{3w}+\frac{b^2}{\Omega_{\rm de}},
\ea
where
\ba \label{eq:rsw1}
w'&=&(1-\Omega_{\rm de})\left(b^2-\frac{\Omega_{\rm de}^{3/2}}{3C}\right) \left[\frac{1}{\Omega_{\rm de}}-\frac{3b^2}
{\Omega_{\rm de}(1-\Omega_{\rm de})}+\frac{2}{C\sqrt{\Omega_{\rm de}}}\right].
\ea

\begin{figure*}
\centering
\includegraphics[width=0.8\textwidth]{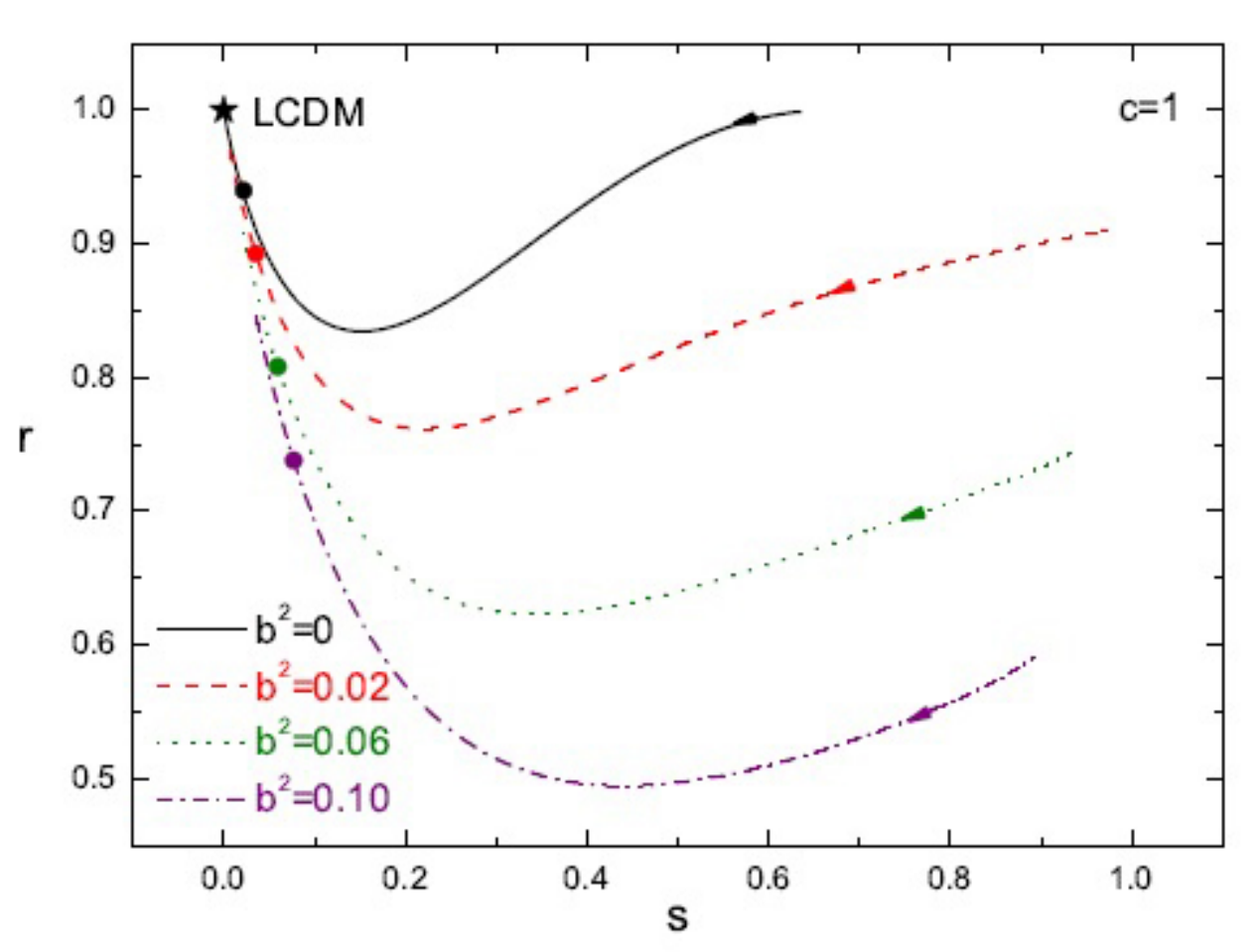}
\caption{The statefinder diagrams $r(s)$ for the IHDE with a fixed parameter $C$ and different interaction strength $b^2$.
The arrows in the diagram denote the evolution directions of the statefinder trajectories.
The star denotes the $\Lambda$CDM fixed point $(0, 1)$,
and the dots show today's values for the statefinder parameters $(s_0, r_0)$.
From \cite{Zhang:2007uh}.}
\label{fig15}
\end{figure*}

Fig. \ref{fig15} shows the statefinder diagrams $r(s)$ of the IHDE model,
for the cases of $C=1$ with various interaction strength such as $b^2=0$, $0.02$, $0.06$ and $0.10$,
meanwhile the present density parameter of DE is chosen as $\Omega_{de0}=0.73$.
The star denotes the $\Lambda$CDM fixed point $(0, 1)$,
and the dots show today's values for the statefinder parameters $(s_0, r_0)$.
One can see that the evolution trajectories with different interaction strengths exhibit different features in the statefinder plane.
When the interaction is absent, the $r(s)$ curve for HDE ends at the $\Lambda$CDM fixed point $(0, 1)$.
However, after taking the interaction into account, the endpoints of the $r(s)$ curves could not arrive at the $\Lambda$CDM fixed point,
though all of the evolution trajectories still tend to approach this point.
Moreover, it can be seen that stronger interaction results in longer distance to the $\Lambda$CDM fixed point.
Therefore, the interaction between HDE and DM makes the statefinder evolutionary trajectories with the same $C$ tremendously different.

\begin{itemize}
 \item Alleviation of Cosmic Age Problem in The IHDE Model
\end{itemize}

The age of the universe at redshift $z$ is given by
\be \label{eq:cosmicage}
t(z)=\int_z^\infty\frac{dz'}{(1+z')H(z')}.
\ee
It is convenient to introduce a dimensionless cosmic age
\be \label{eq:cosmicage2}
T_{cos}(z)\equiv H_0 t(z)=\int_z^\infty \frac{dz'}{(1+z')E(z')}.
\ee
In cosmology, there is a very basic principle that the universe cannot be younger than its constituents.
In other words, at any redshift $z$,
the age of the universe should be larger than, or at least equal to, the age of all the old objects,
namely $T_{cos}(z)\geq T_{obj}(z)\equiv H_0 t_{obj}(z)$, where $t_{obj}(z)$ is the age of the old object at redshift $z$.
One can also define a dimensionless quantity
\be \label{eq:agetau}
\tau(z)\equiv {T_{cos}(z)\over T_{obj}(z)}.
\ee
Then, the condition $T_{cos}(z)\geq T_{obj}(z)$ is translated into $\tau(z)\geq 1$.

As mentioned in subsection \ref{sec:2.3.4},
the existence of an extremely old quasar APM 08279+5255 is still a mystery,
because it has an age lower bound $t_{obj}(3.91)=2.0$ Gyr at $z=3.91$ \cite{Hasinger:2002wg},
which is larger than the cosmic age given by almost all the mainstream cosmological models.
For example, Friaca et al. demonstrated that the $\Lambda$CDM model cannot give a result of $\tau(3.91)\geq 1$,
and thus it is fail to accommodate this extremely old quasar \cite{Friaca:2005ba}.
In addition, Wei and Zhang found that the original HDE model cannot accommodate this quasar, too \cite{Wei:2007ig}.

Cui and Zhang revisited this cosmic age problem in the framework of the IHDE model \cite{Cui:2010dr}.
They parameterized the interaction term as $Q=3H(\Gamma_1\rho_{de}+\Gamma_2\rho_m)$,
and then considered the following three cases:
(i) $\Gamma_2=0$, and thus $Q=3\Gamma_1 H\rho_{de}$;
(ii) $\Gamma_1=\Gamma_2=\Gamma$, and thus $Q=3\Gamma H(\rho_{de}+\rho_m)$;
(iii) $\Gamma_1=0$, and thus $Q=3\Gamma_2 H\rho_m$.
The cosmic age for these three IHDE models (with $C=0.8$, $\Omega_{m0}=0.28$, $h=0.64$ and different interaction strength)
are shown in table \ref{tab:7}.
From this table we see that, along with the increase of the interaction strength,
the cosmic age $T_{cos}$ also increases.
It is clear that, for all the cases,
the value of $\tau(3.91)$ can be greater than 1 when the value of interaction strength is large than $0.1$.
This implies that the interaction between dark sector may be a crucial factor
to alleviate the cosmic age problem \cite{Wang:2008te,Wang:2010su,Cui:2010dr}.

\begin{table}
\begin{center}
\caption{\label{tab:7}
The values of $T_{cos}(3.91)$ and $\tau(3.91)$ in the IHDE models
with $C=0.8$, $\Omega_{m0}=0.28$, $h=0.64$ and different interaction strength. From \cite{Cui:2010dr}.}
\begin{tabular}{cccccc}
  \hline
  Case~I  ~$(\Gamma_2=0)$&  ~$\Gamma_1$  & ~ $0.02$  &  ~$0.06$  &  ~$0.10$  &  ~$0.15$\\
  \cline{2-6}
  &$T_{cos}(3.91)$   &   $0.1172$   &   $0.1246$   &   $0.1335$   &   $0.1475$ \\\cline{2-6}
  &  ~$\tau(3.91)$  &  ~$0.894$  &  ~$0.951$  &  ~$1.019$  &  ~$1.126$  \\
  \hline
  Case~II~$(\Gamma_1=\Gamma_2)$  &  ~$\Gamma$  &  ~$0.01$  &  ~$0.02$  &  ~$0.03$  &  ~$0.05$\\
  \cline{2-6}
  &$T_{cos}(3.91)$   &   $0.1194$   &   $0.1253$   &   $0.1316$   &   $0.1456$ \\\cline{2-6}
  &  ~$\tau(3.91)$  &  ~$0.912$  &  ~$0.957$  &  ~$1.005$  &  ~$1.111$  \\
  \hline
  Case~III~$(\Gamma_1=0)$  &  ~$\Gamma_2$  &  ~$0.01$  &  ~$0.03$  &  ~$0.05$  &  ~$0.07$\\ \cline{2-6}
  &$T_{cos}(3.91)$   &   $0.1177$   &   $0.1259$   &   $0.1346$   &   $0.1440$ \\\cline{2-6}
  &  ~$\tau(3.91)$  &  ~$0.899$  &  ~$0.961$  &  ~$1.028$  &  ~$1.099$  \\
  \hline
\end{tabular}
\end{center}
\end{table}

\

\section{HDE in Various Modified Gravity Theories}
\label{sec:6}

HDE is a great arena for modified gravity (MG) for a few reasons.
\begin{itemize}
  \item As a non-renormalizable theory, and currently the only known non-renormalizable theory, gravity is UV sensitive. In HDE the UV cutoff of the theory depends on the IR cutoff. The IR cutoff is set at the cosmological scale for the concern of DE. As a result, the UV cutoff is much affected. As we do not have a firm RG equation for quantum gravity, the best we can do is to take a modified gravity theory with the presence of HDE as a candidate of the IR gravity theory. For this reason, actually, even if a MG theory is ruled out on earth, solar system or galactic scale experiments, the MG theory may still be considered together with HDE because cosmological scales is a completely different scale.
  \item For many MG models to work as DE, they still have to solve the old cosmological constant problem. HDE solves the problem for this MG theories. Thus the MG theory in HDE can focus on the naturalness from first principle, dynamics of DE, agreement with observations, etc.
  \item Observationally, HDE has two parameters, which is relatively few compared with most DE models (although the cosmological constant has one parameter only). Thus MG on top of HDE has stronger predictability compared to those on top of DE scenarios with more parameters (or even free functions).
\end{itemize}
There has been rich literature in HDE. In the below subsections we review some of them.

\subsection{Brans-Dicke Theory}
\label{sec:6.1}

The Jordan-Fierz-Brans-Dicke theory (often referred to as the Brans-Dicke theory) \cite{Jordan1955, Fierz1956, Brans:1961sx} is one of the earliest theories of modified gravity which is still viable now. The Brans-Dicke theory in the framework of HDE is studied in \cite{Gong:2004fq, Banerjee:2007zd, Setare:2006yj, Xu:2008sn}.

In Brans-Dicke theory, the action of gravity is modified into
\be
  S = \int d^4 x \sqrt{-g} \left [
    - \frac{1}{8\omega} \phi^2 R
    + \frac{1}{2} g^{\mu\nu} \partial_\mu\phi \partial_\nu\phi
    + L_M
  \right ]~,
\ee
where $\phi$ is the Brans-Dicke scalar field. The effective Newton's gravitational constant is
\be
  G_\mathrm{eff} = \frac{\omega}{2\pi \phi^2}~.
\ee
The generalized Friedmann equation, generalized second Friedmann equation and the continuity equation for $\phi$ take the forms
\be
  \frac{3}{4\omega} \phi^2 \left ( H^2 + \frac{k}{a^2} \right )
  - \frac{1}{2} \dot\phi^2 + \frac{3}{2\omega} H \dot\phi \phi = \rho~,
\ee
\be
  - \frac{1}{4\omega}\phi^2 \left(\frac{2\ddot a}{a} + H^2 + \frac{k}{a^2} \right)
  - \frac{1}{\omega} H \dot\phi \phi - \frac{1}{2\omega} \ddot\phi \phi
  - \left ( \frac{1}{2} + \frac{1}{2\omega}  \right ) \dot\phi^2 = p~,
\ee
\be
  \ddot\phi + 3 H \dot\phi - \frac{3}{2\omega} \left (
    \frac{\ddot a}{a} + H^2 + \frac{k}{a^2}
  \right ) \phi = 0~.
\ee
One can then decompose $\rho$ into matter part $\rho_m$ and HDE part $\rho_{DE}$.

We now proceed to solve those equations. For simplicity, flat space model with $k=0$ is reviewed.
It is possible and convenient to parameterize the time evolution of $\phi$ as power law
\be
  \frac{\phi}{\phi_0} = \left ( \frac{a}{a_0}  \right )^\zeta~.
\ee
The Friedmann equation is thus
\be
  H^2 = \frac{2 \rho}{(6+6\zeta-\omega \zeta^2)\phi} ~.
\ee
Taking the $\phi>0$ branch of solution, the consistency requirement for $\zeta$ is that $6+6\zeta-\omega \zeta^2 > 0$. In this regime, the equations can be solved and the equation of state of HDE is
\be
  w = - \frac{1}{3} \left ( 1 + \zeta + \frac{2}{c}\sqrt{\Omega_{de}}  \right )~.
\ee
The ``deceleration parameter'' $q$ can be calculated as
\be
  q \equiv - \frac{\ddot a}{aH^2}  = \frac{1}{2} + \zeta
  + \frac{\zeta}{8+2\zeta} + \frac{6w\Omega_{de}}{4+\zeta}~.
\ee
The standard HDE can be recovered in the $\zeta\rightarrow 0$ limit.
Observationally, given the constraint on Newton's gravitational constant, one needs $\zeta < 0.14$.

\

\subsection{Braneworld Theory}
\label{sec:6.2}

HDE in braneworld theories are extensively studied. There are many versions of brane world theories
\cite{Wang:2004nqa, Kim:2007dp, Saridakis:2007cy, Saridakis:2007ns, Wu:2007tp, Lepe:2008ka, Liu:2010am, Ghaffari:2014pxa, Farajollahi:2016lrk}.
In Section~\ref{sec:4.7}, we have discussed how a Randall-Sundrum \cite{Randall:1999ee} braneworld model avoids the doomsday in a $C<1$ HDE scenario.
Here, we shall focus on the Dvali-Gabadadze-Porrati (DGP) \cite{Dvali:2000hr} braneworld. Although DGP gravity itself does not fit current dark energy data well, with HDE, DGP can agree with up-to-date observations \cite{Farajollahi:2016lrk}.

In a DGP braneworld, it is postulated that spacetime have 5 dimensions (5d) in total (or the additional dimensions are compactified as usual). Gravity lives in both 5d and 4d, and matter lives only in 4d. The corresponding action is
\be
  S = \frac{M_5^2}{2} \int d^5 X \sqrt{-G} R_5
  + \frac{M_p^2}{2} \int d^4 x \sqrt{-g} R
  + \int d^4 x \sqrt{-g} \mathcal{L}_m + S_{GH}~,
\ee
where the capital $X$ and $G$ denote 5d coordinate and metric, and small $x$ and $g$ denote 4d coordinate and metric, respectively. The 4d metric $g_{\mu\nu}$ is the induced metric on the brane, embedded in the 5d. The $S_{GH}$ is the Gibbons-Hawking boundary term.

The Friedmann equation in DGP gravity is
\be
  H^2 - \epsilon \frac{H}{r_c} = \frac{\rho}{3M_p^2}~,
\ee
where $H$ is the 4d Hubble parameter, and $r_c$ is a characteristic distance arising from different dimensional Planck constants, $r_c \equiv M_p^2 / (2M_5^3)$. At distances $r\gg r_c$, the 4D Einstein's gravity applies. And at distances $r\ll r_c$, one recovers the 5d gravity. Here $\epsilon=\pm 1$ denotes two branches of solutions of the DGP cosmology: the $\epsilon = 1$ branch is the self-accelerating branch and the $\epsilon = -1$ branch is the branch where the DGP gravity itself does not have acceleration.

Although the $\epsilon=1$ branch can have self-acceleration, the solution does not look like a DE component \cite{Fang:2008kc, Lombriser:2009xg}. Thus in either branch of DGP, additional components are needed to behave as DE. To avoid the cosmological constant problem, HDE is an ideal component to add to the DGP scenario.

There are two approaches to apply HDE to DGP braneworld. Namely, one can study the 4d HDE or the 5d HDE, where the UV/IR cutoff relation is imposed in 4d and 5d respectively.

In most studies, the 4d HDE is studied. In this approach, one adds the $\rho_{DE}$ to the energy density of the 4d components. By defining that $\Omega_{r_c} = 1/ (4r)$, one can calculate the evolution of $\Omega_{de}$ for the HDE component, namely, for the cutoff chosen to be the future event horizon, one gets \cite{Wu:2007tp}
\be
  \Omega'_{de} = \frac{2}{C(1+z)}\Omega_\Lambda^{3/2}
  \left (
    \frac{-1}{\sqrt{\Omega_m+\Omega_{de}+\Omega_{r_c}}
    + \epsilon \sqrt{\Omega_{r_c}}}
    + \frac{C}{\sqrt{\Omega_{de}}}
  \right ) ~.
\ee
Similar results can be obtained for particle horizon and Hubble horizon. Namely, for the particle horizon,
\be
  \Omega'_{de} = \frac{2}{C(1+z)}\Omega_\Lambda^{3/2}
  \left (
    \frac{1}{\sqrt{\Omega_m+\Omega_{de}+\Omega_{r_c}}
    + \epsilon \sqrt{\Omega_{r_c}}}
    + \frac{C}{\sqrt{\Omega_{de}}}
  \right ) ~,
\ee
and for the Hubble horizon,
\be
  \Omega'_{de}
  =
  \frac{3\Omega_mC^2}{(1-C^2)(1+z)\sqrt{\Omega_{r_c}+\Omega_m(1-C^2)}}
  \left [
    \epsilon \sqrt{\Omega_{r_c}}
    + \sqrt{\Omega_{r_c} + \Omega_m(1-C^2)}
    \right ]
    ~.
\ee
The evolution of HDE in DGP can thus be solved with those cutoff possibilities.

Some efforts are also made in implementing the 5d HDE in the DGP braneworld \cite{Saridakis:2007cy, Saridakis:2007ns, Farajollahi:2016lrk}. The HDE in 5d can be derived from a similar observation as in 4d. For the 5d vacuum fluctuation not to exceed the energy density of a black hole, one requires $\rho_{\Lambda 5} V(S^3) < M_{BH}$, where the volume of a sphere in 5d is $V(S^3) = \pi^2 r^4/2$, the Schwarzschild black hole mass in 5d is
$M_{BH} = 3\pi M_5^3 r_s^2 /8$. Thus,
\be
  \rho_{\Lambda 5} = \frac{3C^2M_5^3}{4\pi L^2}~,
\ee
Here $L$ is the IR cutoff in the DGP theory. One can then write down the Friedmann equation in 4d. One finds that the 4d effective DE can be written as
\be
  \rho_{de}^\mathrm{eff} = \frac{3M_p^2\epsilon}{8\pi r_c}
  \sqrt{H^2 - C^2 L^{-2} }~.
\ee
This equation can be considered as the defining feature of \cite{Saridakis:2007cy, Saridakis:2007ns, Farajollahi:2016lrk} as a model of dark energy. A choice on $L$ can be made (for example $L=1/H$) and then the model can be fit with observations \cite{Farajollahi:2016lrk}.

\

\subsection{Scalar-Tensor Theory}
\label{sec:6.3}

The Brans-Dicke gravity is a simple type of scalar-tensor theory. It can be further generalized into $f(R, \phi)$ gravity, and further a more general scalar-tensor theory. HDE in a scalar-tensor theory is studied in \cite{Bisabr:2008gu}, with the action
\be
  S = \frac{1}{2} \int d^4 x \sqrt{-g}
  \left  [
    F(\phi) R
    + U(\phi) g^{\alpha\beta} \nabla_\alpha \phi \nabla_\beta \phi
    + V(\phi)
  \right ]
  + S_m~.
\ee
HDE with flat and curved spatial section are studied within this framework. Actually, this scalar-tensor theory can be further generalized to be \cite{Horndeski:1974wa}
\ba \label{eq:lgginf}
  \mathcal{L} &=&  P(\phi,X)-G_3(\phi,X) \Box \phi + G_4(\phi,X) R
  + G_{4,X} \left[ (\Box \phi)^2 - (\nabla_\mu\nabla_\nu\phi)^2\right]
  \nonumber\\
  &&   + G_5(\phi,X) G_{\mu\nu}\nabla^\mu\nabla^\nu\phi
  - \frac{1}{6} G_{5,X} \left[ (\Box \phi)^3 - 3\Box \phi(\nabla_\mu\nabla_\nu\phi)^2
                       + 2(\nabla_\mu\nabla_\nu\phi)^3 \right]~.
\ea
This is known as the Horndeski's theory. Interesting DE candidates (as tuning mechanism) are studied by \cite{Charmousis:2011bf}. And the theory can actually be further generalized. It remains interesting to study HDE in those contexts, including how holography requirements (such as Schwarzschild radius is modified and thus the UV-IR relation would be modified), how existing DE theories from \ref{eq:lgginf} would be altered with the presence of HDE, and so on. Alternatively, those studies can also be performed in the framework of effective field theory of dark energy \cite{Gubitosi:2012hu}.

\

\subsection{Horava-Lifshitz Theory}
\label{sec:6.4}

The key issue in quantum gravity is that gravity is a non-renormalizable theory. There have been many approaches to attack this problem. One simple approach is proposed by \cite{Horava:2009uw}. The observation is as follows.

In a Feynman diagram, each gravitational propagator contributes a factor of $1/k^2$. In loop integrals, this $1/k^2$ behavior is not enough to control the UV divergence and thus the theory becomes not only divergent in the UV, but non-renormalizable. If the propagator had higher powers in $k$, namely $1/k^n$ where $n\geq 4$, one would obtain a normalizable theory.

It is straightforward to construct theories with $1/k^n$ ($n\geq 4$) propagators. Simply, one can add higher derivatives to the gravitational Lagrangian. As the propagator is the inverse of the quadratic Lagrangian in momentum space, one gets $1/k^n$ where $n\geq 4$ behavior.

However, an additional problem arises in such high (finite) derivative theories, namely the ghost problem. Note that a propagator like
\be
  \frac{1}{k^2 G_N k^4} = \frac{1}{k^2} - \frac{1}{k^2 - 1/G_N}
\ee
can be decomposed into two propagators, one with positive propagator and the other with negative propagator. The negative propagator is problematic because such a term either breaks the optical theorem of quantum field theory and thus destroy the probability interpretation of quantum field theory as a quantum theory, or its energy is not bounded from below and thus unstable (which case to appear depends on the choice of the $i\epsilon$ prescription).

\cite{Horava:2009uw} proposed a solution to this problem, by giving up the Lorentz invariance in the UV. For example, a scalar with anisotropic scaling in space and time directions (known as the Lifshitz scaling) can have an action
\be
  S = \int dt d^3x \left \{
    \dot\phi^2 - (\nabla^2 \phi)^2
  \right \}~,
\ee
where $\nabla$ denotes covariant derivative in the spatial direction. This action is obviously not Lorentz invariant. But it generates a propagator which scales as $1/k^4$ in the UV.

Similarly, a gravitational theory with such anisotropic scaling can be constructed. One can write the gravitational section
\be
  S_g = \int dt d^3 x \sqrt{-g} \left  \{
    \frac{2}{\kappa^2} (K_{ij}K^{ij}-\lambda K^2) + \ldots
  \right \}~,
\ee
where $\ldots$ denotes the high derivative terms that will not affect our review. Here $\kappa\equiv 8\pi G$, and  $K_{ij}$ is the extrinsic curvature on a 3-dimensional spatial hypersurface
\be
  K_{ij} = \frac{1}{2} \left (
    \dot h_{ij} - \nabla_i N_j - \nabla_j N_i
  \right ) ~,
\ee
where $h_{ij}$ is the induced metric on the spatial hypersurface and $N_i$ is the shift vector. They are defined in the ADM decomposition of the metric as
\be
  ds^2 = -dt^2 + h_{ij}
  \left ( dx^i + N^i dt \right ) \left ( dx^j + N^j dt \right )~,
\ee
and we have set the lapse function $N=1$ as an assumption of the (projectable version of) the Horava-Lifshitz gravity.  Here $\lambda$ is a free parameter.

The Friedmann equations can be found to be
\be
  H^2 = \frac{\kappa^2}{6(3\lambda -1)} \rho
  + \frac{\beta k}{a^2}~,
\ee
\be
  \dot H + \frac{3}{2}H^2 = - \frac{\kappa^2}{4(3\lambda -1)} p
  + \frac{\beta k}{2a^2}~,
\ee
where $\beta \equiv \kappa^4 \mu^2\Lambda / [8(3\lambda-1)^2]$. The continuity equation is not modified by the Horava-Lifshitz gravity.

The implication for DE for Horava-Lifshitz gravity is studied in \cite{Saridakis:2009bv}. The implication for HDE is studied in \cite{Setare:2010wt}. In Horava-Lifshitz gravity, the evolution equation of HDE can be derived as
\be
  \frac{\Omega'_{de}}{\Omega_{de}}  =
  \frac{2}{\Omega_{de}}
  \left  [
    2 + \frac{\sqrt{\Omega_{de}}}{2C}
    + \frac{\Omega_{de}}{4(3\lambda -1)}
    + \frac{\Omega^{3/2}_{de}}{2C(3\lambda -1)}
  \right ]  ~.
\ee
Compared with the original HDE scenario, it is clear that the $\lambda$ parameter enters the evolution equation for $\Omega_{de}$. This $\lambda$ parameter is from the defining feature of Horava-Lifshitz gravity as possible anisotropic scaling of spacetime. The cosmological evolution can be deduced from here. It remains interesting to constrain $\lambda$ together with $c$ with the observational data.

\

\subsection{Other MG Theories}
\label{sec:6.5}

We here review a few other approaches of HDE in MD.

In \cite{Zhang:2007an}, the impact of HDE in loop quantum gravity and braneworld scenarios with timelike extra dimension are considered. In both theories, the Friedmann equation is modified into (see \cite{Ashtekar:2006uz, Randall:1999ee})
\be
  H^2 = \frac{8\pi G}{3} \rho
  \left ( 1 - \frac{\rho}{\rho_c}  \right )~.
\ee
This Friedmann equation is similar to that with spacelike extra dimension, but differ by a sign in front of the $\rho/\rho_c$ term. The different sign makes significant difference in the time evolution. This is because, with the minus sign, the Hubble parameter has a chance to reach 0 when $\rho$ approaches to $\rho_c$. Once crossing $H=0$, the universe turns between expansion and contraction by a big bounce. In such cosmology, the equation of state of HDE is not modified
\be
  w = - \frac{1}{3} \left ( 1 + \frac{2}{C} \sqrt{\Omega_{de}}  \right )~,
\ee
while the evolution equation of $\Omega_{de}$ is modified into
\be
  \Omega'_{de} = 2\Omega_{de} (\Omega_{de}-1)
  \left ( \frac{1}{C}\sqrt{\Omega_{de}} -1  \right )~.
\ee
The universe experience cyclic evolution in this model. It is noted that one has to replace the future event horizon into the future event horizon at the bouncing time, for the model to be self-consistent.

As another example, HDE is studied in induced gravity (\cite{Zee:1978wi, Zee:1979hy}) by \cite{Sun:2007rh}. The idea of induced gravity is that the Einstein's gravity is a result of spontaneous symmetry breaking. The action of induced gravity is
\be
  S = \int d^4 x \sqrt{-g} \left [
    - \frac{1}{2} \epsilon \phi^2 R
    - \frac{1}{2} g^{\mu\nu}\partial_\mu\phi\partial_\nu\phi
    - V(\phi)
    + \mathcal{L}_{\omega}
  \right ]~,
\ee
where $\epsilon$ is a dimensionless coupling constant. The effective Newton's gravitational constant in induced gravity is
\be
  G_\mathrm{eff} = \frac{1}{8\pi\epsilon\varphi^2}~.
\ee

The $V(\phi)$ is taken to be of a shape of spontaneous symmetry breaking
\be
  V(\phi) = \frac{1}{8} \lambda \left ( \phi^2 - v^2 \right )^2~.
\ee
At low energy, the spontaneous symmetry breaking drives gravity to the Einstein's gravity. The evolution of HDE can be solved in induced gravity as shown in \cite{Sun:2007rh}.

Although HDE with modified gravity is already a huge literature \cite{Setare:2010md,Landim:2015hqa}, there are many possibilities left unexplored.
Especially, recently there have been fast development on massive gravity and bimetric theories
(see for example \cite{Hinterbichler:2011tt} and the references therein).
It is interesting to combine those theories and HDE, and compare such theories with observations.

\

\section{Reconstruction Scalar Field DE and MG from HDE}
\label{sec:7}

\subsection{Reconstructing Scalar Field DE}
\label{sec:7.1}

HDE is usually considered to be a very different model of DE compared to scalar field models. However, it is interesting to note that with non-trivial potentials, one can use scalar field to reconstruct HDE \cite{Guberina:2005fb, Kim:2005gk, Zhang:2006av, Zhang:2006qu, Setare:2007eq, Setare:2007jw, Setare:2007hq, Zhang:2007es, Zhao:2007qy, Setare:2008pc, Cruz:2008cwa, Karami:2009we, RozasFernandez:2009bz}.

Although the parameter regime $C>1$ is ruled out in the simplest HDE models observationally by more than 5$\sigma$, technically the $C>1$ case is more conventional for the purpose of construction. This is because with $C>1$, HDE does not evolve across the phantom divide and can be simply reconstructed by a single scalar field model with well-defined kinetic term. This is done in \cite{Zhang:2006av} numerically.

To study the $C<1$ region, the reconstruction of HDE using phantom is studied in \cite{Setare:2007eq}. A phantom scalar field \cite{2002PhLB..545...23C} is a scalar field with wrong sign kinetic term
\be
  S = \int d^4 x \sqrt{-g} \left [ g^{\mu\nu} \partial_\mu\phi \partial_\nu\phi - V(\phi) \right ]~,
\ee
where the metric convention is $(-, +, +, +)$. Thus the kinetic term has negative kinetic energy $-\dot\phi^2$. For a phantom field, the energy density and pressure take the forms
\be
  \rho_{\phi} = - \frac{1}{2} \dot\phi^2 + V(\phi)~,
  \qquad
  p_{\phi} = - \frac{1}{2} \dot\phi^2 - V(\phi)~.
\ee
Note the additional minus sign in front of $\dot\phi^2/2$, which is the key difference between phantom and a conventional scalar field. The equation of state is thus
\be
  w_{\phi} = \frac{- \frac{1}{2} \dot\phi^2 + V(\phi)}{- \frac{1}{2} \dot\phi^2 - V(\phi)}~.
\ee
This equation of state satisfies $w<-1$ for positive $V$. Note that with $C<1$, HDE also have $w<1$ when it dominates. This is why phantom is used to mimic HDE at late times.

One can then solve $V(\phi)$ and $\dot\phi^2$ as
\be
  V(\phi) = \frac{1}{2} (1-w_\phi) \rho_\phi~, \qquad \dot\phi^2 = - (1-w_\phi)\rho_\phi~.
\ee
Note that the equation of state of $\phi$ can be written as
\be
  w_\phi = - \frac{1}{3\Omega_\phi H^2} \left ( 2\dot H + 3H^2 + \frac{k}{a^2}  \right ) ~.
\ee
One can thus solve $V(\phi)$ and $\dot\phi^2$ in terms of the HDE properties
\be \label{eq:reconv}
  V(\phi) = \frac{M_p^2}{2}  \left [ 2\dot H + 3H^2(1+\Omega_\phi ) + \frac{k}{a^2} \right ]~,\qquad
  \dot\phi^2 = M_p^2  \left [ 2\dot H + 3H^2(1-\Omega_\phi ) + \frac{k}{a^2} \right ]
\ee
We yet have to replace $H$ and $\dot H$ with functions of $\phi$. It is noticed that the following ansatz simplifies the construction problem:
\be\label{eq:reconans}
  \phi = t~, \qquad H = f(t)~.
\ee
Using slow roll approximated equation of motion of the scalar,
\be
  -3H\dot\phi + V'(\phi) = 0~,
\ee
one gets
\be
  3f(\phi) = V'(\phi)~.
\ee

With this ansatz \ref{eq:reconans}, Eq. \ref{eq:reconv} can be written as
\be \label{eq:reconsfinal}
  V(\phi) = \frac{M_p^2}{2}  \left [ 2f'(\phi) + 3f^2(\phi)(1+\Omega_\phi ) + \frac{k}{a^2} \right ]~,\qquad
  1 = \dot\phi^2 = M_p^2  \left [ 2 f'(\phi) + 3f^2(\phi)(1-\Omega_\phi ) + \frac{k}{a^2} \right ]
\ee

One can compare those equations with the scalar field with the corresponding equations of HDE. We make the correspondence
\be
  \rho_\phi \leftrightarrow \rho_{de}~,
  \quad
  p_\phi \leftrightarrow p_{de}~,
  \quad
  w_\phi \leftrightarrow w_{de}~.
\ee
Using the corresponding equations of HDE, one can thus write
\be
  V(\phi) = \frac{1}{2} (1-w) \rho_{de} = \frac{3H^2 \Omega_{de}}{16\pi G}
  \left  [
    \frac{4}{3} + \frac{2\sqrt{\Omega_{de}-C^2 \Omega_k}}{3C} + \frac{b^2(1+\Omega_k)}{\Omega_{de}}
  \right ] ~,
\ee
\be
  1 = \dot\phi^2 = - (1-w)\rho_{de} = \frac{H^2 \Omega_{de}}{4\pi G}
  \left  [
    -1 + \frac{\sqrt{\Omega_{de}-C^2 \Omega_k}}{C} + \frac{3b^2(1+\Omega_k)}{2\Omega_{de}}
  \right ] ~.
\ee
Matching those equations with Eq. \ref{eq:reconsfinal}, one can solve that
\be
  V = 3M_p^2 f^2(\phi)
  \left  [
    1 + \frac{2}{
        (6M_p^2f^2(\phi)-1)
        \pm
        \sqrt{
            (1-6M_p^2 f^2(\phi))^2 + 24 M_p^2 f^2(\phi)
        }
    }
  \right ]~.
\ee

It should be noticed that nowadays, the phantom instability problem may be avoided, thanks to the Galileons \cite{Deffayet:2010qz, Kobayashi:2010cm}, and more generally the Horndeski theory \cite{Horndeski:1974wa}. In those theories, $w<-1$ can be achieved without any quantum instabilities, because the fluctuations of those Galileon fields behaves very differently from the background evolution thanks to the apparently higher derivative action. It would be interesting to reconstruct HDE using those healthier models.

\

\subsection{Reconstructing MG}
\label{sec:7.2}

Modified gravity (MG) theories have rich dynamics \cite{Sotiriou:2008rp,Nojiri:2010wj,SaezGomez:2011yp,Zhang:2012jsa}.
It is thus interesting to study reconstructing those modified gravity theories using the dynamics of HDE.
This topic is studied by \cite{Wu:2007tn,Setare:2008hm,BouhmadiLopez:2011xi,Daouda:2011yf,Houndjo:2011fb,Karami:2011np,Chattopadhyay:2012eu,Jawad:2012xy,Borah:2013mna}.
Different MG theories has different degree of freedoms and dynamics. But the methodology of reconstruction is similar.
Here we shall review an example of $f(R)$ gravity to demonstrate this process \cite{Wu:2007tn}.
For simplicity, the spatial curvature is set to be $k=0$.

Consider the $f(R)$ gravity theory
\be
  S = \int d^4 x \sqrt{-g} \left [
  f(R) + \mathcal{L}_m
  \right ]~,
\ee
where $M_p$ is set to one for simplicity. The modified field equations can be written as
\be
  G_{\mu\nu} =
  T_{\mu\nu}^\mathrm{(curv)} +
  T_{\mu\nu}^\mathrm{(m)}~,
\ee
where
\be
  T_{\mu\nu}^\mathrm{(curv)}
  \equiv
  \frac{1}{f'(R)} \left  \{
    \frac{1}{2} g_{\mu\nu} \left  [
      f(R) - R f'(R)
    \right ]
    + f'(R)^{;\alpha\beta}
    \left  (
      g_{\mu\alpha}g_{\nu\beta} - g_{\mu\nu}g_{\alpha\beta}
    \right )
  \right \} ~,
\ee
and the stress tenor of matter is rescaled by
\be
  T_{\mu\nu}^\mathrm{m} \equiv
  \frac{\tilde T_{\mu\nu}^\mathrm{m}}{f'(R)}~,
\ee
where $\tilde T_{\mu\nu}^\mathrm{m}$ is the stress tensor in the case of Einstein's gravity.

The continuity equation for total energy remains the same with this modification, while the Friedmann equations take the form
\be
  H^2 + \frac{k}{a^2} = \frac{1}{3}
  \left [
    \rho_\mathrm{curv} + \frac{\rho_m}{f'(R)}
  \right ] ~,
\ee
\be
  2 \frac{\ddot a}{a} + H^2 + \frac{k}{a^2} = - ( p_\mathrm{curv} + p_m )~.
\ee

Those equations can be rewritten into
\be
  \dot H = - \frac{1}{2f'(R)} \left  \{
    3H_0^2 \Omega_m(1+z)^3 + \ddot R f''(R)
    + \dot R \left  [
      \dot R f'''(R) - H f''(R)
    \right ]
  \right \} ~.
\ee
One can future use redshift as cosmic time by noting that $d\cdot/dt = -(1+z)Hd\cdot/dz$. Then the terms are reorganized according to the derivatives on $f(R)$ as
\be\label{eq:frcn}
  \mathcal{C}_3 (z) \frac{d^3f}{dz^3} +
  \mathcal{C}_2 (z) \frac{d^2f}{dz^2} +
  \mathcal{C}_1 (z) \frac{df}{dz}
  =
  -3H_0^2\Omega_{m0}(1+z)^3~.
\ee
Those $\mathcal{C}_n$ coefficients can be written in terms of $\Omega_{DE}$ and its derivatives. One can thus solve the differential equation \ref{eq:frcn} numerically to obtain the expression of $f$. This accomplishes the reconstruction.

\

\section{Other DE Models inspired by Holographic Principle}
\label{sec:8}

The original HDE scenario by \cite{Li:2004rb} uses the future event horizon as the IR cutoff of the theory. Although this choice is so far working well with the experiments, many alternative choices of the IR cutoff exist. So far, we are still lack of a first principle to pin down the nature of the IR cutoff. Thus we survey different IR cutoffs and their implications in this section.

\

\subsection{Agegraphic Dark Energy}
\label{sec:8.1}

It is proposed in \cite{Cai:2007us, Wei:2007ty, Wei:2007xu, Chen:2011rz} the one can use time of the FRW universe to be the IR cutoff. This model is known as the agegraphic dark energy (ADE).

There have been two major versions of the ADE. The first version of ADE by \cite{Cai:2007us} made use of the physical time $t$ as the IR cutoff. But soon it is found that such a component cannot evolve from a sub-dominate component to a dominate component. This does not behave as the DE that we see today. We thus shall not review in detail this version of ADE.

To find a realistic model of ADE, it is noted by \cite{Wei:2007ty} that once the IR cutoff is replaced by the conformal time, the domination problem can be solved. The improvement in this second version of ADE is similar to the original HDE by \cite{Li:2004rb}, where it is proposed to use the future event horizon instead of the Hubble horizon as the IR cutoff (note the similar relation between the Hubble horizon and future event horizon, between the cosmic physical time and the conformal time).

In this new version, the energy density of ADE is
\be \label{eq:adedef}
  \rho_{de} = \frac{3n^2M_p^2}{\eta^2}~,
\ee
where $\eta$ is the conformal time
\be
  \eta = \int \frac{dt}{a} = \int \frac{da}{a^2 H}~.
\ee
The fractional energy density thus takes the form
\be
  \Omega_{de} = \frac{n^2}{H^2\eta^2}~.
\ee
The evolution equation of $\Omega_{de}$ can be calculated to be
\be
  \Omega'_{de} = \Omega_{de} (1-\Omega_{de})
  \left ( 3- \frac{2}{n} \frac{\sqrt{\Omega_{de}}}{a}  \right )~.
\ee
The equation of state is
\be \label{eq:adeeos}
  w = -1 + \frac{2}{3n} \frac{\sqrt{\Omega_{de}}}{a}~.
\ee

Unlike the original HDE, in ADE, $\Omega_{de}$, $n$ and $a$ has to satisfy a consistency relation. In a matter dominated universe, $\eta \propto \sqrt{a}$. From the defining equation of ADE \ref{eq:adedef}, $\rho_{de}\propto 1/a$. From the continuity equation, this implies $w=-2/3$. Compare it with the solved equation of state  \ref{eq:adeeos}, one get that, in matter dominated era,
\be\label{eq:adecons}
  \Omega_{de} = \frac{n^2a^2}{4}~.
\ee
In other words, the amount of ADE in matter dominated era is determined. As a result, in terms of observations, the ADE model is a one-parameter model instead of a two-parameter model. This is different from HDE, as in HDE there are two free parameters, namely $C$ and $\Omega_{de}$.

ADE also solves the cosmic coincidence problem, but in a different way compared to HDE. In ADE, the fractional DE density is determined by Eq. \ref{eq:adecons} and thus there is no longer coincidence.

\begin{figure*}
\centering
\includegraphics[width=0.8\textwidth]{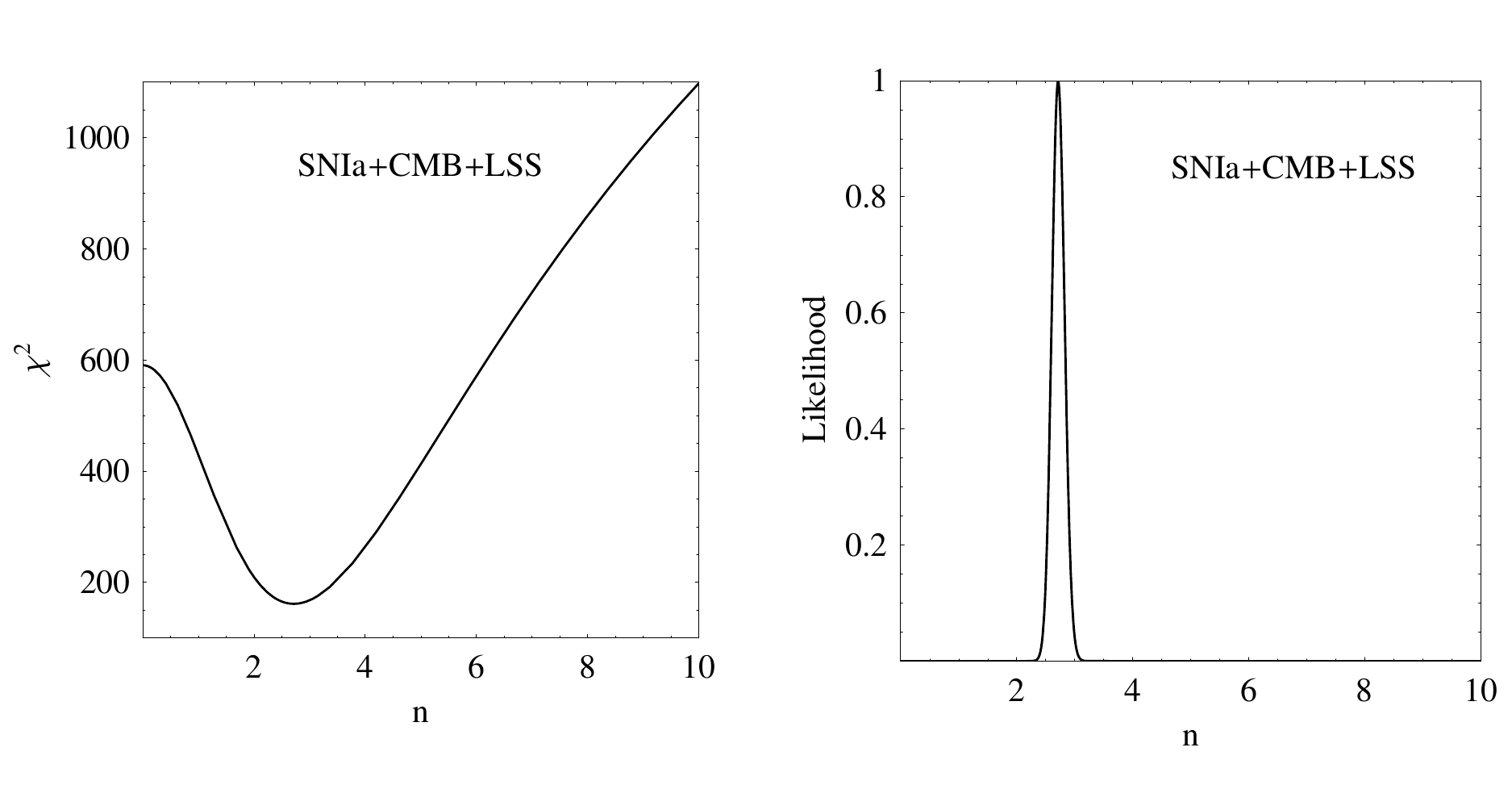}
\caption{The $\chi^2$ and the corresponding likelihood for the ADE model,
from a joint analysis of the SNIa, CMB, and BAO observations.
From \cite{Wei:2007xu}.}
\label{fig16}
\end{figure*}

In \cite{Wei:2007xu}, ADE is compared to data. Using the SNIa data, it is shown that $n = 2.954^{+0.264}_{-0.245}$ at $1\sigma$ CL.
Using the combined SNIa, CMB and LSS data, $n$ is constrained to be $n = 2.716 ^{+0.111}_{-0.109}$ (See Fig.~\ref{fig16}). As we shall review in the next section, compared with $\Lambda$CDM and HDE, ADE is strongly disfavored by current data.

\

\subsection{Ricci Dark Energy}
\label{sec:8.2}

Geometrically, an important IR scale of our universe is the curvature of spacetime. The scalar quantity built from the spacetime curvature, the Ricci scalar, is thus a good candidate for a covariant choice of IR cutoff \cite{Nojiri:2005pu, Gao:2007ep, Zhang:2009un,delCampo:2013hka}.

\begin{figure*}
\centering
\includegraphics[width=0.8\textwidth]{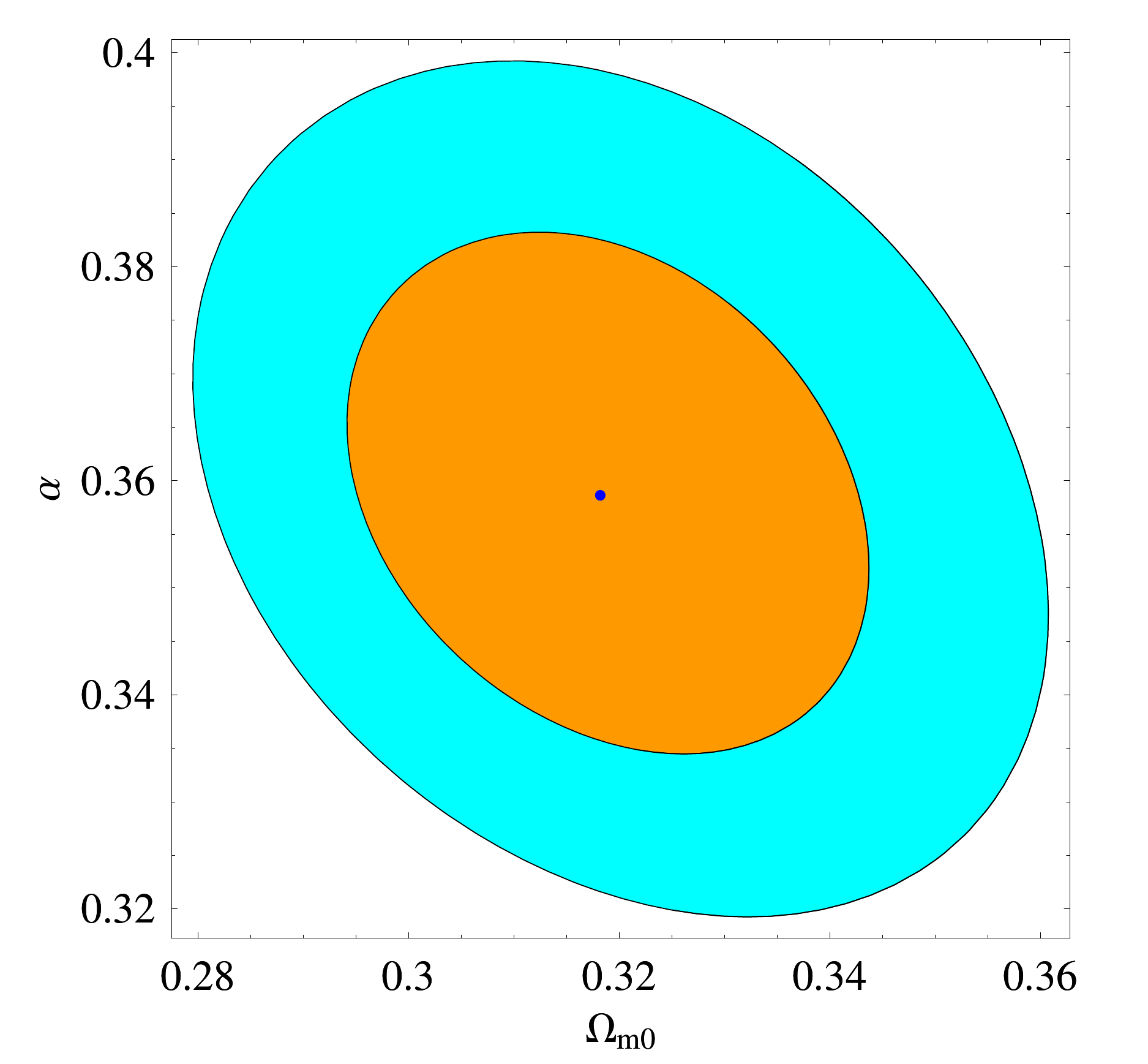}
\caption{Probability contours at $1\sigma$ and $2\sigma$ CL in the $(\Omega_{\rm m0},~\alpha)$ plane for the RDE model,
from a joint analysis of the SNIa, CMB, and BAO observations.
A point denotes the best fit; at the best fit $\chi^2_{\rm min}=324.317$.
From \cite{Zhang:2009un}.}
\label{fig17}
\end{figure*}

In FRW cosmology, the Ricci scalar takes the form
\be
  R = -6 \left ( \dot H + 2H^2 + \frac{k}{a^2}  \right )~.
\ee
To use the Ricci curvature as the cutoff of the UV-IR correlation, one thus obtains the energy density of Ricci dark energy (RDE)
\be
  \rho_{de} = - \frac{\alpha}{16\pi} R
  = \frac{3\alpha}{8\pi}
  \left (\dot H + 2H^2 + \frac{k}{a^2}  \right )~.
\ee
Here the numerical factor $16\pi$ is chosen to ease the calculation. With this definition, the Friedmann equation can be written as
\be
  H^2 = \frac{8\pi G}{3} \rho_{m0}e^{-3x}
  + (\alpha-1)k e^{-2x}
  + \alpha \left  (
    \frac{1}{2} \frac{dH^2}{dx} +2H^2
  \right )~,
\ee
where $x\equiv \ln a$. This equation can be solved as
\be\label{eq:rdesol}
  E^2(a) = \Omega_{m0} a^{-3}
  + \Omega_{k0} a^{-2}
  + \frac{\alpha}{2-\alpha} \Omega_{m0} a^{-3}
  + f_0 a^{-(4-\frac{2}{\alpha} )}~,
\ee
where $f_0$ is the integration constant, which can be fixed by noting $E_0=1$:
\be
  f_0 = 1-\Omega_{k0}-\frac{2}{2-\alpha} \Omega_{m0}~.
\ee
In Eq.~\ref{eq:rdesol}, one can identify the RDE density
\be
  \Omega_{de} = \frac{\alpha}{2-\alpha}\Omega_{m0}a^{-3}
    + f_0 a^{-(4-\frac{2}{\alpha} )}~.
\ee
The value $\alpha = 1/2$ is of special interest, where the RDE behaves as a cosmological constant plus a component of ``dark matter''. When $1/2 \leq \alpha < 1$, the RDE has EoS $-1 \leq w \leq < -1/3$. When $\alpha < 1/2$, the RDE start from quintessence-like and evolves to phantom-like. Thus the behavior is like quintom \cite{Feng:2004ad}. The EoS for the RDE can be written as
\be
  w = -1 + \frac{(1+z)}{3} \frac{d\ln\Omega_{de}}{dz}~.
\ee
The observational constraint for RDE is performed by \cite{Zhang:2009un}, with $\alpha = 0.394^{+0.152}_{-0.106}$ from SNIa only ($1\sigma$).
A joint analysis of the SNIa, CMB, and BAO observations gives a much tighter constraint $\alpha = 0.359^{+0..024}_{-0.025}$ (See Fig.~\ref{fig17}).

\

\subsection{Hubble Horizon as Characteristic Length Scale}
\label{sec:8.3}

As we have discussed, the Hubble parameter itself, though must natural, cannot be the IR cutoff of HDE
\footnote{not only in Einstein gravity, but also in simple Brans-Dicke gravity \cite{Gong:2008br}.}.
Nevertheless, there are lots of efforts in generalizing this idea.

The simplest generalization is to note that one can not only use Hubble but also is derivative \cite{Granda:2008dk}. One can than impose
\be
  \rho_{de} = 3 \left ( \alpha H^2 + \beta \dot H \right )~,
\ee
where $\alpha$ and $\beta$ are constants. The Friedmann equation in this case can be written as
\be
  M_p^2 H^2 = \frac{1}{3}
  \left ( \rho_{m0} e^{-3x} + \rho_{r0} e^{-4x} \right )
  + \alpha H^2 + \frac{\beta}{2} \frac{dH^2}{dx} ~,
\ee
where $x\equiv \ln a$. This form is similar to the Ricci-DE. This is not surprising because in flat spatial sections without spatial curvature, the Ricci scalar is just a linear combination of Hubble parameter and its time derivative. The equation can be solved similarly to Ricci-DE. The study is generalized to non-flat spatial sections by \cite{Karami:2009je}.

\cite{Gong:2009dc} proposed a different approach to use the Hubble scale as the IR cutoff. The authors studied a DGP type model. Instead of imposing the UV-IR relation completely in four spacetime dimensions, the UV cutoff is chosen to be the black hole formation bound with extra dimensions. In this case, the Hubble IR cutoff gives a working model of HDE.

Another approach to make sense of the Hubble IR cutoff is to make the $C$ parameter time-dependent (and thus redshift-dependent) \cite{Xu:2009ys}. It is also noted that HDE with Hubble cutoff can be made to work in Brans-Dicke cosmology with a potential \cite{Liu:2009ha}, or in Einstein gravity with interacting HDE \cite{Duran:2012yr}.

\

\subsection{Other Characteristic Length Scales}
\label{sec:8.4}

Apart from the above considerations, there are a lot of other choices as the IR length scale.

In \cite{Guberina:2005mp}, an energy scale dependent Newton's gravitational constant is discussed. This is similar to the time varying  gravitational constant but of a different origin -- the time variation is induced by an energy scale variation, which originates from the time dependence of the IR cutoff.

Thus at IR energy scale $\mu$, the energy density for HDE is written as
\be
  \rho_{de}(\mu) = \kappa \mu^2 G_N^{-1}(\mu)~.
\ee

To proceed, \cite{Guberina:2005mp} postulate that the stress tensors $G_N T^{\mu\nu}_\mathrm{total}$ and $T^{\mu\nu}_\mathrm{matter}$ are separately conserved. The resulting continuity equation of the  $G_N T^{\mu\nu}_\mathrm{total}$ conservation takes the form
\be
  \dot G_N (\rho_{de}+\rho_m) + G_N \dot\rho_{DE} = 0~.
\ee
Inserting it to the definition of HDE with scale dependence, one get the RG running equation of the Newton's gravitational constant
\be
  \frac{d G_N}{d\mu} = - \frac{2\kappa\mu}{\rho_m}~.
\ee
Solutions can be found for this equation, and the cosmological implications are studied in \cite{Guberina:2005mp}.

Alternatively, \cite{Sheykhi:2009zv} considered the possibility to use the apparent horizon as the IR cutoff. In spatial flat FRW spacetime, the apparent horizon is just the Hubble horizon. While with spatial curvature, the apparent horizon takes the form
\be
  r_A = \frac{1}{\sqrt{H^2 + k/a^2}}~.
\ee
This is a different form compared to the Hubble horizon in open or closed universes. It is noticed that this modification alone cannot accelerate the universe. Interaction between DE and dark matter is considered to further make this HDE component consistent with observations.

In \cite{Huang:2012nz}, a conformal-age-like length is used for the IR cutoff. This approach is similar to ADE.
However, a length scale instead of a time scale is proposed:
\be
  L = \frac{1}{a^4(t)} \int_0^t dt' a^3(t')~.
\ee
The cosmological implication and fitting with data is studied in \cite{Huang:2012gd}. It is interesting to note that from similar reasoning one can easily define two families of horizons
\be
  L = \frac{1}{a^{n+1}(t)} \int_0^t dt' a^n(t')~,
\ee
and
\be
  L = \frac{1}{a^{n+1}(t)} \int_t^\infty dt' a^n(t')~,
\ee
facing to the past and the future, respectively. The cosmological implications can be studied.

As another choice of IR cutoff, \cite{Huang:2012xma} considered the possibility of the total comoving horizon of the universe
\be
  \eta = \int_0^t \frac{c dt'}{a(t')} = \int_0^a \frac{da'}{H' (a')^2}~.
\ee

It would be interesting to see whether other cosmological quantities can behave as an IR cutoff. For example, the growth of structure defines characteristic time scales (time at which the universe becomes nonlinear at small scales) and spatial scales (length scales beyond which the universe is homogeneous and isotropic). Those fluctuation quantities may be relevant because the vacuum energy fall into black hole argument seem to apply better in a universe with structures instead of perfectly homogeneous and isotropic. Also, the FRW metric defines a preferred time direction. Thus it makes sense to study spatial curvature quantities such as the extrinsic curvature of space, etc. in the ADM formalism. Those possibilities have not been fully studied.

\

\subsection{Entropy-Corrected HDE}
\label{sec:8.5}

The HDE UV/IR relation, as part of holography, is closely related to the black hole entropy. To see this, one can rewrite the energy density of HDE as
\be
  \rho_{de} = 3 C^2 L^{-4} \times (M_p^2 L^2)
  \propto 3 C^2 L^{-4} S_0~,
\ee
where $S_0 \equiv A/(4G)$ is the black hole entropy formula following the area law.

It is well known that the area law of the black hole entropy is an approximate relation. The corrections in general take the following form (see, for example, \cite{2003gr.qc.....3030C, Cai:2008ys, 2009arXiv0901.1302C})
\be
  S = \frac{A}{4G} + \tilde\alpha \ln \frac{A}{4G} + \tilde\beta~.
\ee
Here $\tilde\alpha$ and $\tilde\beta$ can in principle be computed but their values are now not yet agreed on. Thus here they are considered as free parameters.

With the entropy correction, \cite{Wei:2009kp} generalized
the energy density of HDE as
\be
  \rho_{de} = 3C^2 M_p^2 L^{-2}
  + \alpha L^{-4} \ln(M_p^2L^2) + \beta L^{-4}~.
\ee
Here $C$, $\alpha$ and $\beta$ are dimensionless parameters, which naturally take values of order one. It is also noted by \cite{Wei:2009kp} that ADE can be similarly generalized. More topics about entropy correction and their cosmological consequences for HDE are studied in \cite{Sadjadi:2010az, Jamil:2010sk, Setare:2010dr, Sheykhi:2010pa}.

\

\section{Comparisons of Dark Energy Models}
\label{sec:9}

It is impossible to make a fair comparison for different DE models by directly comparing their values of $\chi^2_{min}$.
This is because different models may have different numbers of parameters,
while adding extra parameters will reduce the predictive power of the model.
Therefore, to establish the validity of the model, one must take the factor of parameter number into account.

The most commonly used model selection criteria are
the Akaike Information Criterion (AIC) \cite{1974ITAC...19..716A} and the Bayesian Information Criterion (BIC) \cite{1978AnSta...6..461S}, defined as
\be \label{eq:AICBIC}
{\rm AIC}=-2\ln{\mathcal{L}_{\rm{max}}}+2k, \quad {\rm BIC}=-2\ln{\mathcal{L}_{\rm{max}}}+k\ln{N},
\ee
where $\mathcal{L}_{\rm max}$ is the maximum likelihood,
$k$ is the number of parameters and $N$ is the number of data points used in the fit.
Note that for Gaussian errors, $\chi^{2}_{\rm{min}}=-2\ln{\mathcal{L}_{\rm{max}}}$.
Information criteria penalize the introduction of new parameters that do not significantly improve the quality of the fit \cite{Liddle:2004nh}.
It is clear that a model favored by the observations should give a small ${\rm AIC}$ and a small ${\rm BIC}$.
In practice, it is convenient to choose the $\Lambda $CDM model as the reference model, then
\be \label{eq:AICBIC2}
  \Delta {\rm AIC}={\rm AIC}_{model}-{\rm AIC}_{\Lambda CDM}, \quad \Delta {\rm BIC}={\rm BIC}_{model}-{\rm BIC}_{\Lambda CDM}.
\ee

We will also use the Bayesian Evidence (BE) as a model selection criterion.
The BE of a model takes the form \cite{Mukherjee:2005wg}
\be \label{eq:BE}
\mathrm{BE}=\int {\mathcal{L}(\mathbf{d|\theta
},M)\mathbf{p}(\mathbf{\theta }|M)d\mathbf{\theta }},
\ee
where $\mathcal{L}(\mathbf{d|\theta },M)$ is the likelihood function given by the data $\textbf{d}$, model $M$ and parameters $\mathbf{\theta }$,
and $\mathbf{p}(\mathbf{\theta}|M)$ is the priors of model parameters.
BE is the average of the likelihood of a model over its prior of the parameter space,
and thus automatically includes the penalties of the number of parameters and data.
It is clear that a model favored by the observations should give a large BE.
In practice, it is convenient to use the logarithm of BE as a guide for model comparison, then
\be \label{eq:BE2}
  \Delta \ln \mathrm{BE}=\ln \mathrm{BE}_{model}-\ln \mathrm{BE}_{\Lambda CDM}.
\ee

\

\subsection{Comparisons among Various HP-Inspired DE models}
\label{sec:9.1}

The earliest comparisons among the HP-inspired DE models was made in \cite{Li:2009bn}.
By using a combination of the Constitution SNIa sample, the shift parameter $R$ from the WMAP5, and the BAO measurement from the SDSS,
Li et al. constrained the parameter spaces of three holographic DE models,
including the original HDE model \cite{Li:2004rb}, the ADE model \cite{Wei:2007ty}, and the RDE model \cite{Gao:2007ep}.
It is found that,
for the HDE model, $\Omega_{m0}=0.277^{+0.022}_{-0.021}$ and $C=0.818^{+0.113}_{-0.097}$, corresponding to $\chi^2_{min}=465.912$;
for the ADG model, $n=2.807^{+0.087}_{-0.086}$, corresponding to $\chi^2_{min}=481.694$;
for the RDE model, $\Omega_{m0}=0.324^{+0.024}_{-0.022}$ and $\alpha=0.371^{+0.023}_{-0.023}$, corresponding to $\chi^2_{min}=483.130$.
Moreover, the results of $\Delta \ln \mathrm{BE}$ for the three holographic DE models are listed in table \ref{tab:8}.
It is seen that although the HDE model performs a little poorer than the $\Lambda$CDM model,
it performs much better than the ADE model and the RDE model.
In other words,
among these three HP-inspired DE models, the HDE model is more favored by the observational data.
This conclusion is consistent with the results given by other observations \cite{Cardenas:2013moa,Guo:2015gpa}.

\begin{table}
\caption{The results of $\Delta \ln \mathrm{BE}$ for three holographic DE models. From \cite{Li:2009bn}.}
\begin{center}
\label{tab:8}
\begin{tabular}{cccc}
  \hline\hline
  ~~~Model~~~ & ~~~HDE~~~ & ~~~ADE~~~ & ~~~RDE~~~ \\
  \hline
  ~~~$\Delta \ln \mathrm{BE}$~~~ & ~~~$-0.86$~~~ & ~~~$-5.17$~~~ & ~~~$-8.14$~~~ \\
  \hline\hline
\end{tabular}
\end{center}
\end{table}

\begin{figure*}
\centering
\includegraphics[width=0.8\textwidth]{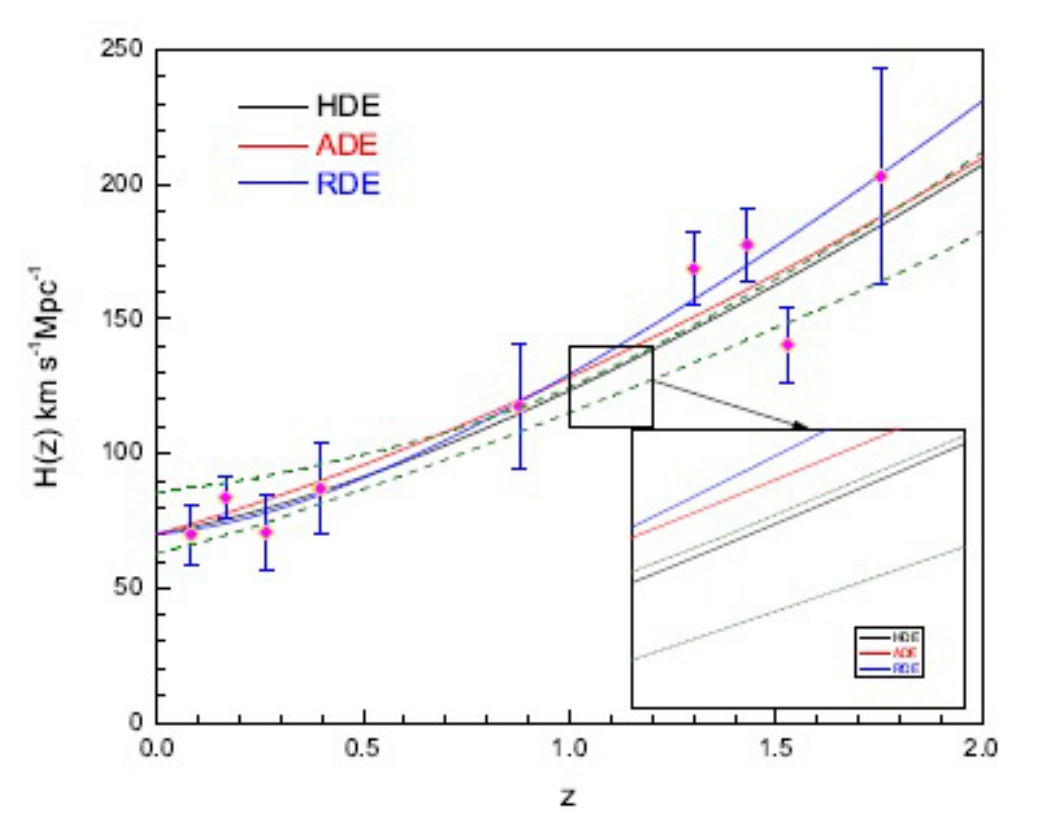}
\caption{Comparison of the observed $H(z)$, as square dots, with the predictions from the three holographic DE models.
From \cite{Li:2009bn}.}
\label{fig18}
\end{figure*}

In Fig.\ref{fig18}, Li et al. further compared the observed expansion rate $H(z)$ \cite{Simon:2004tf}
with that predicted by these three holographic DE models.
Notice that the area surrounded by two dashed lines shows the $1\sigma$ confidence interval \cite{Spergel:2003cb},
and a DE model would be disfavored by the observation
if it gives a curve of $H(z)$ that falls outside this area.
It is seen that among these three holographic DE models,
only the curve of $H(z)$ predicted by the HDE model falls inside this confidence interval.
This result verifies the conclusion of table \ref{tab:8} from another perspective.

\begin{figure*}
\centering
\includegraphics[width=0.8\textwidth]{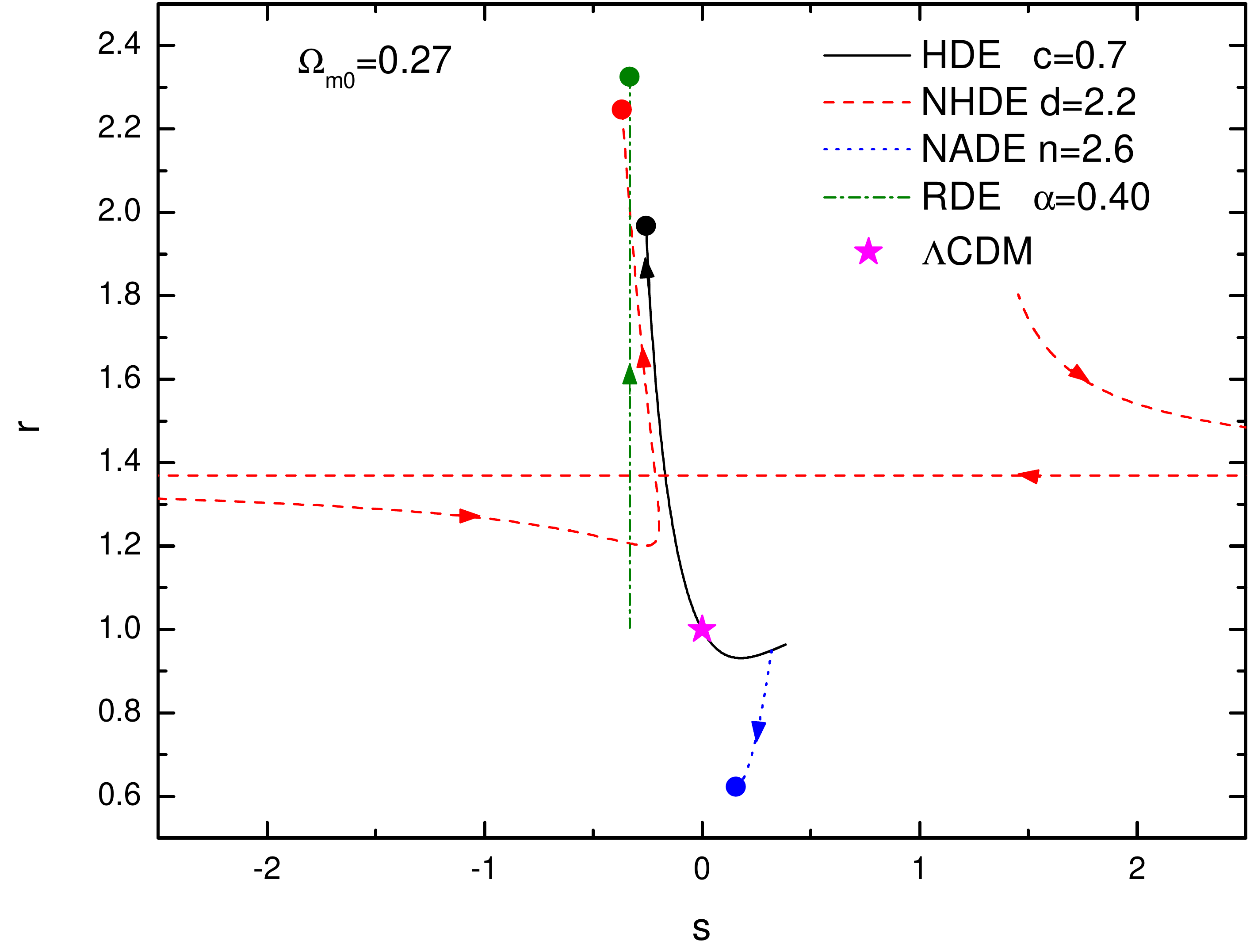}
\caption{Comparison of the evolutionary trajectories $r(s)$ of the four holographic DE models in the $r$--$s$ plane.
The present values $\{r_0,~s_0\}$ of the four holographic DE models are marked by the round dots.
The $\Lambda$CDM model, denoted by a star, is also shown for a comparison. The arrows indicate the evolution directions of the models.
From \cite{Cui:2014sma}.}
\label{fig19}
\end{figure*}

In addition to data fitting, it is also popular to distinguish these HP-inspired DE models with various diagnostic tools.
For example, in \cite{Cui:2014sma}, Cui and Zhang applied the statefinder diagnostic to the four holographic DE models,
including the original HDE model, the new holographic dark energy (NHDE) model \cite{2012arXiv1210.0966L}, the ADE model, and the RDE model.
The comparison of the evolutionary trajectories $r(s)$ of the four holographic DE models are plotted in Fig. \ref{fig19}.
From this figure one can see that, in the low-redshift region,
the difference between the HP-inspired DE models and the $\Lambda$CDM model can be easily distinguished,
which is quite different from the cases of $H(z)$ and $q(z)$ \cite{Cui:2014sma}.
This implies that the statefinder diagnostic is very useful in breaking the low-redshift degeneracies of various holographic DE models.

\begin{figure*}
\centering
\includegraphics[width=0.8\textwidth]{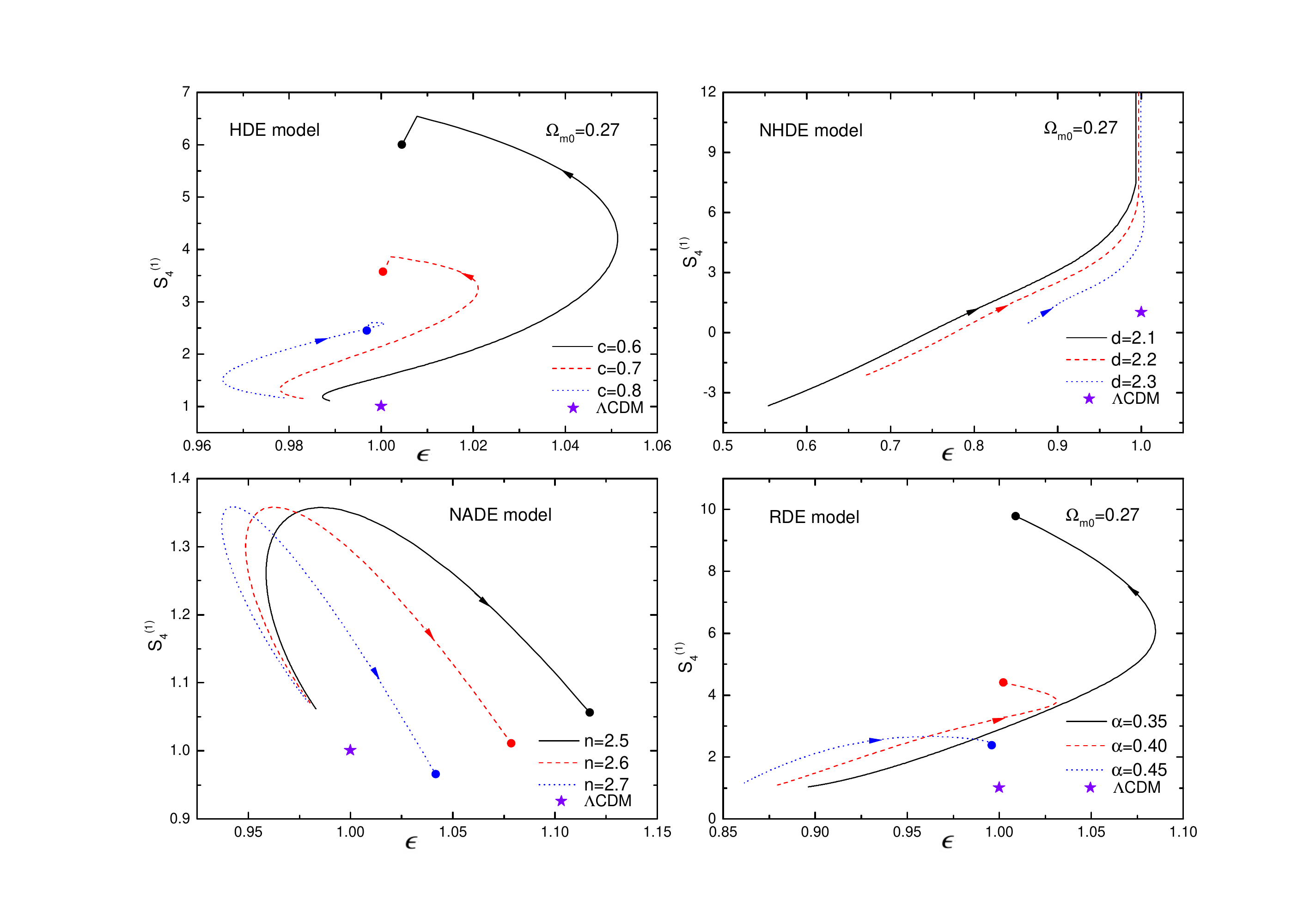}
\caption{The evolutionary trajectories of $S^{(1)}_4(\epsilon)$ for the HDE, NHDE, NADE and RDE models.
The current values of $\{S^{(1)}_4, \epsilon\}$ of the holographic DE models are marked by the round dots.
$\{S^{(1)}_4, \epsilon\}=\{1,1\}$ for the $\Lambda$CDM model is also shown as a star for a comparison.
The arrows indicate the evolution directions of the models.
Note that the dots for current values of the NHDE model are not shown in this plot
owing to the fact that the present-day $S_4^{(1)}$ values are too large compared to that of $\Lambda$CDM.
From \cite{Zhang:2014sqa}.}
\label{fig20}
\end{figure*}

In \cite{Zhang:2014sqa}, Zhang et al. further diagnosed these four HP-inspired DE models
with statefinder hierarchy $S^{(1)}_3$, $S^{(1)}_4$
\footnote{The statefinder hierarchy satisfy \cite{Arabsalmani:2011fz} $S^{(1)}_3=A_3$, $S^{(1)}_4=A_4+3(1+q)$,
where $A_3$ and $A_4$ are given in Eq.\ref{eq:An}.
For more studies about statefinder hierarchy, see \cite{Hu:2015bpa,Zhou:2016rtz}.}
and fractional growth parameter $\epsilon$
\footnote{The fractional growth parameter is defined as \cite{Acquaviva:2008qp,Acquaviva:2010vr} $\epsilon \equiv f_{model}(z)/f_{\Lambda CDM}(z)$,
where $f(z)=d\ln\delta/d\ln a$ describes the growth rate of the linear density perturbation.}.
The evolutionary trajectories of $S^{(1)}_4(\epsilon)$ for HDE, the NHDE, the ADE and the RDE models are plotted in Fig. \ref{fig20}.
From this figure, one can also clearly see that,
there are significant differences among the evolution of $S^{(1)}_4(\epsilon)$ given by the four HP-inspired DE models.
In other words, employing the composite null diagnostics $\{S^{(1)}_4, \epsilon\}$,
all the holographic DE models can be differentiated quite well.

\

\subsection{Comparisons Between HP-Inspired DE models and Other Cosmological Models}
\label{sec:9.2}

\begin{figure*}
\centering
\includegraphics[width=0.8\textwidth]{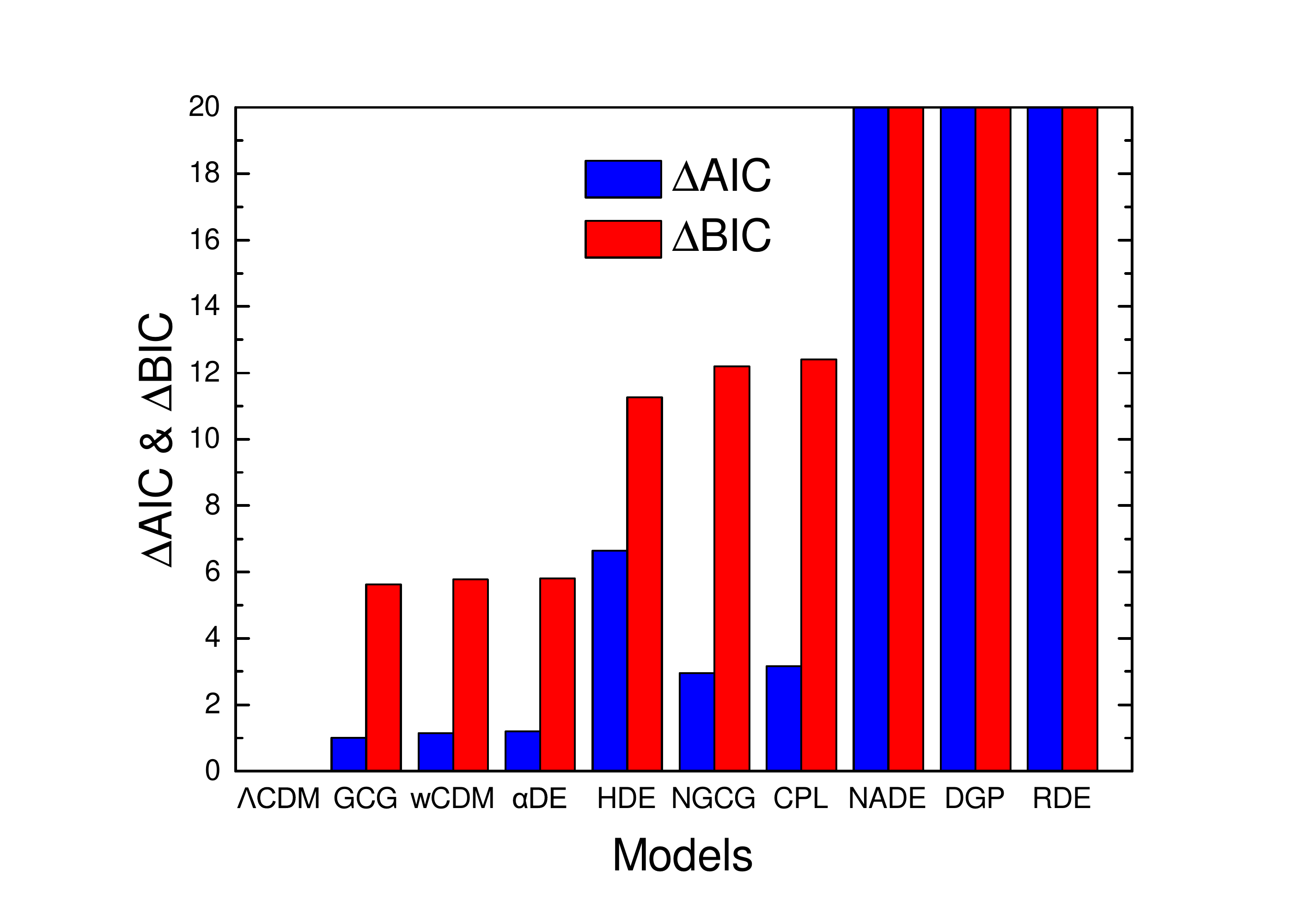}
\caption{Graphical representation of the model comparison result.
The order of models from left to right is arranged according to the order of increasing $\Delta {\rm BIC}$.
From \cite{Zhang:2014sqa}.}
\label{fig21}
\end{figure*}

A lot of efforts have also been paid to compare the HP-inspired DE models with other cosmological models \cite{Li:2009jx,Wei:2010wu}.
A latest result was given in \cite{2016arXiv160706262X},
where Xu and Zhang made a comparison for ten popular cosmological models according to their capabilities of fitting the current observational data.
These ten cosmological models include $\Lambda$CDM model, $w$CDM model,
Chevalliear-Polarski-Linder (CPL) model \cite{Chevallier:2000qy,Linder:2002et},
generalized Chaplygin gas (GCG) model \cite{Bento:2002ps}, new generalized Chaplygin gas (NGCG) model \cite{Zhang:2004gc},
HDE model, ADE model, RDE model,
Dvali-Gabadadze-Porrati (DGP) model \cite{Dvali:2000hr}, and a theoretical variant of DGP called $\alpha$DE model \cite{Dvali:2003rk}.
The model comparison results are shown in table \ref{tab:9} and Fig. \ref{fig21}.

\begin{table*}\tiny
\caption{Summary of the information criteria results of ten cosmological models. From \cite{2016arXiv160706262X}.}
\label{tab:9}
\small
\setlength\tabcolsep{40.5pt}
\renewcommand{\arraystretch}{1.5}
\centering
\begin{tabular}{cccccccccccc}
\\
\hline
Model &$\chi^{2}_{{\rm min}}$  &$\Delta$AIC& $\Delta$BIC \\ \hline

$\Lambda$CDM
                   &$ 699.375$
                   &$ 0$
                   &$ 0$
                   \\
GCG
                   &$ 698.381$
                   &$ 1.006$
                   &$ 5.623$
                   \\

$w$CDM
                   &$ 698.524$
                   &$ 1.149$
                   &$ 5.766$
                   \\
$\alpha$DE
                   &$698.574$
                   &$ 1.199$
                   &$ 5.816$
                   \\
HDE
                   &$ 704.022$
                   &$ 6.647$
                   &$ 11.264$
                   \\
NGCG
                   &$698.331$
                   &$2.956$
                   &$12.191$
                   \\
CPL
                   &$ 698.543$
                   &$ 3.199$
                   &$ 12.401$
                   \\
NADE
                   &$ 750.229$
                   &$ 50.854$
                   &$ 50.854$
                   \\
DGP
                   &$ 786.326$
                   &$ 86.951$
                   &$ 86.951$
                   \\
RDE
                   &$987.752$
                   &$290.337$
                   &$294.994$
                   \\
\hline
\end{tabular}
\end{table*}

One can see that, the $\Lambda$CDM model is still the best one among all the cosmological models.
The GCG model, the $w$CDM model, and the $\alpha$DE model are worse than the $\Lambda$CDM model, but still are good models compared to others.
The HDE model, the NGCG model, and the CPL model can still fit the current observations well,
but from an economically feasible perspective, they are not so good.
The ADE model, the DGP model, and the RDE model are excluded by the current observations.

\

\section{Concluding Remarks}
\label{sec:10}

Since discovered in 1998, DE has become one of the biggest puzzles in theoretical physics and modern cosmology.
In essence, the DE problem may be an issue of quantum gravity.
Therefore, as the most fundamental principle of quantum gravity,
HP may play an important role in solving the DE problem.

In this review article, based on the HP and the dimensional analysis, we derived the general formula of HDE $\rho_{de} = 3C^2 M_p^2L^{-2}$.
Then, we described the idea of HDE, in which the future event horizon is chosen as the characteristic length scale,
i.e. $L=a\int_{t}^{\infty}\frac{dt'}{a}=a\int_{a}^{\infty}\frac{da'}{Ha'^{2}}$.
This model has an EoS $w = -{1\over 3}-{2\sqrt{\Omega_{de}}\over 3C}$,
and provides an explanation of the coincidence problem via inflation.
We discussed several theoretical mechanisms of explaining the origin of HDE,
such as entanglement entropy, holographic gas, Casmir energy, entopic force and action principle.
In addition, we showed that the current observational data mildly favor the case of $C < 1$,
which corresponds to a phantom universe with big rip.

In the framework of HDE, we discussed various topics,
such as spatial curvature, neutrino, instability of perturbation, time-varying gravitational constant, inflation, black hole and big rip singularity.
In addition, we introduced the studies of exploring the interaction between DM and HDE,
from both the theoretical and the observational aspects.
The current cosmological observations do not support the existence of DM/DE interaction in the HDE cosmology.

We also discussed the HDE scenario in various MG theories,
such as Brans-Dicke theory, braneworld theory, scalar-tensor theory, Horava-Lifshitz theory, and so on.
Besides, we showed that HDE can be used to reconstruct various scalar-field DE and MG models.

Moreover, we introduced other DE models inspired by the HP, such as ADE, RDE, and so on.
We also made comparisons among these HP-inspired DE models by using cosmological observations and diagnostic tools,
and showed that the original HDE model is more favored by the current observational data.

Finally, let us conclude with outlook for the future developments of HDE and related research area.

Theoretically, HP is a fast-developing research direction in string theory and related areas. The recent theoretical developments in holographic entanglement entropy has refreshed our understanding of HP in the context of AdS/CFT -- spacetime may be understood as emergent from the entanglement entropy of the dual theory (see \cite{Ryu:2006bv} for original work and \cite{Rangamani:2016dms} for a review of recent development). It remains interesting to see how the refreshed understanding of HP in string theory would bring new insight to the study of HDE. There have also been many new ideas and proposals to apply HP to cosmology (see, for example, \cite{Verlinde:2016toy, Bao:2017iye}). Those new theoretical ideas and technologies may enable more understandings of HDE, including the origin of HDE, theoretical prediction of $C$, detailed study of HDE perturbations and a more predictable fate of our universe.

Observationally, as is well known, astronomical observation is the key to determine the nature of DE.
Moreover, the future for DE observations is very exciting.
In table \ref{tab:10}, we list some most representative DE projects of Stage IV, which is classified by the DE Task Force \cite{2006astro.ph..9591A}.

\begin{table}
\caption{Future Larger-Scale DE Projects.}
\begin{center}
\label{tab:10}
\begin{tabular}{cccc}
  \hline\hline
  ~~~Survey~~~ & ~~~Location~~~ & ~~~Description~~~ & ~~~Probes~~~ \\
  \hline
  ~~~LSST~~~ & ~~~Cerro Pachon (Chile)~~~ & ~~~Optical, $8.4m$~~~ & ~~~SN, BAO, WL~~~ \\
    \hline
  ~~~SKA~~~ & ~~~Australia and South Africa~~~ & ~~~Radio, $km^2$~~~ & ~~~BAO, WL~~~ \\
      \hline
  ~~~Euclid~~~ & ~~~Sun-Earth L2 orbit~~~ & ~~~Optical/NIR, $1.2m$~~~ & ~~~BAO, WL~~~ \\
        \hline
  ~~~WFIRST~~~ & ~~~Sun-Earth L2 orbit~~~ & ~~~Infrared, $2.4m$~~~ & ~~~SN, BAO, WL~~~ \\
  \hline\hline
\end{tabular}
\end{center}
\end{table}

Large Synoptic Survey Telescope (LSST) \cite{2012arXiv1211.0310L} is an 8.4 meter ground-based optical telescope,
which is located in Cerro Pachon of Chile.
Square Kilometre Array (SKA) \cite{Johnston:2008hp} is a global next-generation radio telescope
that will be built in Australia and South Africa.
Euclid \cite{2011arXiv1110.3193L} is a 1.2 meter optical and near-infrared space telescope,
which is a European space agency mission selected for launch in 2020.
In addition, Wide Field Infrared Survey Telescope (WFIRST) \cite{2013arXiv1305.5422S} is a 2.4 meter infrared space telescope
that is a national aeronautics and space administration mission selected for launch in mid-2020s.
In the future ten years,
these projects will investigate the expansion history of the universe and the growth of LSS with unprecedented accuracy.
This means that cosmology will enter in a very interesting and challenging era.

\

\section*{Acknowledgement}

We are grateful to prof. Marc Kamionkowski for supporting us to write this review.
We also thank the referee for a large number of useful suggestions those help us to improve this work.
We would like to thank Yi-Fu Cai and Rong-Xin Miao for very helpful discussions.
SW is supported by the National Natural Science Foundation of China under Grant No. 11405024
and the Fundamental Research Funds for the Central Universities under Grant No. 16lgpy50.
YW is supported by grants HKUST4/CRF/13G and ECS 26300316 issued by the Research Grants Council of Hong Kong.
ML is supported by the National Natural Science Foundation of China (Grant No. 11275247, and Grant No. 11335012)
and a 985 grant at Sun Yat-Sen University.

\

\

\bibliographystyle{elsarticle-num}
\bibliography{ref}

\end{document}